\documentclass[useAMS,usenatbib,babel]{mn2e}
\usepackage[usenames,dvipsnames]{color}

\usepackage[english,english]{babel}
\usepackage{amsmath}
\usepackage{amssymb,amsfonts,textcomp}
\usepackage{array}
\usepackage{supertabular}
\usepackage{hhline}
\usepackage{hyperref}
\usepackage[usenames]{color}
\hypersetup{dvips, colorlinks=true, linkcolor=black, citecolor=black, filecolor=blue, urlcolor=blue}
\usepackage[dvips]{graphicx}

\author[Y. Dubois et al. ]{
\parbox[t]{\textwidth}{
Yohan Dubois$^{1,2}$\thanks{E-mail: dubois@iap.fr}, Rapha\"el Gavazzi$^{1}$, S\'ebastien Peirani$^{1}$ and Joseph Silk$^{1,2,3}$ }
\vspace*{6pt} \\
$^{1}$ Institut d'Astrophysique de Paris, UMR 7095, CNRS, UPMC Univ. Paris VI, 98 bis boulevard Arago, 75014 Paris, France\\
$^{2}$ Sub-department of Astrophysics, University of Oxford, Keble Road, Oxford OX1 3RH\\
$^{3}$ Department of Physics and Astronomy, The Johns Hopkins University Homewood Campus, Baltimore, MD 21218, USA\\
}
\date{Accepted 2013 June 1.  Received 2013 April 24; in original form 2013 January 14}

\title[Shaping ETGs with AGN feedback]
{AGN-driven quenching of star formation: morphological and dynamical implications for early-type galaxies}

\pagerange{\pageref{firstpage}--\pageref{lastpage}} \pubyear{2010}

\def\LaTeX{L\kern-.36em\raise.3ex\hbox{a}\kern-.15em
    T\kern-.1667em\lower.7ex\hbox{E}\kern-.125emX}

\begin{document}

\label{firstpage}

\maketitle

\begin{abstract}
In order to understand the physical mechanisms at work during the formation of massive early-type galaxies, we performed six zoomed hydrodynamical cosmological simulations of halos in the mass range $4.3 \times 10^{12} \le M_{\rm vir} \le 8.0 \times 10^{13}\, \rm M_\odot$ at $z=0$, using the Adaptive Mesh Refinement code {\sc ramses}. 
These simulations explore the role of Active Galactic Nuclei (AGN), through jets powered by the accretion onto supermassive black holes on the formation of  massive elliptical galaxies.
In the absence of AGN feedback, large amounts of stars accumulate in the central galaxies to form overly massive, blue, compact and rotation-dominated galaxies.
Powerful AGN jets transform the central galaxies into red extended and dispersion-dominated galaxies.
This morphological transformation of disc galaxies into elliptical galaxies is driven by the efficient quenching of the in situ star formation due to AGN feedback, which transform these galaxies into systems built up by accretion.
For galaxies mainly formed by accretion, the proportion of stars deposited farther away from the centre increases, and galaxies have larger sizes.
The accretion is also directly responsible for randomising the stellar orbits, increasing the amount of dispersion over rotation of stars as a function of time.
Finally, we find that our galaxies simulated with AGN feedback better match the observed scaling laws, such as the size-mass, velocity dispersion-mass, fundamental plane relations, and slope of the total density profiles at $z\sim0$,  from dynamical and strong lensing constraints.
\end{abstract}


\section{Introduction}

Early-type galaxies (ETG) are amongst the most massive galaxies observed in our Universe.
They are red, almost dead, and ellipsoidal-shaped objects lying in the densest regions of the Universe at redshift zero, predominantely in groups and clusters of galaxies.

In the standard $\Lambda$CDM paradigm, progenitors of massive collapsed structures such as groups and clusters of galaxies form at high redshift by the accretion of cold filamentary gas.
At low redshift, the gas cooling rate at the virial radius becomes insufficient to efficiently evacuate the energy of infall transformed into heat due to shocks, and leads to the formation of large, diffuse and hot quasi-spherical regions of gas around bright galaxies~\citep{rees&ostriker77, white&rees91, birnboim&dekel03, keresetal05, ocvirketal08, dekeletal09, vandevoort11}.
However, in the absence of any strong source of feedback from galaxies,  theory fails to produce this population of red massive galaxies~\citep[e.g.][]{boweretal06}, as the gas cooling flows proceed uninterrupted in the cores of halos.
This leads to overly massive and actively star-forming galaxies at redshift zero, in stark contradiction with observations.

Supermassive black holes (BHs) are commonly observed in the centres of galaxies with bulges.
The BH mass $M_{\rm BH}$ scales with its host bulge stellar mass $M_{\rm b}$, and stellar velocity dispersion $\sigma_{\rm b}$~\citep{magorrianetal98, tremaineetal02, haring&rix04, gueltekinetal09, graham&scott13}.
This has led several authors to suggest that a self-regulation process is taking place between BHs and their host galaxy~\citep{silk&rees98, king03, wyithe&loeb03}, as accretion onto supermassive BHs can release tremendous amounts of energy that can potentially drive large-scale outflows, and quench the star formation.
This mechanism is also supported by the detection of strong Active Galactic Nuclei (AGN) activity in groups and clusters in the form of radio jets~\citep[e.g.][]{boehringeretal93}, or in the distant Universe as seen through quasar spectra~\citep[e.g.][]{chartasetal03}.

Semi-analytical models of galaxy formation using cosmological N-body simulations have demonstrated that AGN feedback can be responsible for creating a population of red and dead massive galaxies~\citep{crotonetal06, boweretal06, cattaneoetal06, somervilleetal08}.
Hydrodynamical cosmological simulations which are dynamically and spatially resolved where AGN feedback sub-grid prescriptions are implemented also confirm that this mechanism is able to self-regulate the baryon content in massive halos to get realistic galaxy masses and colours~\citep{sijacki&springel06, puchweinetal08, khalatyanetal08, mccarthyetal10, mccarthyetal11, duboisetal10, duboisetal11, teyssieretal11}, and that it can reproduce the relatively tight $M_{\rm BH}-M_{\rm b}$, $M_{\rm BH}-\sigma_{\rm b}$ relationships~\citep{sijackietal07, dimatteoetal08, booth&schaye09, duboisetal12}.

Studies of strong gravitational lensing by massive ETGs at low redshift ($z\sim 0.2$) have provided a wealth of information about their mass distribution thanks to sizeable sample of lenses that have been built up in the last few years. 
The most noticeable result from the SLACS survey \citep[e.g.][]{koopmansetal09, augeretal10} is that the density profile within  one half of the projected effective radius is found to be close to isothermal ($\rho \propto r^{-2}$) with little ($\sim10\%$) scatter around this value (see also~\citealp{rusinetal03}).
It is noteworthy to mention that these conclusions given by strong lenses apply to low ($z\sim0.2$) redshift ETGs since most of this work builds on the SDSS survey.
More recent results have suggested that the isothermal profile is also present at higher redshift ($z\sim 0.5$) but flatter than at low redshift~\citep{ruffetal11, boltonetal12}.

Deep observations of distant galaxies have revealed that massive galaxies were more compact in the past~\citep{daddietal05, trujilloetal06} than they are today.
These observations challenge the scenario where galaxies are built from a monolithic collapse, and favour a scenario where mergers drive the formation of massive ETGs at $z=0$.
The size increase of these objects due to mergers is expected in semi-analytical theory and in numerical simulations~\citep{khochfar&silk06, boylankolcinetal06, malleretal06, naabetal06, naabetal07, delucia&blaizot07, bournaudetal07, guo&white08, hopkinsetal09, nipotietal09, feldmannetal10, shankaretal13}, and is supported by the direct evidence of merger remnants of what are the progenitors of today's most massive galaxies.
Major mergers have been proposed as a solution  for the evolution of the sizes of galaxies~\citep[e.g.][]{khochfar&silk06size}, however they are rare events (one major merger on average since $z=2$) that cannot explain  the rapid size-mass evolution with redshift~\citep[e.g.][]{bundyetal09}.
Dry minor mergers seem to be appropriate candidates for growing the galaxy sizes as fast as found in the observations~\citep{naabetal09, lackner&ostriker10, oseretal10, oseretal12, hilzetal12}.

Hydrodynamical cosmological simulations following the formation of elliptical galaxies by~\cite{oseretal10}, using Smoothed Particle Hydrodynamics (SPH), have shown that these galaxies can be formed even in the absence of AGN feedback, because the fraction of accreted stellar mass dominates the fraction of in situ-formed stars for low redshift massive galaxies.
This result  has been challenged recently by~\cite{lackneretal12}, using Adaptive Mesh Refinement (AMR), where they find that in the absence of AGN feedback, in-situ star formation dominates at all times, even at $z=0$.
The aim of this paper is to probe the role of AGN feedback on  galaxy morphology, and on the  history of mass accretion onto galaxies.

The paper is organised as follow.
In section~\ref{section:simulations}, we describe our numerical model for galaxy formation and our zoom initial conditions.
In section~\ref{section:results}, we detail the results obtained with our simulated galaxies and the particular role of AGN feedback on the mass build-up of massive ETGs, and we check the consistency of dynamical properties with observations.
In section~\ref{section:comparison}, we compare our work to previous numerical studies solving for gas hydrodynamics. 
Finally, in section~\ref{section:conclusion}, we summarise the main results of this work and discuss the possible implications.

\section{Simulation set-up}
\label{section:simulations}

\begin{table*}
\caption{Simulations performed with different sub-grid galactic models and different resolutions. (a) Name of the simulation. (b) Halo virial mass at $z=0$. (c) Mass resolution of the high resolution DM particles. (d) Stellar mass of the central galaxy at $z=0$. (e) Minimum cell size. (f) Star formation efficiency. (f) Presence of AGN feedback (Y), or without AGN feedback (N). (h) g$-$r relative colour magnitude of the central ETG accounting for dust extinction. (i) g$-$r$^*$ relative colour magnitude of the central ETG  without dust extinction. (j) Star Formation Rate (SFR) of the central galaxy at $z=0$. }
\label{tab:names}
\begin{tabular}{@{}|c|c|c|c|c|c|c|c|c|c|}
  \hline
  (a) & (b) & (c) & (d) & (e) & (f) & (g) & (h) & (i) & (j)\\
  Name & $M_{\rm vir}$ & $M_{\rm res, DM}$ & $M_{\rm *}$ & $\Delta x$ & $\epsilon_*$ & AGN & g$-$r & g$-$r$^*$ & SFR \\
   &($10^{13}\,\rm M_{\odot}$) & ($10^7\, \rm M_{\odot}$) & ($10^{11}\, \rm M_{\odot}$) & ($\rm kpc$) & ($\%$) & & & & ($\rm M_\odot . yr^{-1}$)\\
  \hline
  G1 & $0.43$ & $1.0$ & $5.6$ & 0.5 &  2 & N & 0.63 & 0.44 & 20\\
  G1A & $0.38$ & $1.0$ & 0.7 & 0.5 &  2 & Y & 0.71 & 0.68 & 0.2 \\
  \hline
  G2 & $0.63$ & $1.0$ & 7.3 & 0.5 & 2 & N & 0.66 & 0.49 & 20 \\
  G2A & $0.53$ & $1.0$ & 0.6 & 0.5 &  2 & Y & 0.73 & 0.73 & $<0.1$ \\
  \hline
  G3 & $1.0$ & $1.0$ & 13 & 0.5 &  2 & N & 0.62 & 0.46 & 50\\
  G3A & $1.0$ & $1.0$ & 2.3 & 0.5 &  2 & Y & 0.73 & 0.70 & 0.4 \\
  \hline
  G4 & $1.7$ & $1.0$ & 20 & 0.5 &  2 & N & 0.67 & 0.48 & 50 \\
  G4A & $1.5$ & $1.0$ & 3.2 & 0.5 &  2 & Y & 0.73 & 0.64 & 2  \\
 \hline
  G5 & $2.4$ & $1.0$ & $27$ & 0.5 &  2 & N & 0.59 & 0.42 & 100\\
  G5A & $2.3$ & $1.0$ & 5.4 & 0.5 &  2 & Y & 0.71 & 0.60 & 4 \\
  \hline
  G6 & $8.0$ & $8.2$ & $81$ & 1.1 &  2 & N & 0.66 & 0.45 & 250 \\
  G6A & $8.0$ & $8.2$ & $13$ & 1.1 &  2 & Y & 0.72 & 0.61 & 10 \\
  \hline
\end{tabular}
\end{table*}

\subsection{Initial conditions and simulation parameters}

We assume a flat $\Lambda$CDM cosmology with total matter (baryons$+$DM) density $\Omega_{m}=0.272$, baryon density $\Omega_b=0.045$, dark energy density $\Omega_{\Lambda}=0.728$, fluctuation amplitude at $8 \, h^{-1}.\rm Mpc$ $\sigma_8=0.809$ and Hubble constant $H_0=70.4\, \rm km.s^{-1}.Mpc^{-1}$ consistent with WMAP 7-year data \citep{komatsuetal11}.

The simulations are run with the AMR code {\sc ramses} \citep{teyssier02}. The evolution of the gas is followed using a second-order unsplit Godunov scheme for the Euler equations. The Riemann solver used to compute the flux at the cell interface is the HLLC solver~\citep{toroetal94} and a first-order MinMod Total Variation Diminishing scheme is applied to reconstruct the interpolated variables from their cell-centred values. Collisonless particles (DM, stellar and sink particles) are evolved using a particle-mesh solver with a Cloud-In-Cell interpolation.

Two different box sizes $L_{\rm box}=100\, h^{-1}.\rm Mpc$ and $50\, h^{-1}.\rm Mpc$ have been used to generate the initial conditions. Six groups of galaxies in the mass range $4.3 \times 10^{12} \le M_{\rm vir} \le 8.0 \times 10^{13}\, \rm M_\odot$ have been selected to be resimulated at high resolution using the zoom technique, with only high resolution DM particles ending up in the virial radius of the halos at $z=0$.
Simulations are allowed to refine the initial mesh up to 7 levels of refinement, which reach up to a $\Delta x=1.1$ and $0.5\, \rm kpc$ physical length, respectively, for our low and high resolution initial conditions ($M_{\rm DM,res}=8.2\times 10^7 \, \rm M_\odot$ and $1.0\times 10^7\, \rm M_\odot$). A cell is refined following a quasi-Lagrangian criterion: if more than 8 dark matter particles lie in a cell, or if the baryon mass exceeds 8 times the initial dark matter mass resolution.

Six halos are simulated, five out of the 50 $h ^{-1}.\rm Mpc$ simulation box at the highest resolution, and one (the most massive halo) taken out of the 100 $h ^{-1}.\rm Mpc$ simulation box at lower resolution.
These six halos are simulated with the same physics (gas cooling, star formation, metal cooling, SN feedback, see section~\ref{sec:galphysics}), and with one set without AGN feedback (GX simulations), and one set with AGN feedback as described in section~\ref{sec:agnmodel} (GXA simulations).
Halo masses, DM particle mass resolution, and minimum cell sizes are summarised in table~\ref{tab:names} together with some of the basic properties of the central galaxy at $z=0$ (mass, colour, star formation rate).

\subsection{Physics of galaxy formation}
\label{sec:galphysics}

Gas is allowed to radiate energy away and cool down to a minimum temperature of $T_0=10^4$ K due to atomic collisions in a gas of  primordial composition, i.e. only  H and He~\citep{sutherland&dopita93}.
We also account for the cooling enhancement due to the presence of metals released during Supernovae (SNe) explosions from massive stars.
Metals are passively advected with the gas, and their distribution depends on the history of the gas enrichment through SNe explosions.
A fixed solar composition of heavy elements is assumed for the computation of the gas cooling rates.
Heating from a UV background is considered following \cite{haardt&madau96} during and after the redshift of reionisation $z_{reion}\sim 6$.

Star formation in our model occurs in regions with gas number density $n> n_0= 0.1\, \rm H.cm^{-3}$ using a random Poisson process to spawn star cluster particles, according to a Schmidt-Kennicutt law 
\begin{equation}
\dot \rho_*= \epsilon_* {\rho \over t_{\rm ff} (\rho)}\, ,
\end{equation}
where $\rho$ is the gas mass density, $\dot \rho_*$ the star formation rate mass density, $\epsilon_*$ the star formation efficiency, and $t_{\rm ff} (\rho)$ the gas fee-fall time.
In these simulations, we set the efficiency of star formation to a low value $\epsilon_*=0.02$ that is in good agreement with observational surface density relationships of galaxies \citep{kennicutt98}, and local giant molecular clouds \citep{krumholz&tan07}.
Each star cluster particle has a mass resolution of $m_*=\rho_0\Delta x^3$ ($\rho_0$ is the gas mass density threshold of star formation)  where a random Poisson process is used to create a new star particle~\citep{rasera&teyssier06, dubois&teyssier08winds}.
The stellar mass resolution reaches $5.1 \times10^5\, \rm M_{\odot}$ for our most resolved simulation with $\Delta x=0.5\, \rm  kpc$.
We ensure that no more than 90\% of the gas in a cell is depleted during the star formation process for numerical stability.

We account for the mass and energy release from type II SNe assuming a Salpeter Initial Mass Function (IMF).
Using this IMF, $10\%$ of the massive stars above $8\, \rm M_{\odot}$ end their life into a type II SN releasing $10^{51}$ erg per $10 \, \rm M_{\odot}$.
For each individual SN explosion, after 10 Myr, energy is released spherically in a kinetic mode transferring mass, momentum and energy with a mass loading parameter $f_w=1$ that mimic a Sedov blast wave explosion~\citep[see][for further details]{dubois&teyssier08winds}.
We assume that type II SNe release all their mass into the gas (no stellar remnant forms) with a $y=0.1$ stellar yield, which is the fraction of primordial gas transformed into heavy elements and released back into the gas.
With this prescription, we do not take into account the energy and mass release from stellar winds (AGB stars), nor from long-lived type Ia SNe.

The gas follows an adiabatic equation of state with an adiabatic index $\gamma_{\rm adiab}=5/3$.
In order to take into account the thermal impact of the heating of the gas by SNe, we modify the temperature at high gas density $\rho> \rho_0$ with a polytropic equation of state
\begin{equation}
T=T_0 \left ( {\rho \over \rho_0} \right )^{p-1}\, , 
\label{polytropic_EoS}
\end{equation}
where $p$ is the polytropic index of the gas.
The adopted value of $p=4/3$ is comparable to the value obtained in \cite{springel&hernquist03} considering the multiphase structure of the ISM with stellar heating.
We point out that this value of $p=4/3$ does not rigorously ensure that gas fragmentation is avoided due to numerical instabilities \citep{trueloveetal97}, as the Jeans length is proportional to the gas density
\begin{equation}
\lambda_{\rm J} = 10.7 \left( { \rho \over 0.1\, \rm H.cm^{-3} }\right )^{-1/3} \, \rm kpc.
\end{equation}
This last formula shows that at very high gas densities, the Jeans length can be smaller than our minimum resolution and would provide spurious fragmentation of the gas.
However, the gas cannot indefinitely condense because of the finite force resolution sampling and  the star formation process that removes gas in high density regions.

\begin{figure*}
  \centering{\resizebox*{!}{6.2cm}{\includegraphics{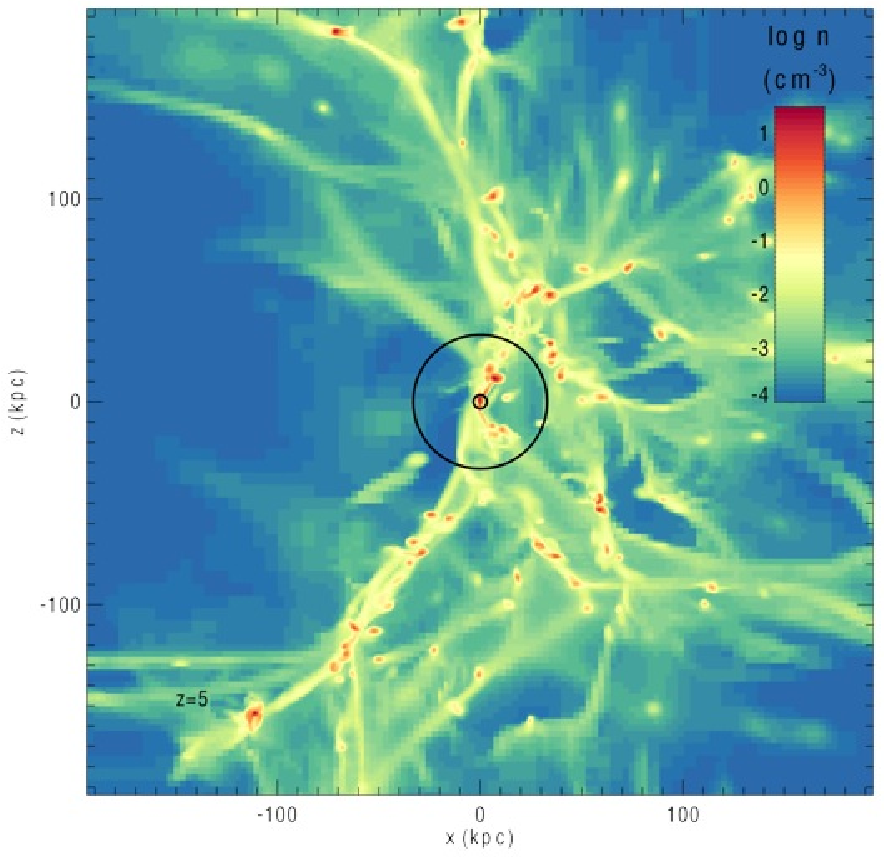}}}\hspace{-.6cm}
  \centering{\resizebox*{!}{6.2cm}{\includegraphics{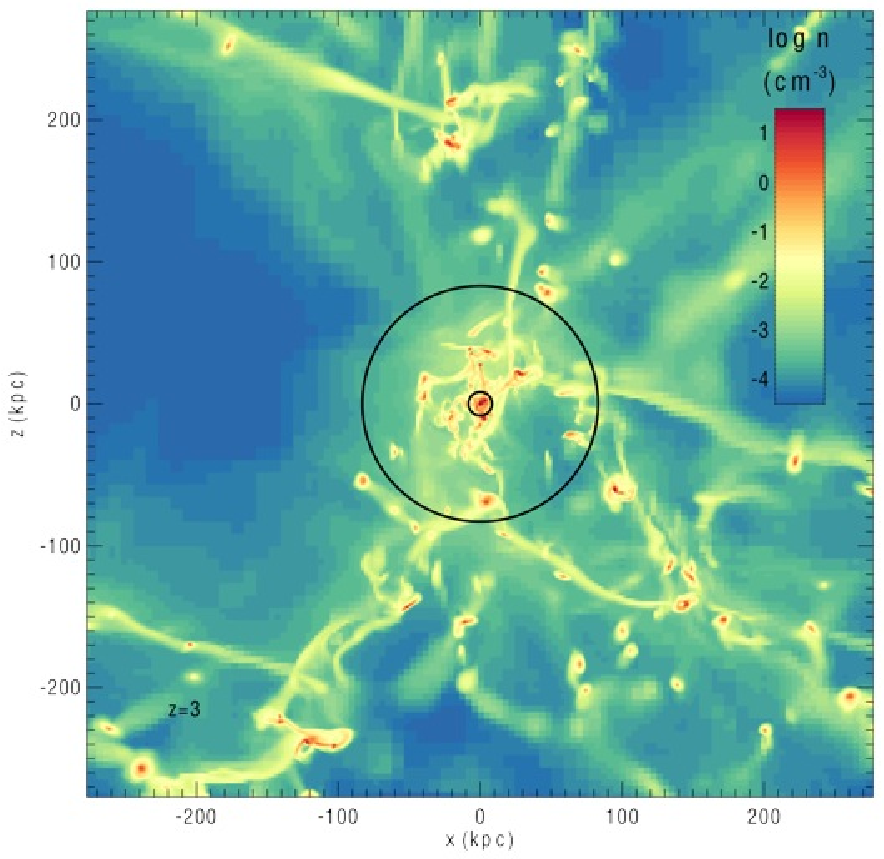}}}\hspace{-.6cm}
  \centering{\resizebox*{!}{6.2cm}{\includegraphics{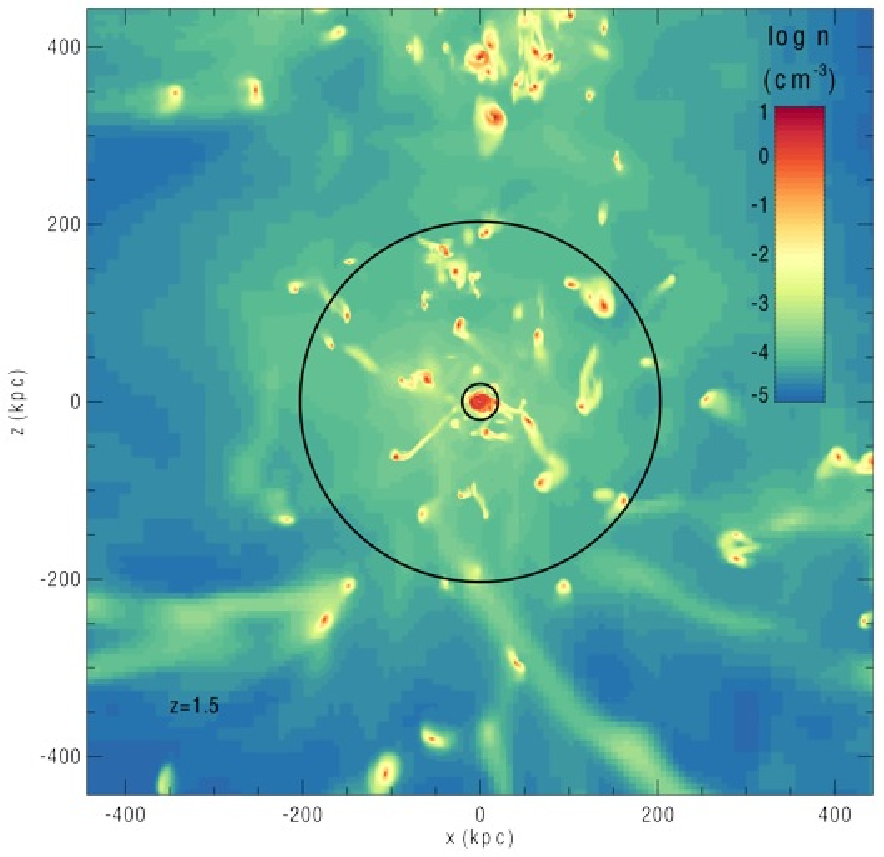}}}\vspace{-.6cm}
  \centering{\resizebox*{!}{6.2cm}{\includegraphics{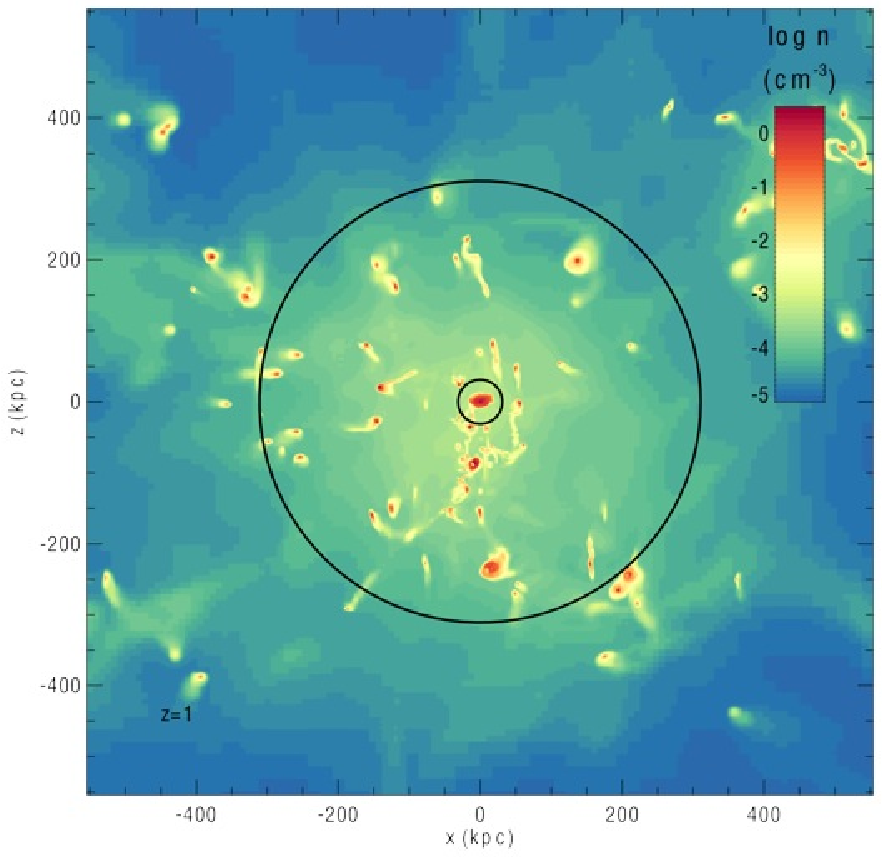}}}\hspace{-.6cm}
  \centering{\resizebox*{!}{6.2cm}{\includegraphics{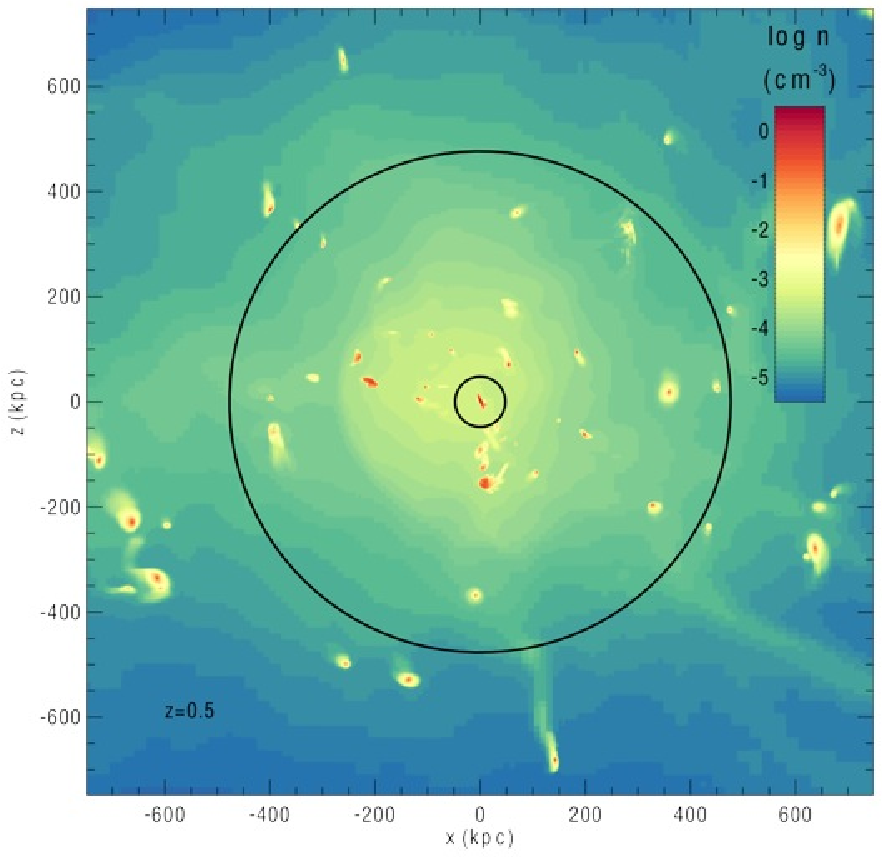}}}\hspace{-.6cm}
  \centering{\resizebox*{!}{6.2cm}{\includegraphics{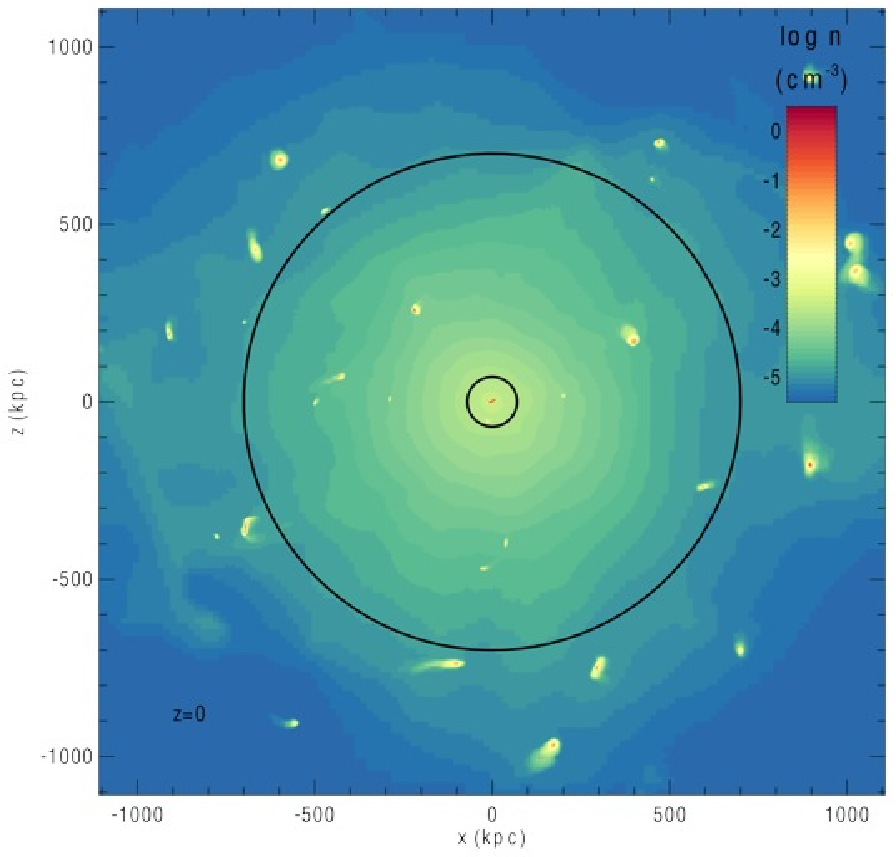}}}\vspace{-.6cm}
  \centering{\resizebox*{!}{9cm}{\includegraphics{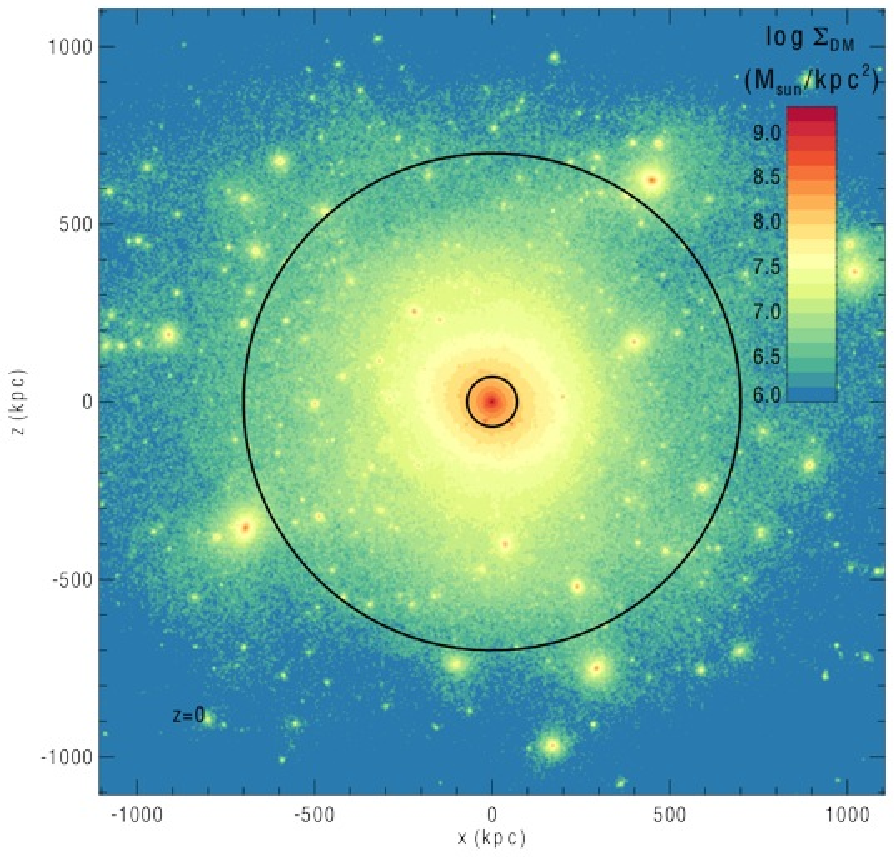}}}\hspace{-.7cm}
  \centering{\resizebox*{!}{9cm}{\includegraphics{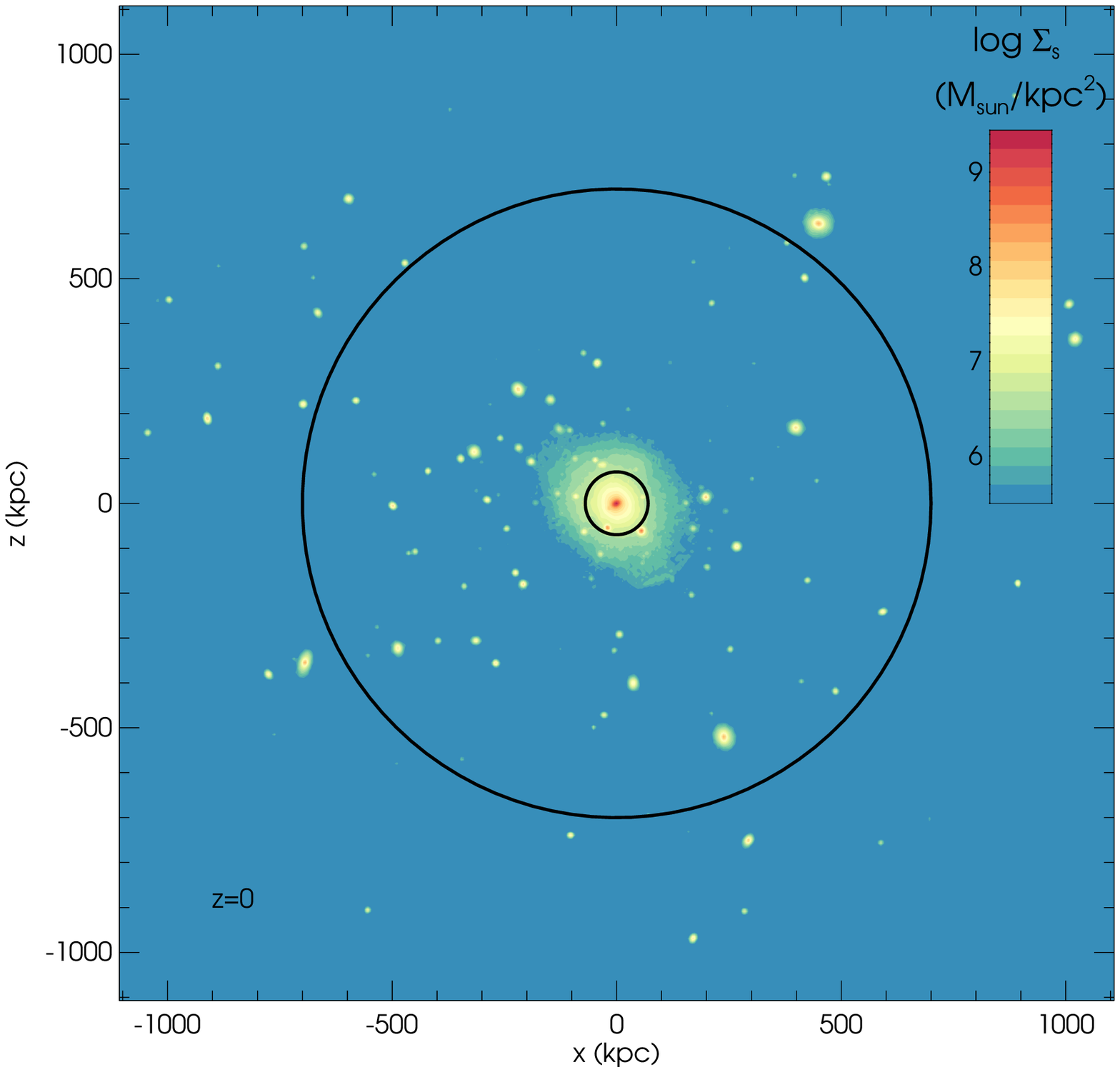}}}
  \caption{\emph{Top and middle rows}: Images of the projected gas number density at $z=5$ (top left), $z=3$ (top centre), $z=1.5$ (top right), $z=1$ (middle left), $z=0.5$ (middle centre), and $z=0$ (middle right) around the most massive galaxy of the G5A simulation; and \emph{bottom:} Images of the projected DM surface density (left), and stellar surface density (right) at $z=0$. The large circles correspond to the virial radius of the host halo and the small circles to 10 per cent of the virial radius. The image sizes are in physical length. The gas distribution around the central galaxy shows a complex filamentary network at high redshift, and a smooth spherical distribution of gas with a large number of satellites at intermediate and low redshift.}
    \label{fig:zoom}
\end{figure*}

\subsection{AGN feedback}
\label{sec:agnmodel}

Our modelling of AGN feedback closely follows the implementation detailed in~\cite{duboisetal12}.
 We recall here the main aspects of this modelling, and refer to~\cite{duboisetal10} and~\cite{duboisetal12} for the technical issues.
Black hole particles are modelled with the sink particles technique from~\cite{bateetal95}, and particularly~\cite{krumholzetal04} for mesh codes.
The BHs are created in the centres of galaxies, with a sufficiently large exclusion radius (100 kpc) to prevent formation of multiple massive BHs per galaxy.
The initial BH seed mass is chosen to be $10^5\,\rm M_\odot$.
BHs can merge when they are sufficiently close to one another ($<4 \Delta x$), determined using a Friend-Of-Friends algorithm on sink particles performed on-the-fly.

BHs can grow by accretion at a Bondi-Hoyle-Lyttleton rate \citep{bondi52}
\begin{equation}
\dot M_{\rm BH}={4\pi \alpha G^2 M_{\rm BH}^2 \bar \rho \over (\bar c_s^2+\bar u^2) ^{3/2}}\, ,
\label{dMBH}
\end{equation}
where $G$ is the gravitational constant, $\bar \rho$ the average gas mass density, $\bar c_s$ the average sound speed, $\bar u$ the average gas velocity relative to the BH, and $\alpha$ a dimensionless boost factor with $\alpha=(\rho/\rho_0)^2$ when $\rho>\rho_0$ and $\alpha=1$ in order to account for our inability to capture the cold and high density regions of the gas at these kpc resolutions that would substantially increase the accretion rate if these regions were resolved as suggested in~\cite{booth&schaye09}.

The accretion rate onto BHs is limited by its Eddington rate
\begin{equation}
\dot M_{\rm Edd}={4\pi G M_{\rm BH}m_{\rm p} \over \epsilon_{\rm r} \sigma_{\rm T} c}\, ,
\label{dMEdd}
\end{equation}
where $\sigma_{\rm T}$ is the Thompson cross-section, $c$ is the speed of light, $m_{\rm p}$ is the proton mass, and $\epsilon_{\rm r}$ is the radiative efficiency, assumed to be equal to $0.1$ for the \cite{shakura&sunyaev73} accretion onto a Schwarzschild BH.

A drag force is applied to the BH particles in order to replicate the action of dynamical friction of the gas on massive BHs that happens on  unresolved parsec scales. 
The dynamical friction force is defined as 
\begin{equation}
F_{\rm DF}=f_{\rm gas} 4 \pi \alpha \bar \rho \left({G M_{\rm BH}\over \bar c_s}\right)^2,
\label{dragforce}
\end{equation}
where $f_{\rm gas}$ is a fudge factor whose value is between 0 and 2 and is a function of the Mach number ${\mathcal M}=\bar u/\bar c_s<1$~\citep{ostriker99, chaponetal11}, and where $\alpha$ is the boost factor already introduced in equation~(\ref{dMBH}).
This extra force term for BH dynamics has been successfully introduced in high redshift simulations with $\sim 10$ pc resolution to maintain BHs close to the centre of the galaxy~\citep{duboisetal12agnhighz}.

In this paper, we assume that AGN feedback proceeds uniquely in the  jet mode powering large bipolar outflows.
The total energy released by the AGN is 
\begin{equation}
\dot E_{\rm AGN}=\epsilon_{\rm f} L_r=\epsilon_{\rm f} \epsilon_{\rm r} \dot M_{\rm BH}c^2\, ,
\label{E_BH}
\end{equation}
where $\epsilon_{\rm f}=1$ is a free parameter chosen to reproduce the $M_{\rm BH}$-$M_{\rm b}$, $M_{\rm BH}$-$\sigma_{\rm b}$, and BH mass density in our local Universe (see \citealp{duboisetal12}). 
For this jet mode of AGN feedback, mass, momentum and kinetic energy are continuously released in the surrounding gas along a jet axis defined by the local angular momentum of the gas, and such that the jet velocity is $10^4 \, \rm km. s^{-1}$, or equivalently the mass loading factor of the jet is $\eta=100$.

\section{Results}
\label{section:results}

\subsection{Changing the baryon content and the morphologies of massive galaxies}

\begin{figure}
  \centering{\resizebox*{!}{6.cm}{\includegraphics{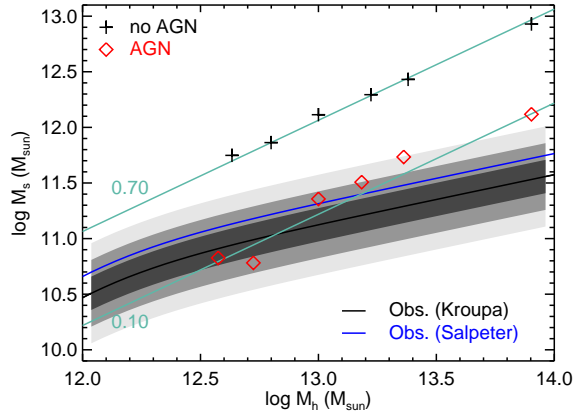}}}
  \caption{Stellar mass of central massive galaxies as a function of their halo mass at $z=0$ for the simulations without AGN (black), and for the simulations with AGN feedback (red). The solid curves are the observational fit from~\citet{mosteretal10}  either assuming a Kroupa (black) or a Salpeter (blue) IMF, with the 1, 2 and 3 $\sigma$ standard deviation (gray shaded areas). Cyan solid lines indicate the amount of stars formed at a constant efficiency $f_{\rm conv}=M_{\rm s} / f_{\rm b}M_{\rm h}$. The presence of AGN feedback reduces the total amount of stars in the central massive galaxy by a factor 7.}
    \label{fig:mstarvsmhalo}
\end{figure}

\begin{figure}
  \centering{\resizebox*{!}{6.cm}{\includegraphics{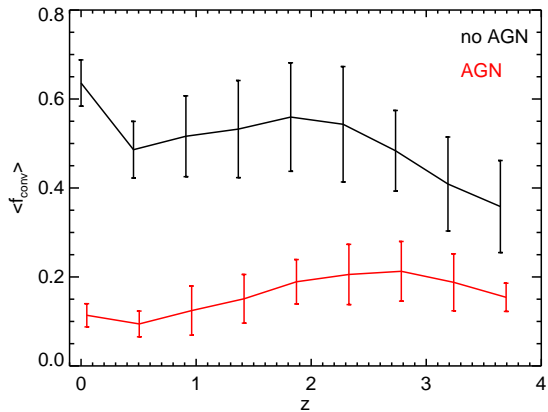}}}
  \caption{Average conversion efficiency of the central galaxies $f_{\rm conv}=M_{\rm s} / f_{\rm b}M_{\rm h}$ as a function of redshift for the simulations including AGN feedback (red) or without AGN feedback (black) with the standard deviation (error bars). The values are averaged over the main progenitor of the central galaxy for the six halos. The conversion efficiency of stars in the central galaxy is strongly reduced by the presence of AGN feedback and the reduction takes place already at high redshift (a factor 2 reduction at $z=4$).}
    \label{fig:fstarvsz}
\end{figure}

\begin{figure}
  \centering{\resizebox*{!}{6.cm}{\includegraphics{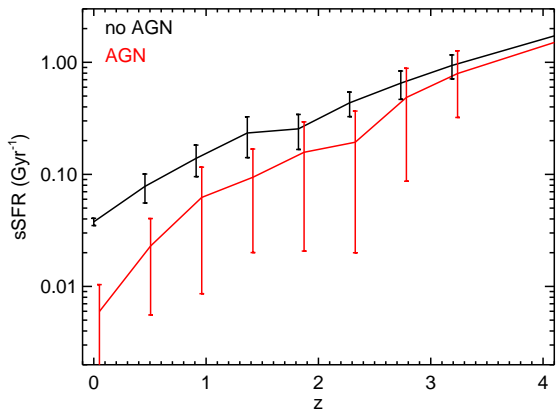}}}
  \caption{Average specific star formation rate as a function of redshift for the central galaxy of the six simulated halos with (red) or without (black) AGN feedback, with the standard deviation (error bars). The presence of AGN feedback diminishes the sSFR by a factor $\sim 5$ at $=z0$.}
    \label{fig:ssfrvsz}
\end{figure}

\begin{figure}
  \centering{\resizebox*{!}{3.25cm}{\includegraphics{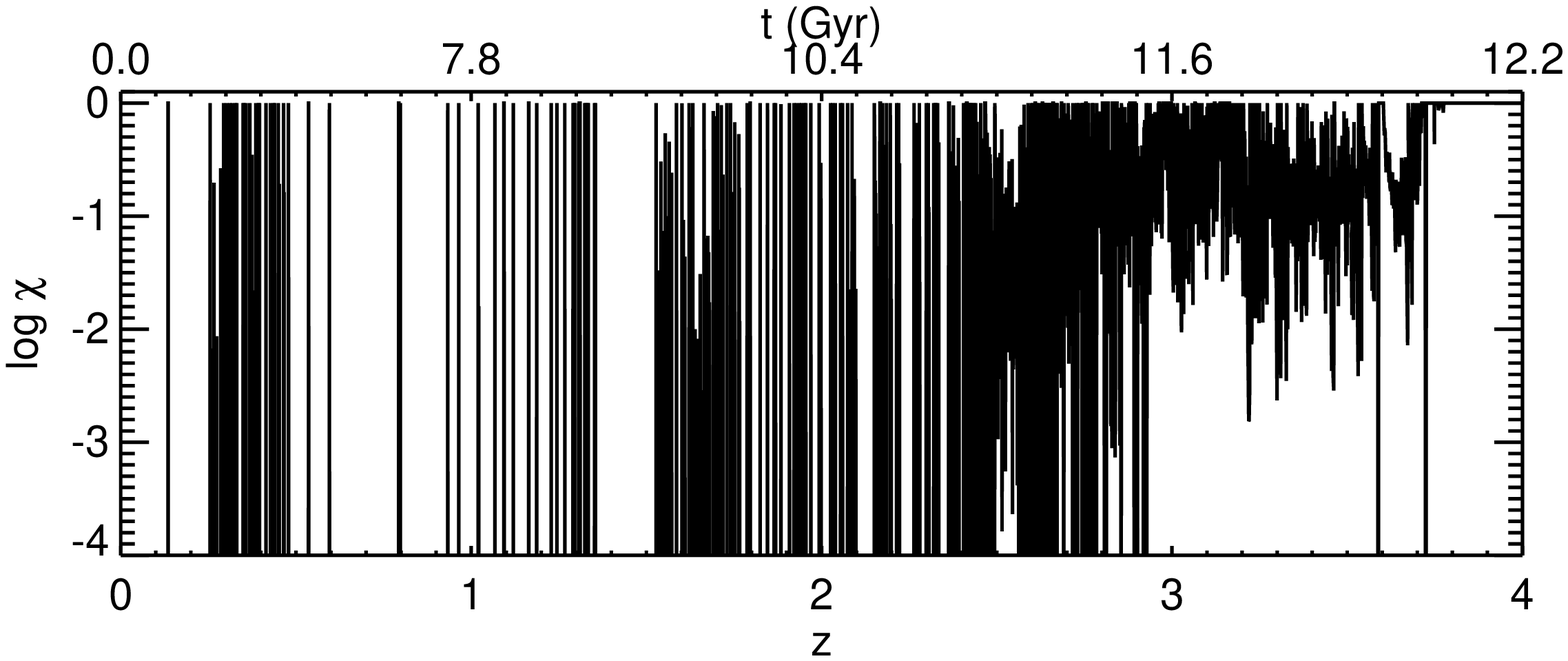}}}
  \centering{\resizebox*{!}{3.25cm}{\includegraphics{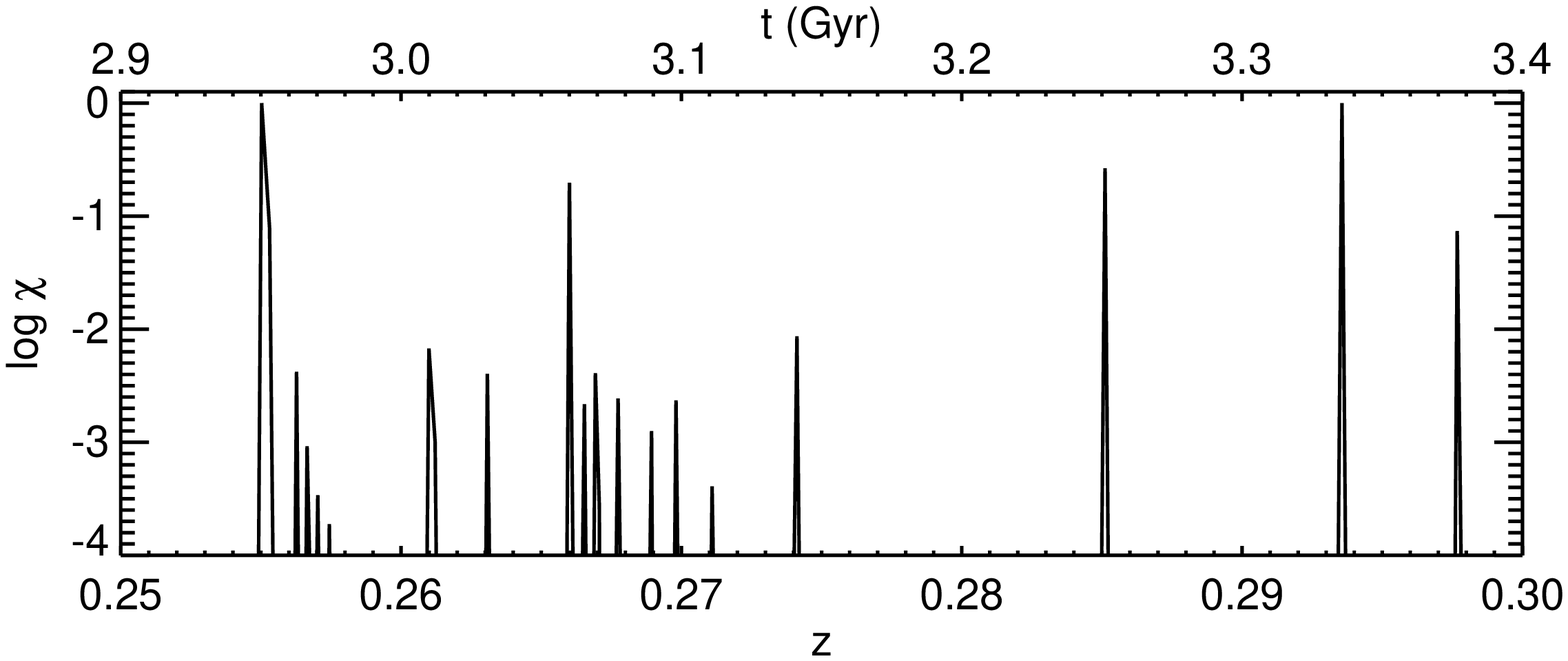}}}
  \caption{Instantaneous accretion rate over Eddington rate $\chi$ as a function of redshift (or lookback time) for the central BH of the G6A galaxy over the redshift range $z=0-4$ (top panel), and zoomed over the redshift range $z=0.25-0.3$. The Eddington ratio is maximum at high redshift with values close to $\sim1$, and with brief episodes of accretion onto BHs (i.e. AGN activity) at low redshift that help to regulate the cold baryon content in the central massive galaxy. }
    \label{fig:lumvsz}
\end{figure}

The basic picture of the formation of massive halos is that they form at high redshift by cold filamentary infall driving high levels of star formation, and at low redshift by a fainter diffuse (and hot) accretion, as illustrated for the halo G5A in Fig.~\ref{fig:zoom}~\citep{rees&ostriker77, white&rees91, birnboim&dekel03, keresetal05, ocvirketal08, dekeletal09, vandevoort11}.
For groups of galaxies, satellites represent a non-negligible fraction of the total stellar mass in the halo with the large number of satellites orbiting inside the virial radius of the halo.
Note that some of these satellites have no dense star-forming gas component.
This gas has been stripped away by the ram pressure of the hot diffuse gas as can be seen for some satellites with tails of gas.

Massive halos usually suffer from over-cooling of their gas content in absence of any strong feedback prescription. 
Vanilla hydrodynamical simulations (i.e. without any strong feedback) produce large amount of stars in massive central galaxies  as shown in Fig.~\ref{fig:mstarvsmhalo}.
The measured stellar mass $M_{\rm s}$ is given by the total stellar mass of the central galaxy as detected by the AdaptaHOP algorithm~\citep{aubertetal04, tweedetal09} performed on stars, and, thus, captures the stellar content in the core of the halo together with its related intra-cluster light, and removes all satellites (considered as ``sub-galaxies'' of the central most massive galaxy).
The halo mass $M_{\rm h}$ is the total mass of gas, stars, DM (and BHs if any) enclosed within the virial radius provided by the halo finder algorithm (using AdaptaHOP on DM particles). 
In the absence of AGN feedback, stars are converted in the central galaxy with a $f_{\rm conv}=0.7$ global efficiency, where $f_{\rm conv}=M_{\rm s} / f_{\rm b}M_{\rm h}$ and $f_{\rm b}=\Omega_{\rm b}/\Omega_{\rm m}=0.165$, while observations (e.g. from~\citealp{mosteretal10}) suggest that the value for groups of galaxies is of 0.1 (for a halo mass of $M_{\rm h}\simeq10^{13}\, \rm M_\odot$) and decreases with increasing halo mass.
AGN feedback manages to reduce the stellar mass of the central massive galaxies at $z=0$ by a factor 7 compared to the no AGN feedback case, but these galaxies still remain too massive, in particular even for those belonging to the most massive halos.

\begin{figure*}
  \centering{\resizebox*{!}{2.8cm}{\includegraphics{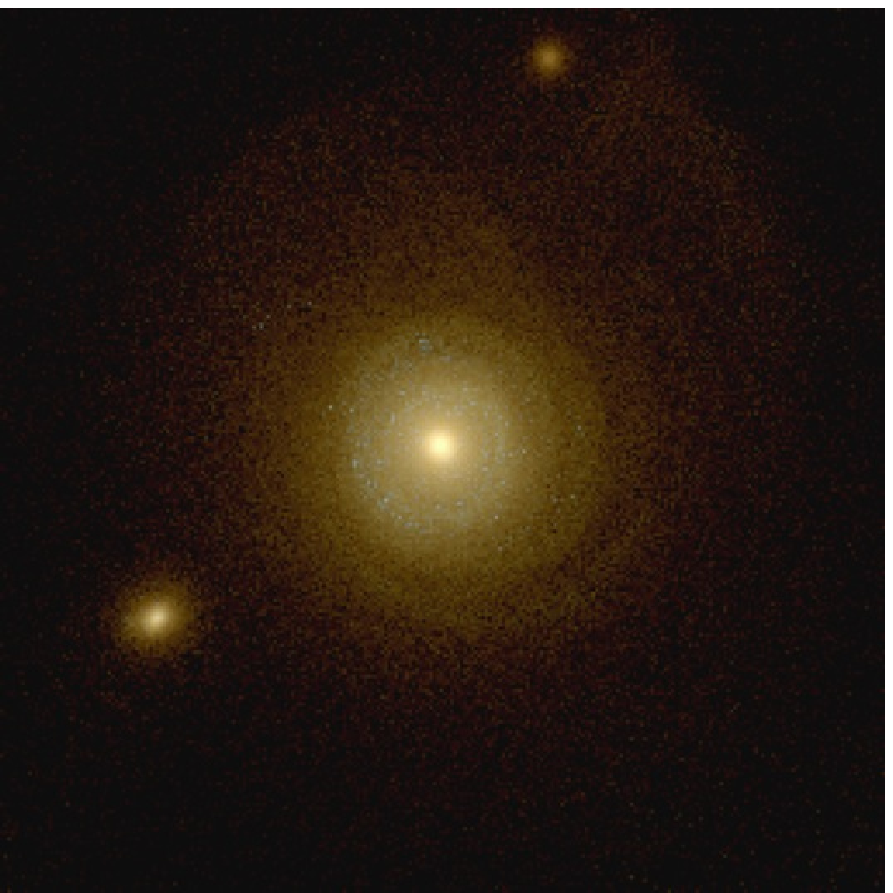}}}
  \centering{\resizebox*{!}{2.8cm}{\includegraphics{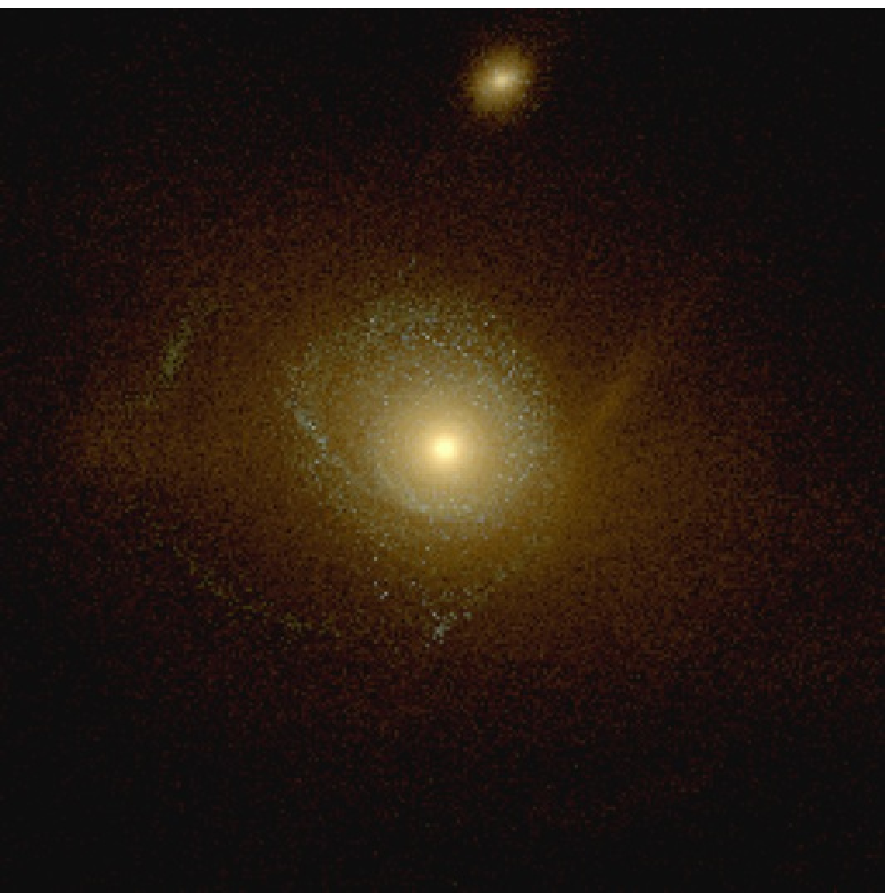}}}
    \centering{\resizebox*{!}{2.8cm}{\includegraphics{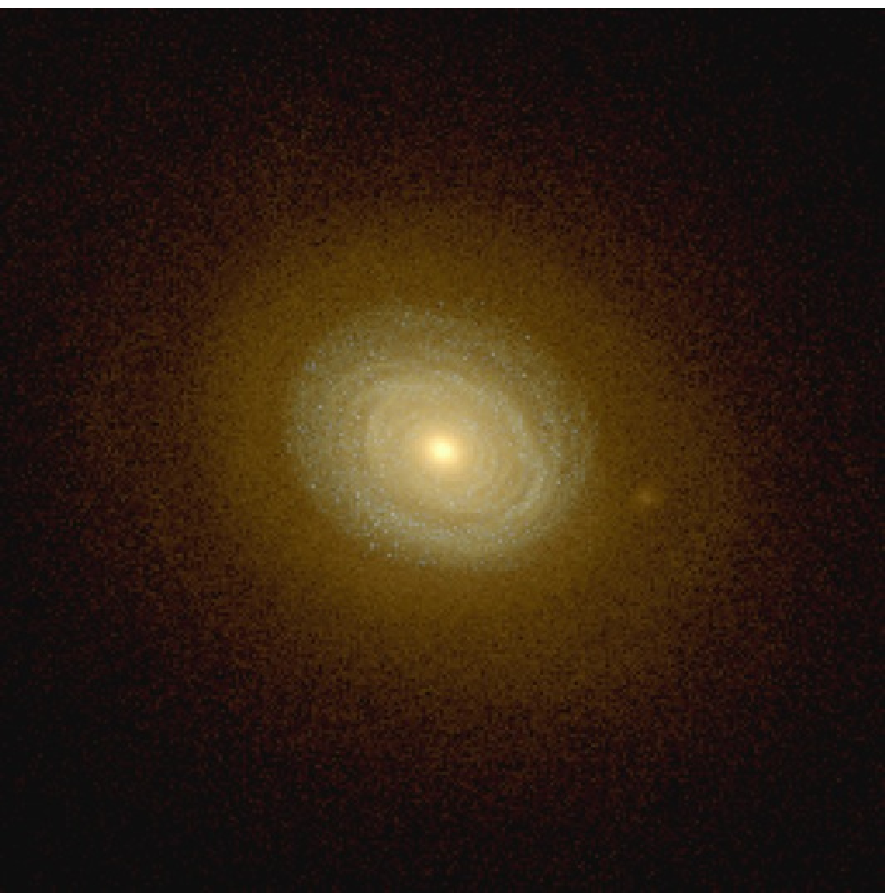}}}
    \centering{\resizebox*{!}{2.8cm}{\includegraphics{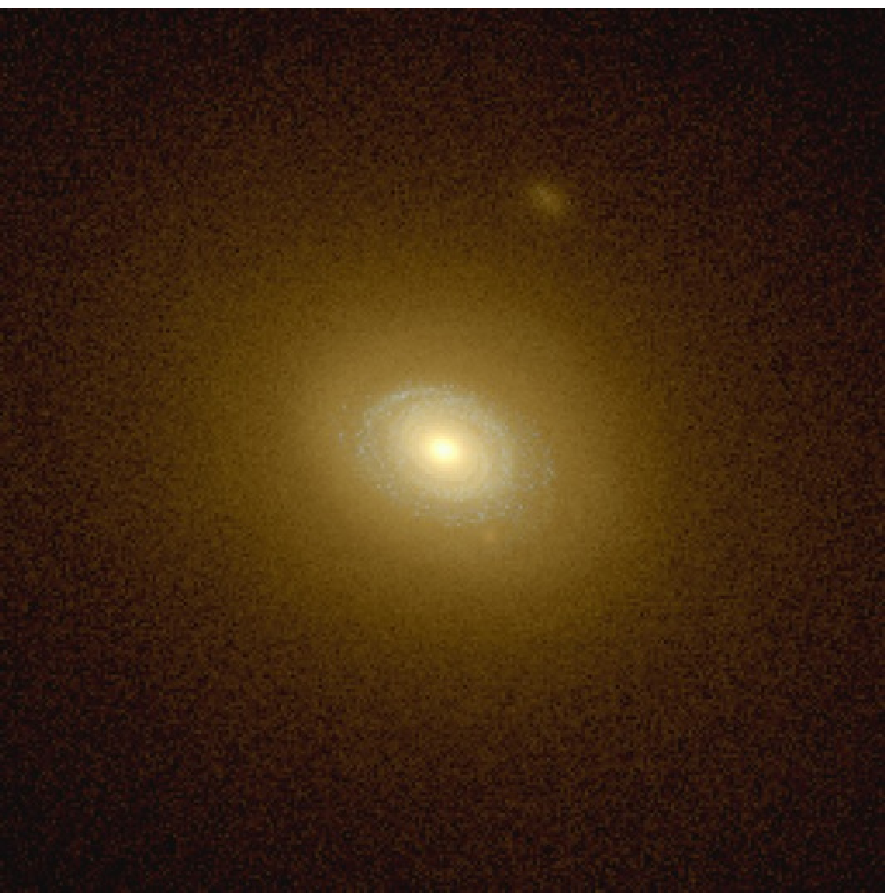}}}
    \centering{\resizebox*{!}{2.8cm}{\includegraphics{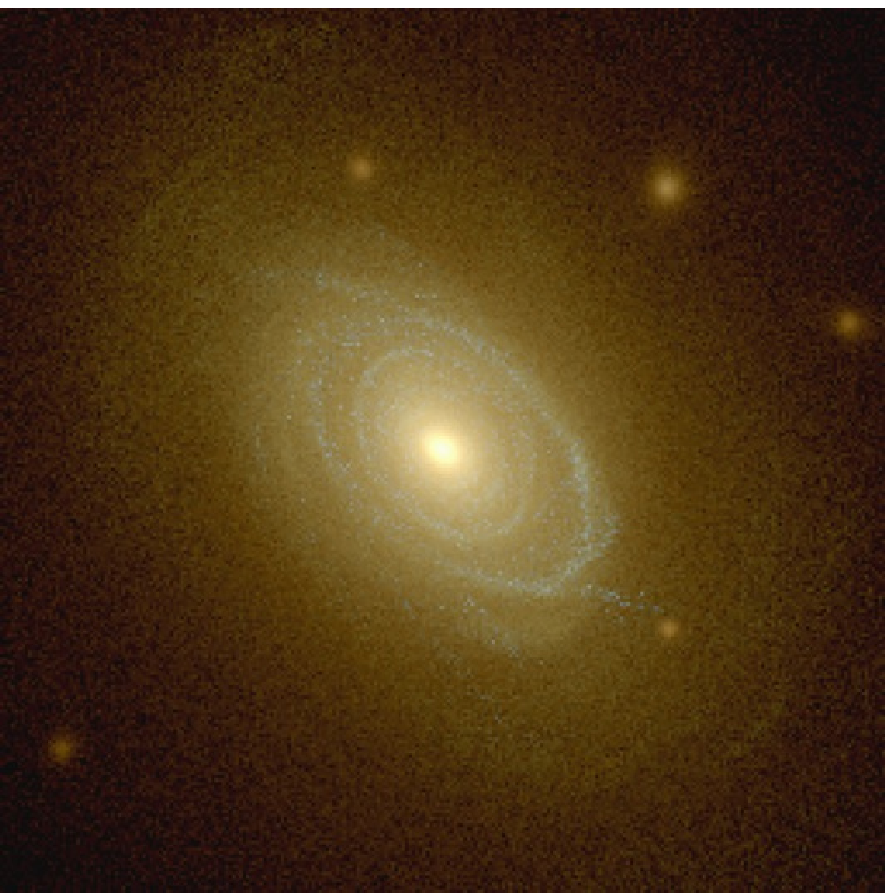}}}
    \centering{\resizebox*{!}{2.8cm}{\includegraphics{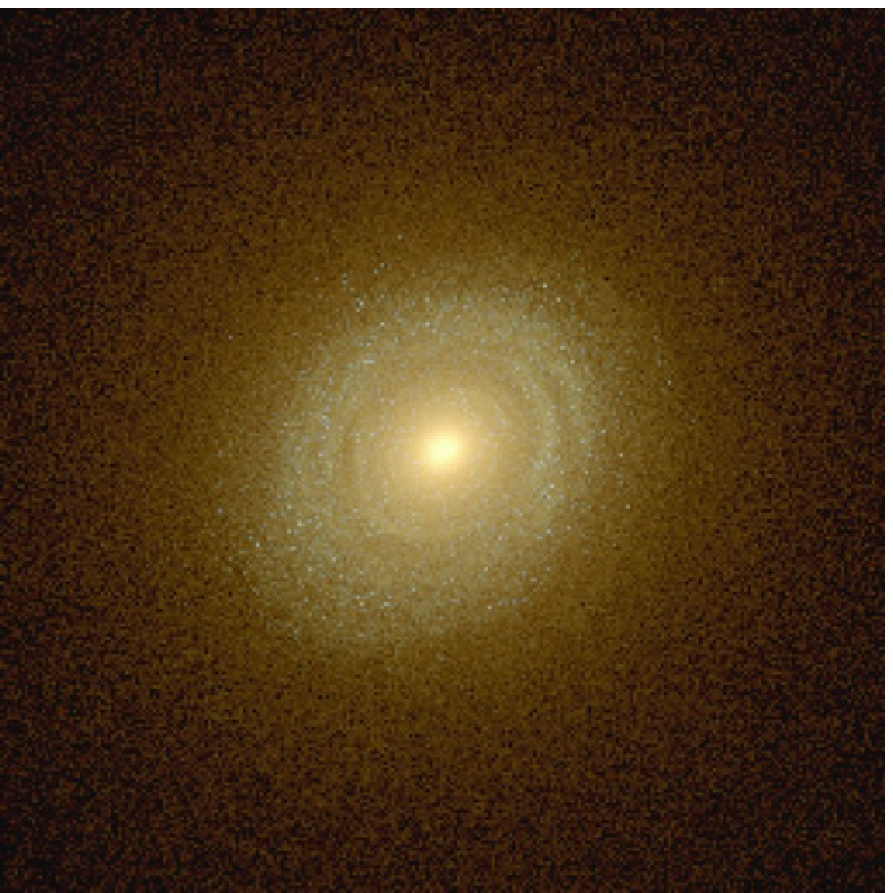}}}\\
  \centering{\resizebox*{!}{2.8cm}{\includegraphics{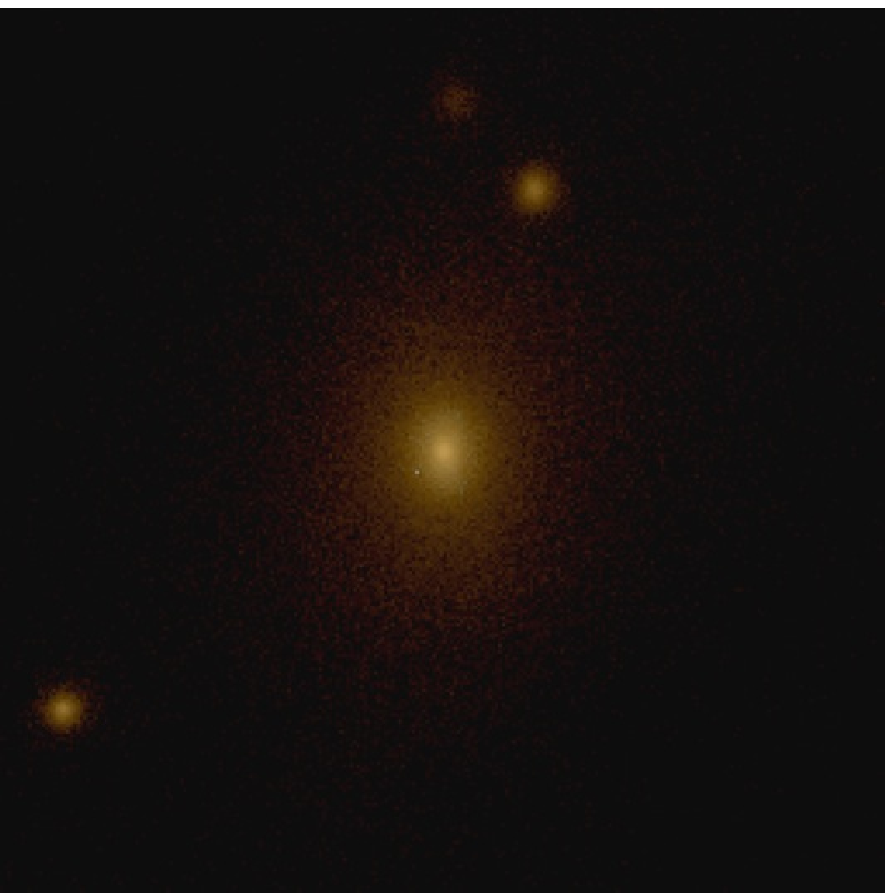}}}
  \centering{\resizebox*{!}{2.8cm}{\includegraphics{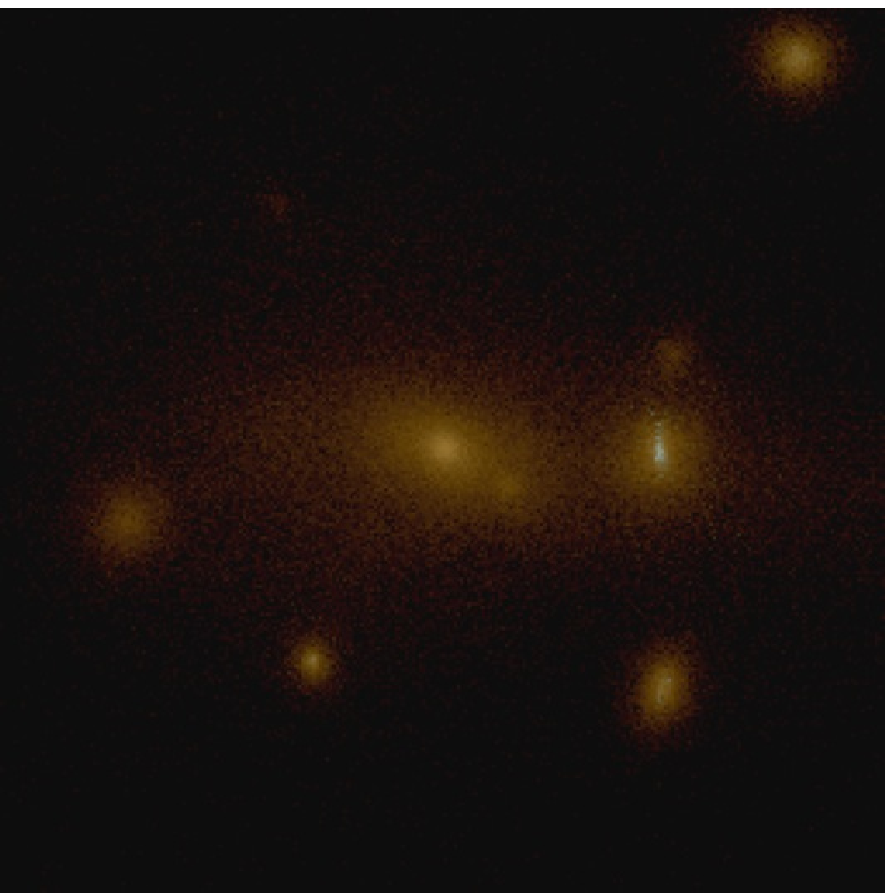}}}
  \centering{\resizebox*{!}{2.8cm}{\includegraphics{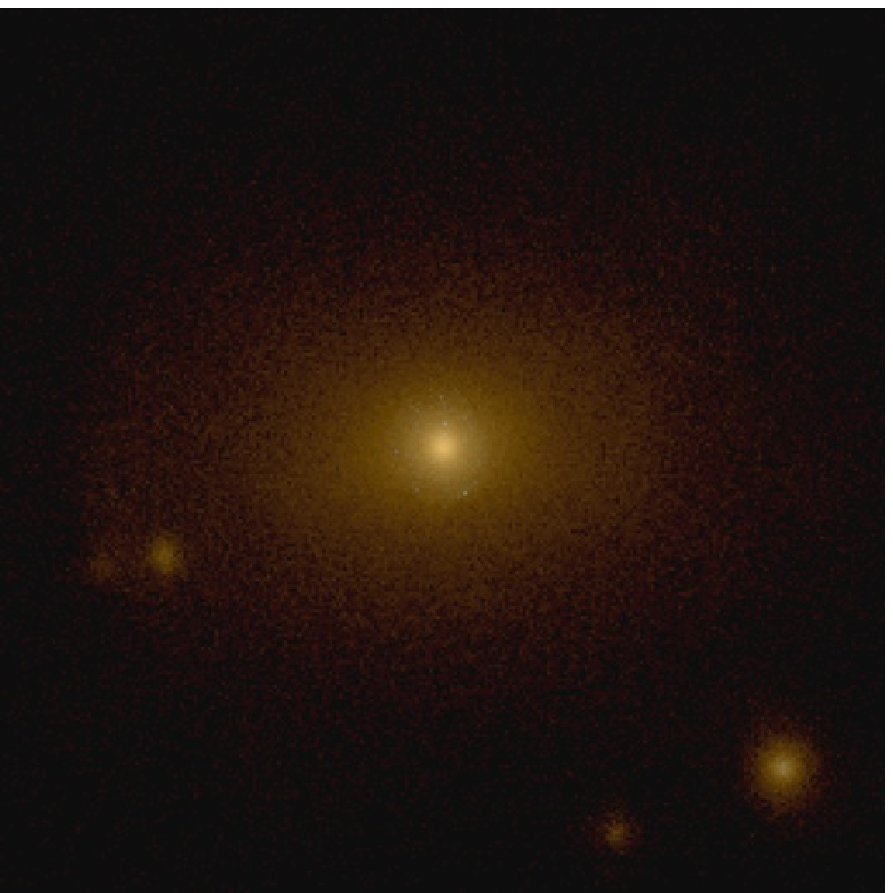}}}
  \centering{\resizebox*{!}{2.8cm}{\includegraphics{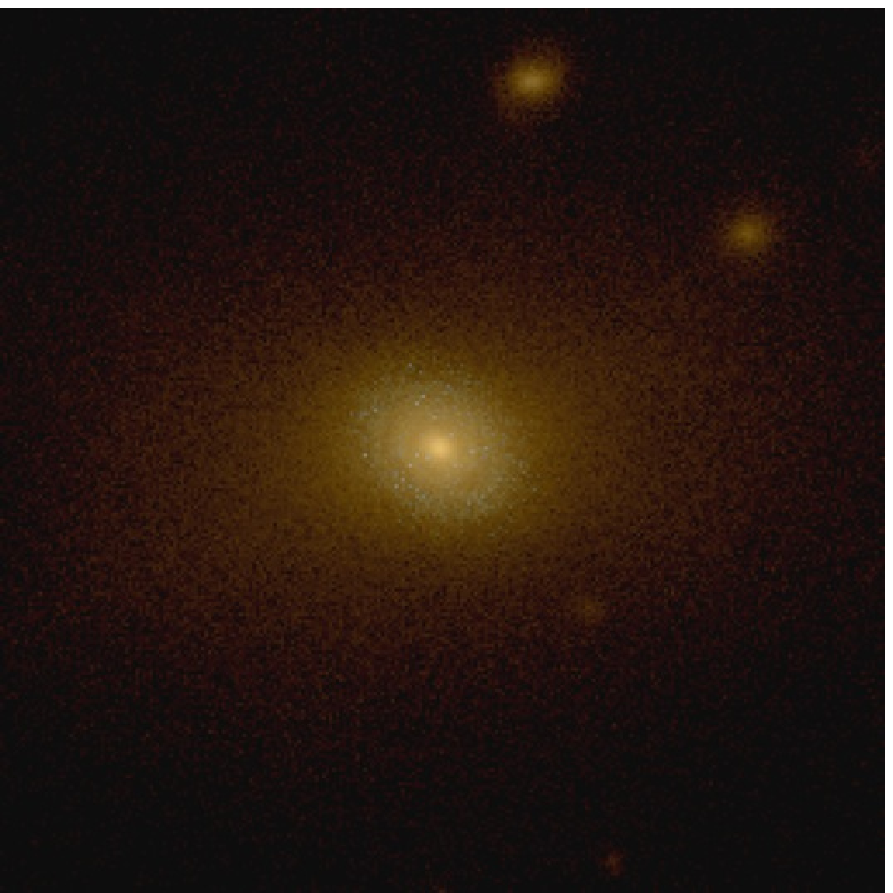}}}
  \centering{\resizebox*{!}{2.8cm}{\includegraphics{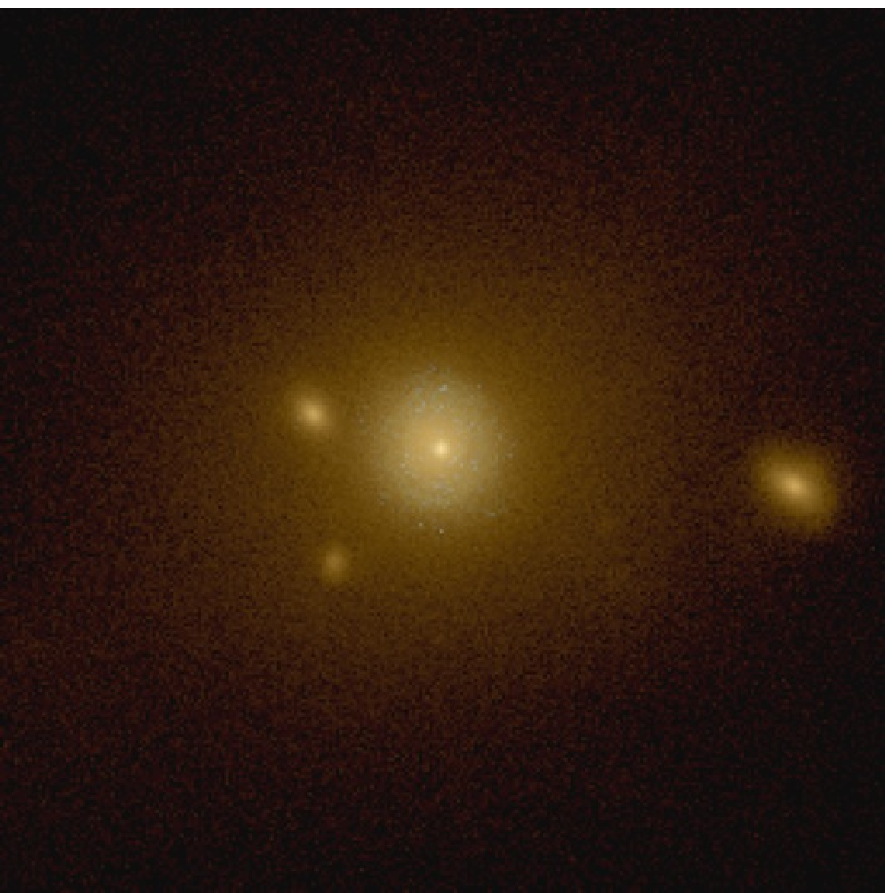}}}
  \centering{\resizebox*{!}{2.8cm}{\includegraphics{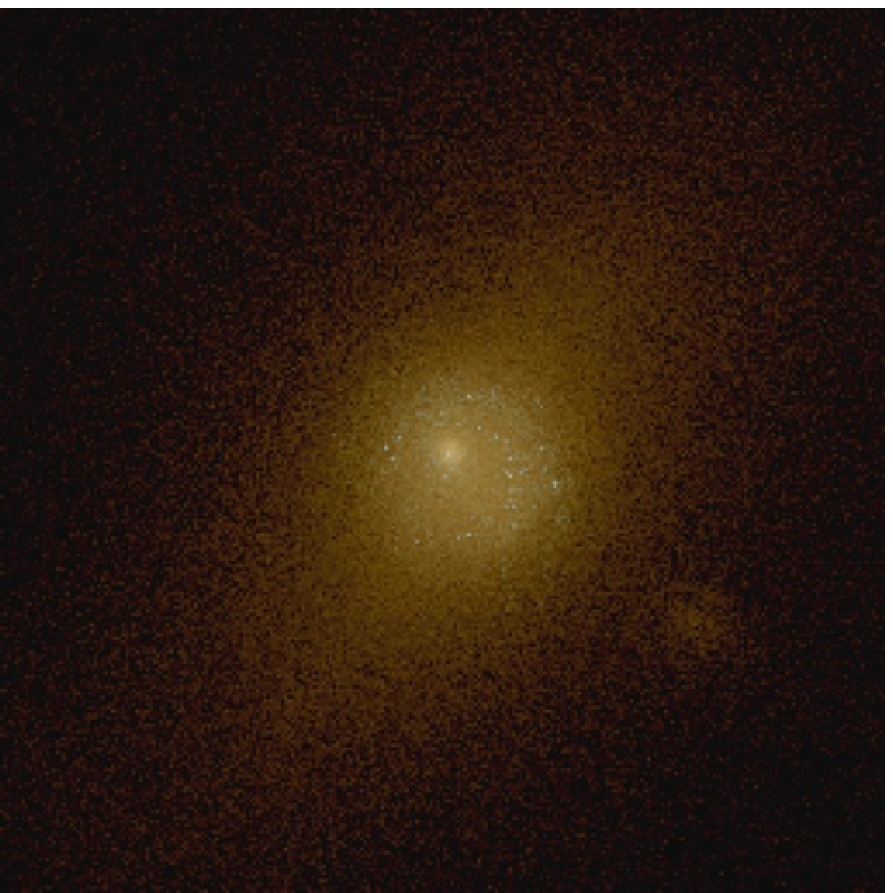}}}
  \caption{Stellar emission of the central galaxies from G1 to G6 (left to right) without (top) or with AGN feedback (bottom) at $z=0$, as they would be observed in u-g-r filter bands. Extinction by dust is not taken into account in these images. Arbitrary units are used for stellar emission but with similar minima and maxima for all the first fifth columns, and with decreased intensities by 0.5 dex for the last column to avoid image saturation. The size of each panel is 140 kpc. Galaxies without AGN feedback exhibit a massive blue disc component with a halo of stars extending to large distance. AGN feedback reduces the total amount of emitted light, and galaxies appear more early type with a weak disc component for some of them.}
    \label{fig:syntheticoptical}
\end{figure*}

Note that if, as suggested by recent observations using lensing (or dynamics) and photometry~\citep{grilloetal08, grillo&gobat10, treuetal10, augeretal10imf, spinielloetal11, thomasetal11, sonnenfeldetal12, cappellarietal12} or spectroscopy~\citep{spinielloetal12, conroy&vdk12}, the IMF is considered to be more bottom-heavy for the most massive galaxies, then the observational relation of $M_{\rm s}$ versus $M_{\rm h}$ should be pushed towards higher values of $M_{\rm s}$.
Assuming a~\cite{salpeter55} IMF instead of a~\cite{kroupa01}~\cite[or][]{chabrier03} IMF,   the ratio $M_{\rm s}/M_{\rm h}$ increases by $+0.3$ dex, which puts our values of the central stellar mass for the four most massive galaxies (with AGN feedback) in better agreement with observations, but in less better agreement for the two least massive galaxies (G1A and G2A).
It exists a possible smooth transition from a Kroupa to a Salpeter IMF for galaxies with velocity dispersion around $100-200\, \rm km.s^{-1}$~\citep{cappellarietal12}, values which correspond to the typical ones of the galaxies at $z=0$ when simulated with AGN feedback (see section~\ref{section:scalinglaws}).

Fig.~\ref{fig:fstarvsz} shows the evolution of the stellar conversion efficiency $f_{\rm conv}$ averaged over the six central galaxies as a function of redshift.
In the absence of AGN feedback, $f_{\rm conv}$ is continuously increasing with decreasing redshift.
Large amounts of gas are piling up in the central galaxy due to over-cooling, but, as it takes several free-fall times ($t_{\rm ff}/\epsilon_*=50 t_{\rm ff}\simeq1$ Gyr for a typical gas number density of $n=10\, \rm cm^{-3}$ in star forming regions, see Fig.~\ref{fig:zoom}) to convert the cold gas content into stars, the value of $f_{\rm conv}$ is lower at high redshift.
The same trend for the evolution of $f_{\rm conv}$ is observed at high redshift above $z\sim2$ for the simulations with AGN feedback.
However, during the early phase of galaxy formation, the presence of AGN feedback allows for a rapid quenching of the star formation with a reduced value of $f_{\rm conv}$ by a factor 2 at $z=3-4$ compared to galaxies simulated without AGN.
From $z\sim2$ to $z=0$, the value of $f_{\rm conv}$ slowly decreases by a factor 2.
While halos continue to grow by diffuse accretion and mergers, the gas is efficiently prevented to cool further down by AGN feedback, and the central galaxy stops actively forming stars (see the values of the SFR obtained at $z=0$ in table~\ref{tab:names}).

The presence of AGN shuts down the star formation in the central galaxies and it starts already at high redshift.
The specific Star Formation Rate (sSFR) (see Fig.~\ref{fig:ssfrvsz}) is continuously decreasing with time due to the starvation of the reservoir of cold gas by the cosmological growth of structures, and the activity of the central AGN that exacerbates the depletion of gas in massive galaxies.
At high redshift, central BHs accrete gas close to the Eddington ratio while at low redshift they experience long periods of time without accreting gas and with intense burst of AGN activity driven by mergers and cooling flows (see~ Fig.\ref{fig:lumvsz}).
Even though there is a lot less gas available in galaxies at low redshift, and that the AGN activity is overall greatly diminished, bursts of AGN activity are necessary to self-regulate the cold baryon content in galaxies at low redshift~\citep{duboisetal11}.

By simulating the stellar light emission (using the SDSS u, g and r filter bands) of the central galaxies, we observe that, because of AGN feedback, they become less luminous, more red and with rounder shapes (see Fig.~\ref{fig:syntheticoptical}).
In the absence of AGN feedback, the central galaxies exhibit a massive blue disc component (with spiral arms for some of them) in their centres, that is confirmed by the measured colours, around $g-r=0.45$ without dust extinction.
For instance some simulations without AGN feedback show a massive blue component, which is associated to a disc with spiral arms
For these over-massive objects, there is a strong dust lane component (not represented in the images of Fig.~\ref{fig:syntheticoptical}), which reddens them by  $\Delta (g-r)_{\rm dust}=0.2$ (see table~\ref{tab:names}).
In contrast, simulated galaxies with AGN feedback have colours between  $0.60<g-r<0.73$ (no dust extinction), and for some of them a contribution is adopted for dust extinction of $\Delta (g-r)_{\rm dust}=0.1$ (closer to zero for some others).
We get better consistency of galaxy colours with observations when AGN feedback is considered~\citep{bernardietal05}.

\begin{figure}
  \centering{\resizebox*{!}{6cm}{\includegraphics{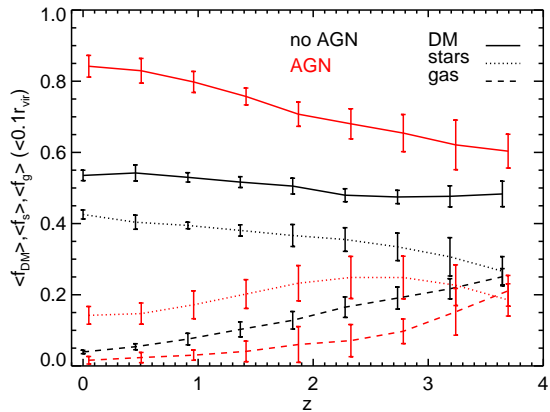}}}
  \caption{Average fraction of DM (solid), stellar (dotted), and gas (dashed) mass within $0.1 r_{\rm vir}$ as a function of redshift for the six halos with (red) or without (black) AGN feedback. The error bars are the standard deviation. The fraction of DM mass is increased, and the fractions of gas and stellar mass are decreased by the effect of AGN feedback.}
    \label{fig:fdmevol}
\end{figure}

We measure the average mass fractions of DM, stars and gas within the $0.1r_{\rm vir}$ of the halo as a function of redshift for the simulations with or without AGN feedback (Fig.~\ref{fig:fdmevol}).
We find that the fraction of DM mass is more important in the core of halos when AGN feedback is present, going from $\langle f_{\rm DM}\rangle=0.53$ without AGN up to $\langle f_{\rm DM} \rangle=0.83$ with AGN at $z=0$.
The DM fraction is constant over time in the absence of AGN feedback, but increases with time for the AGN case.
Note that the DM fraction is already higher in the AGN case at high redshift compared to the no AGN case, but this difference is enhanced at low redshift. 
Without AGN feedback the fraction of stars is increasing with time and the gas fraction is decreasing, while the baryonic fraction is constant (DM fraction is constant).
Thus, galaxies are gas-rich (wet) at high redshift, and gas-poor at low redshift because they consume the gas efficiently in the star formation process.
Both the stellar and the gas mass fractions are reduced due to the presence of AGN feedback, from $\langle f_{\rm s}\rangle= 0.42$ down to $\langle f_{\rm s} \rangle= 0.15$, and from $\langle f_{\rm g} \rangle= 0.04$ down to $\langle f_{\rm g} \rangle= 0.016$ at $z=0$.
The difference in stellar and gas mass fractions is also seen at higher redshift, but in smaller proportions than at low redshift.
Finally, AGN feedback already impacts the mass content in galaxies early on (at $z\sim4$), but its effect is  largest at $z=0$.

\begin{figure}
  \centering{\resizebox*{!}{6.cm}{\includegraphics{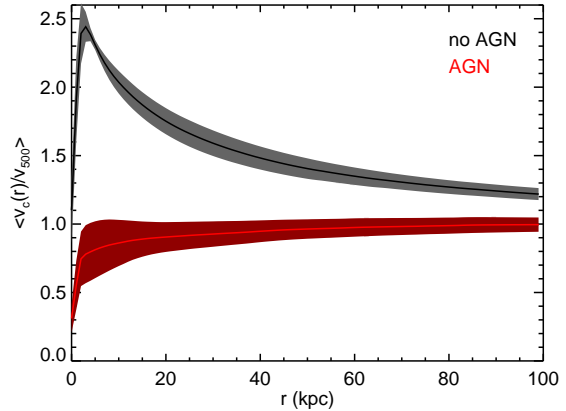}}}
  \caption{Circular velocities averaged over the six halos with (red) or without (black) AGN feedback as a function of radius at $z=0$. Velocities are renormalised by the circular velocities at $r_{500}$. Shaded areas correspond to the standard deviation. Simulations without AGN feedback exhibit a central peak in the circular velocity that is characteristic of too large mass concentration in the centre of the galaxy, while simulations with AGN feedback show flat velocity curves with a value close to the circular velocity at $r_{500}$.   }
    \label{fig:vcircoverv500}
\end{figure}

\begin{figure}
  \centering{\resizebox*{!}{6.cm}{\includegraphics{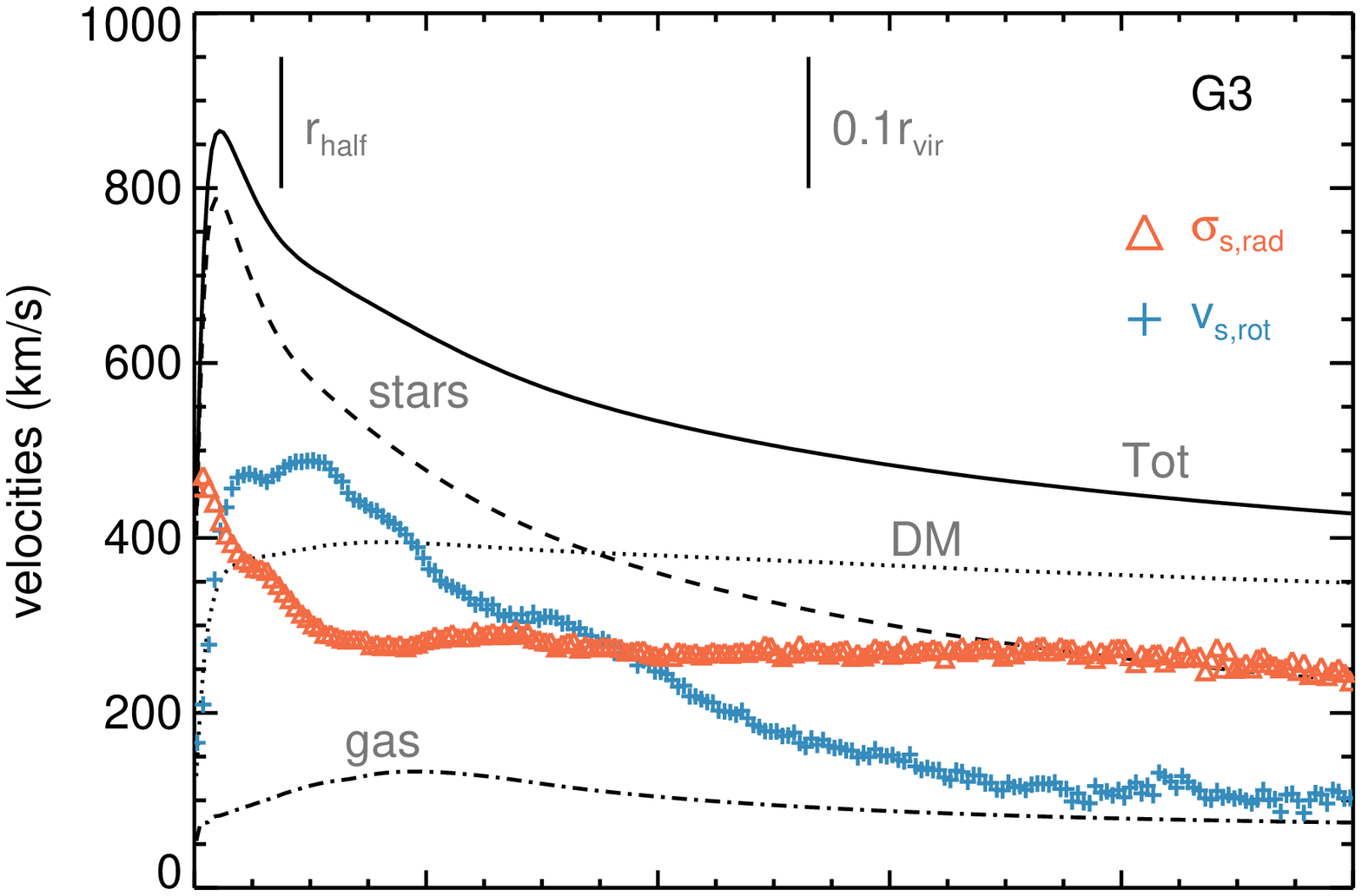}}}\vspace{-1.25cm}
  \centering{\resizebox*{!}{6.cm}{\includegraphics{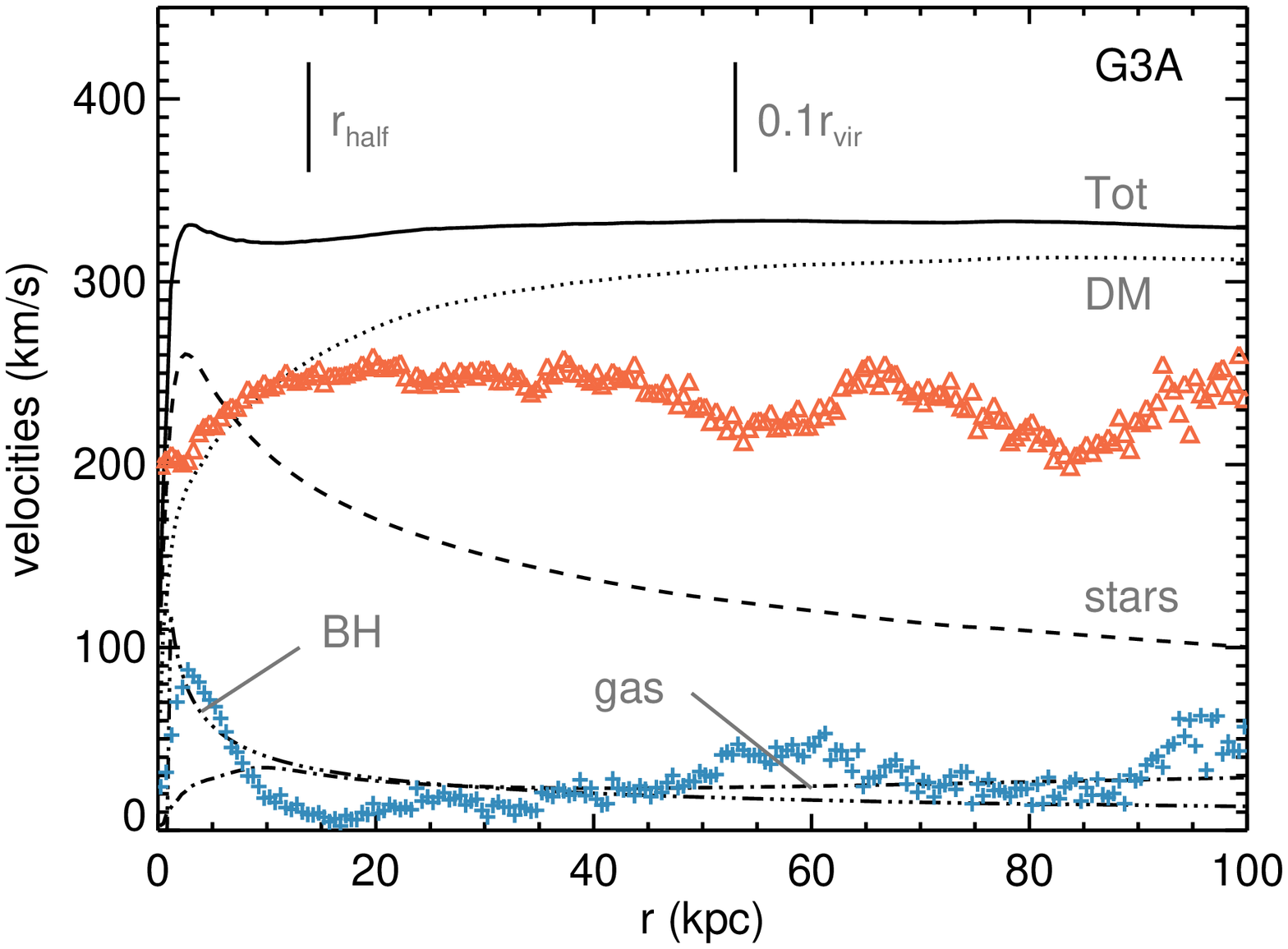}}}
  \caption{Illustration of the morphological transformation of the central massive galaxy of the halo of G3 at $z=0$
   due to AGN feedback. Both panels show the circular velocities as a function of radius (black) of all components (solid), stars (dashed), DM (dotted), gas (dot-dashed), and BHs (triple dot-dashed) without AGN feedback (top panel), or including AGN feedback (bottom panel). The measured rotational velocities for stars $v_{\rm s, rot}$ (blue plus) and radial velocity dispersions $\sigma_{\rm s, rad}$ (orange triangle) are also indicated. The 3D stellar half-mass radius $r_{\rm half}$ and 10 per cent of the virial radius $0.1r_{\rm vir}$ are indicated as vertical bars. The circular velocity is decreased by a factor 2 due to AGN feedback. The circular velocity is dominated by the stellar component up to large distance (30 kpc) in the AGN case, while it dominates only in the very central region (7 kpc) in the AGN case, and the DM is the principal contributor outside these characteristic radius. The galaxy is transformed from a rotating disc (no AGN case) into a velocity dispersion- dominated ellipsoid (AGN case).}
    \label{fig:velvsr_G1}
\end{figure}

Thanks to an exquisite resolution, we are able to go beyond the global properties listed above and investigate the internal dynamical structure of galaxies.
The morphology of galaxies is also transformed by the effect of AGN feedback as shown in~Fig.~\ref{fig:vcircoverv500}.
Circular velocities $v_{\rm c}=\sqrt{GM_{\rm tot}(<r)/r}$ exhibit a peak close to the centre of the galaxy, below 5 kpc, when AGN feedback is not included.
This peak is produced by the large concentration of material in the central region of galaxies, where the total matter distribution is largely dominated by stars (see Fig.~\ref{fig:velvsr_G1}), and is a characteristic feature of the overcooling problem of baryons~\citep[e.g.][]{scannapiecoetal12, fewetal12}.
In the opposite case, for the simulations including AGN feedback, the peak in the circular velocity profiles disappears and profiles become flat (isothermal) up to very large distances away from the centre, with values close to the circular velocity measured far away from the centre (at $r_{500}$).
With the presence of AGN feedback, the maximum of the circular velocity curves decreases, the contribution from stars to circular velocity is strongly attenuated, and replaced by its DM component (Fig.~\ref{fig:velvsr_G1}).
Note that the contribution from gas to the circular velocity curves at $z=0$ is negligible with or without AGN feedback, which suggests that these massive galaxies are relatively gas-poor independently of the presence of AGN.
We notice that the ratio of radial velocity dispersion over the circular velocity is roughly constant and close to a factor $\sqrt{2}$, therefore, it gives further evidence for an isothermal profile.
It assumes that orbits are isotropic, i.e. the velocity tensor is close to isotropic.
Further investigations along this line are left for future work.

\begin{figure}
  \centering{\resizebox*{!}{6.cm}{\includegraphics{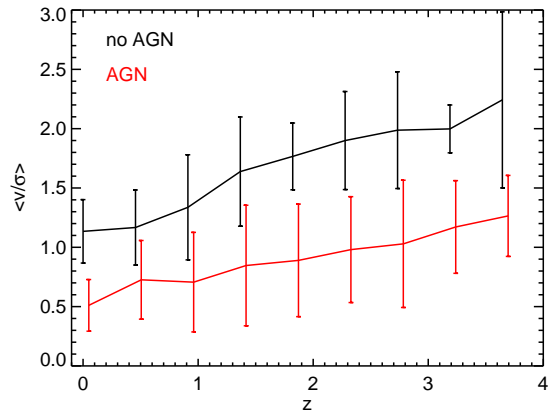}}}
  \caption{Average $v/\sigma$ ratio as a function of redshift for the central galaxy of the six simulated halos with (red) or without (black) AGN feedback, with the standard deviation (error bars). The galaxies simulated without AGN feedback are fast rotators at any redshift, while galaxies simulated with AGN feedback are slow rotators.}
    \label{fig:voversig}
\end{figure}

AGN have also important consequences for the total amount of rotation and velocity dispersion in these massive galaxies.
Without AGN feedback, the stellar component of the massive galaxies at $z=0$ is supported by the rotation of stars, with a comparable but lower contribution from dispersion, particularly within the half mass radius of the galaxy.
Such galaxies would be classified as fast rotators as the $v/\sigma$ ratio for stars approaches one (even though their masses and colours are unrealistic).
At a close distance from the centre, the velocity support is dominated by the dispersion (compact bulge of stars), and also far away, in the intra cluster light.
The picture is changed when AGN feedback is present: velocity dispersion dominates at any distance from the centre at $z=0$.
The disc component has considerably shrank and, now, is barely detectable below 5 kph at $z=0$ (Fig.~\ref{fig:velvsr_G1}).

Fig.~\ref{fig:voversig} shows the average $v/\sigma$ ratio, where $v$ is the mass-weighted rotational velocity of stars and $\sigma$ is the mass-weighted velocity dispersion of stars measured within $r_{\rm eff}$, where $r_{\rm eff}$ is the effective radius at which half of the projected stellar mass is enclosed.
Both quantities are integrated quantities of the projected distribution of stars, and the $v/\sigma$ ratios showed in Fig.~\ref{fig:voversig} are averaged over the six galaxies (with or without AGN feedback).
Without AGN feedback, the galaxies have large values of $v/\sigma$ above 1, i.e. are dominated by the rotation of stars at all redshifts.
Due to the increased proportion of accreted material (see next section), the $v/\sigma$ ratio decreases with time and tends towards 1 at $z=0$.
Thus the massive galaxies at $z=0$ simulated without AGN feedback are disc-like galaxies.
The transformation of rotationally-supported discs into dispersion-dominated ellipsoids by the presence of AGN feedback is observed for all our simulated central galaxies at low redshift ($z<1$).
The same evolution is found for galaxies simulated with AGN feedback as without AGN feedback: the ratio of $v/\sigma$ also decreases with time, but AGN transforms the galaxies into slow rotators ($v/\sigma <1$) where galaxies become essentially supported by the velocity dispersion of the stars, with a value slightly above 1 at $z=4$ down to a value of 0.5 at $z=0$.
The importance of rotation over dispersion is already reduced by a factor 2 at high redshift ($z=3-4$). 

We interpret this morphological change as the signature that massive galaxies are transformed from systems dominated by in situ star formation (without AGN feedback) into systems dominated by accretion of satellites (with the presence of AGN).
We will now clearly demonstrate this mechanism in the following section.

\begin{figure}
  \centering{\resizebox*{!}{6.cm}{\includegraphics{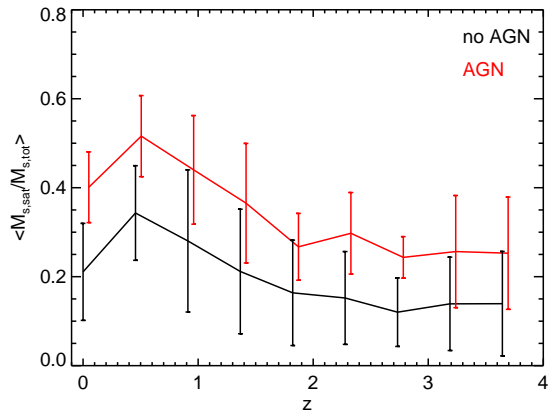}}}
  \caption{Average fraction of the total stellar mass in the halo locked into satellite galaxies without AGN feedback (black) and with AGN feedback (red). The values are averaged over the six simulated halos, and the error bars correspond to the standard dispersion. With the presence of AGN feedback, the fraction of stars locked into the galaxy satellites is increased with almost half of it in the satellites at $z=0$.}
    \label{fig:fsatvsz}
\end{figure}

\subsection{From in situ to accretion-dominated systems}

The fraction of stars locked into satellites $M_{\rm s, sat}/M_{\rm s, tot}$, where $M_{\rm s, sat}$ is the mass of stars in satellites within $r_{\rm vir}$, and $M_{\rm s, tot}$ is the total mass of stars within $r_{\rm vir}$, is increasing with time and reaches values of 20-30 per cent at $z\simeq0$ without AGN and 40-50 per cent with AGN  (see Fig.~\ref{fig:fsatvsz}).
When AGN feedback is active, it efficiently reduces the total stellar mass in the central galaxy (Fig.~\ref{fig:fstarvsz}) by blowing gas away and preventing gas accretion.
The increased fraction of mass locked into galaxy satellites due to AGN is a clear indicator that the mass build-up of central galaxies should have a decreased proportion of stellar mass formed in the main progenitor.
The shapes of the curves are similar for AGN and no AGN cases. This is due to the fact that the presence of AGN does not affect the merger history of halos that are dominated by the DM component.

\begin{figure*}
  \centering{\resizebox*{!}{3.cm}{\includegraphics{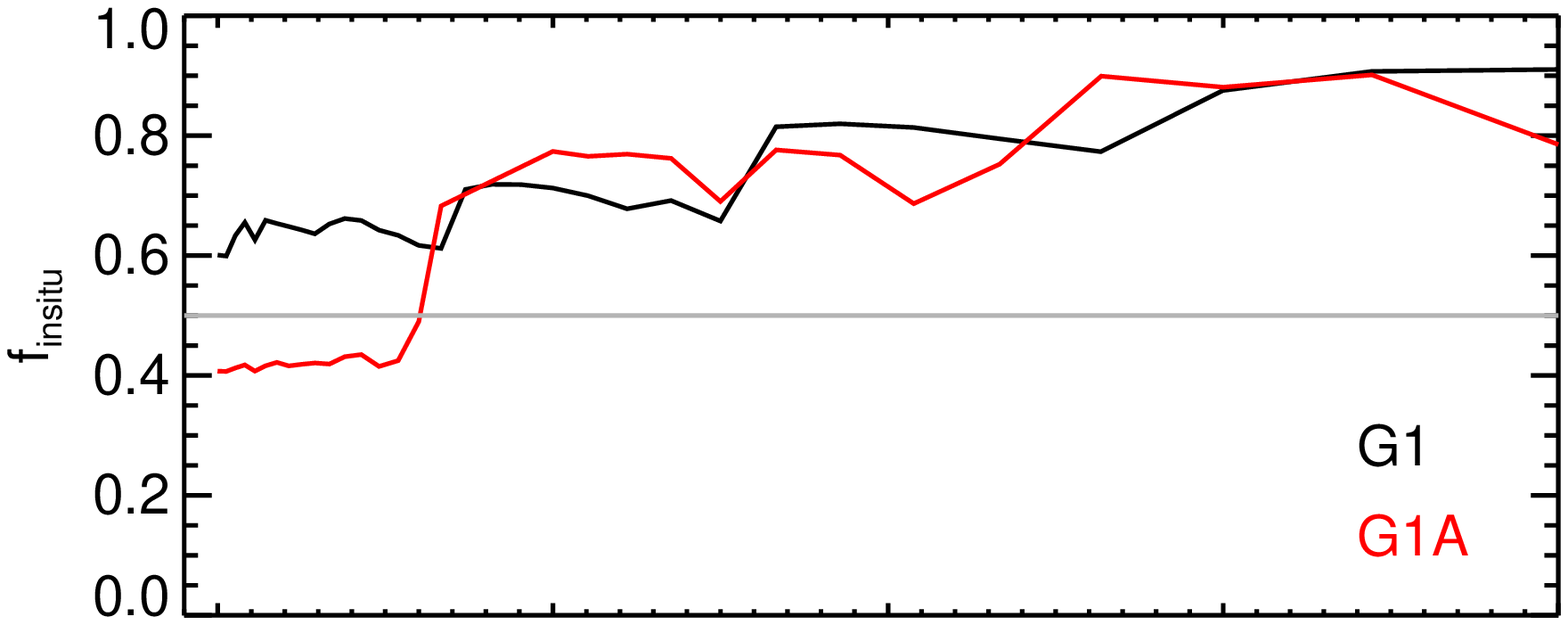}}}\hspace{-1.2cm}
  \centering{\resizebox*{!}{3.cm}{\includegraphics{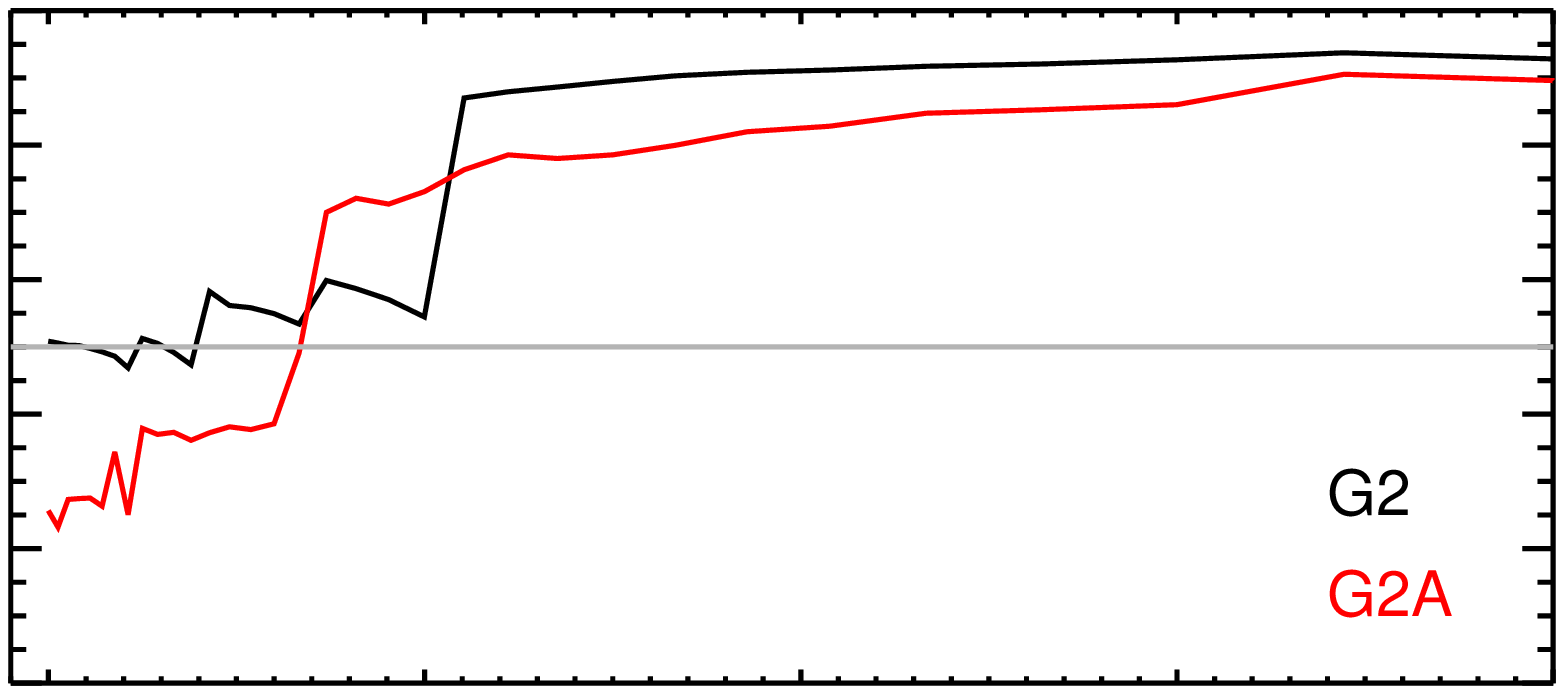}}}\hspace{-1.2cm}
  \centering{\resizebox*{!}{3.cm}{\includegraphics{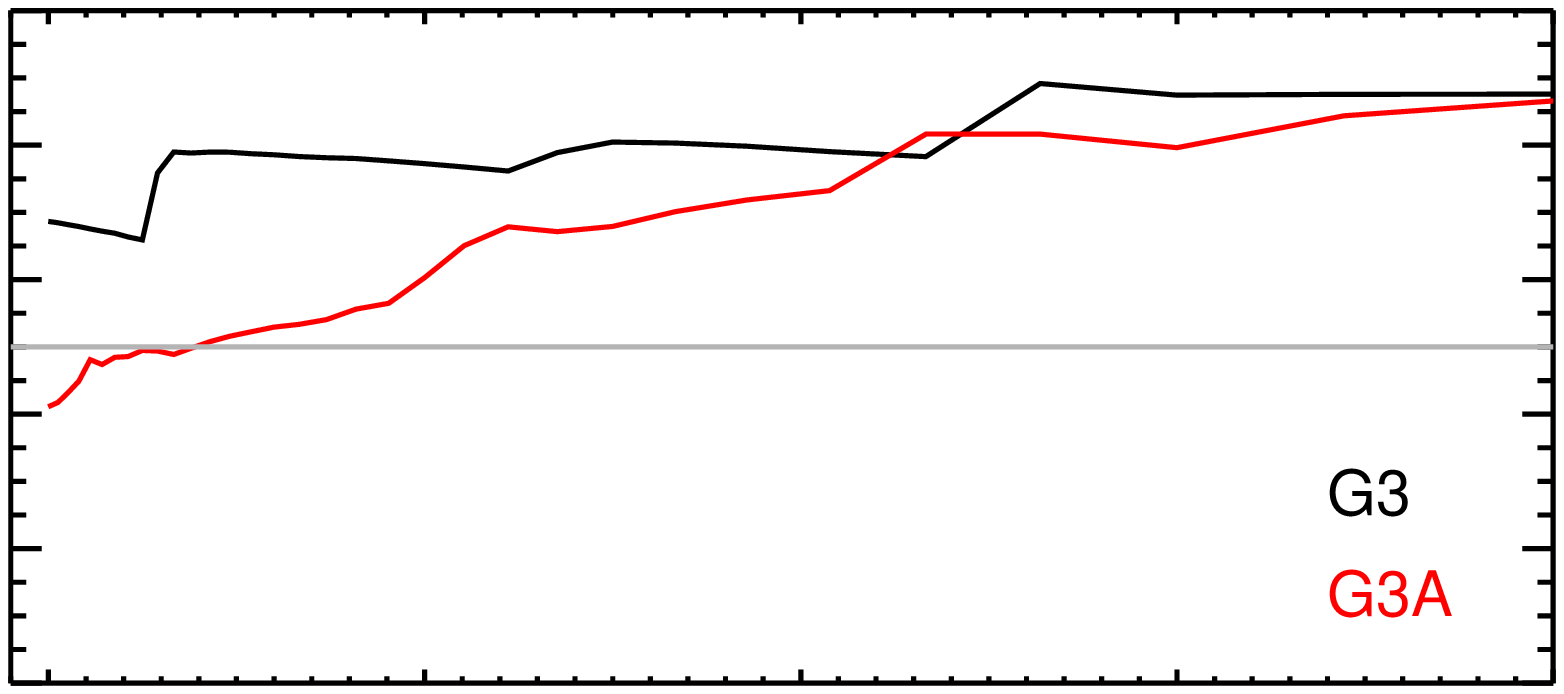}}}\vspace{-.7cm}
  \centering{\resizebox*{!}{4.5cm}{\includegraphics{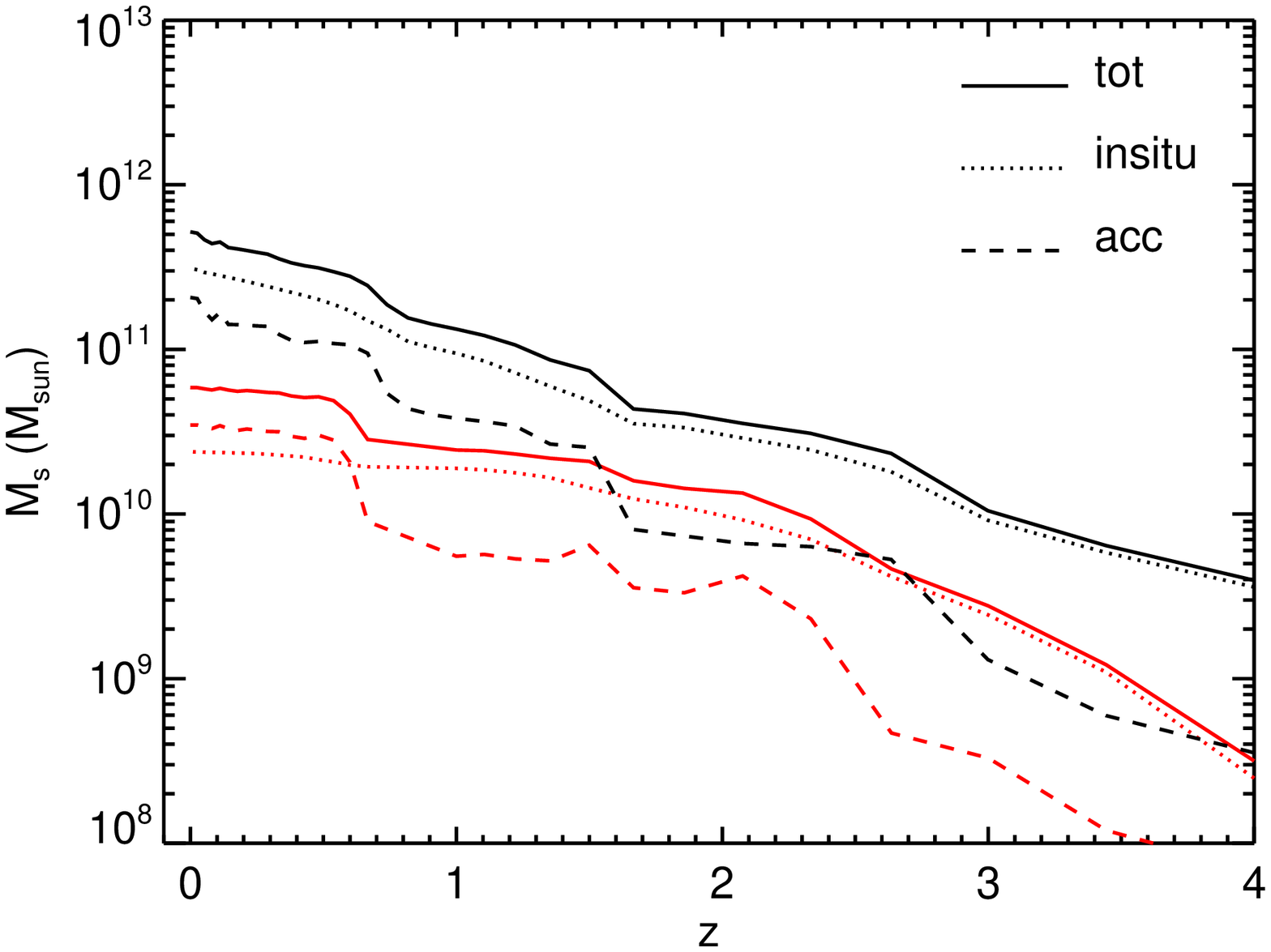}}}\hspace{-1.2cm}
  \centering{\resizebox*{!}{4.5cm}{\includegraphics{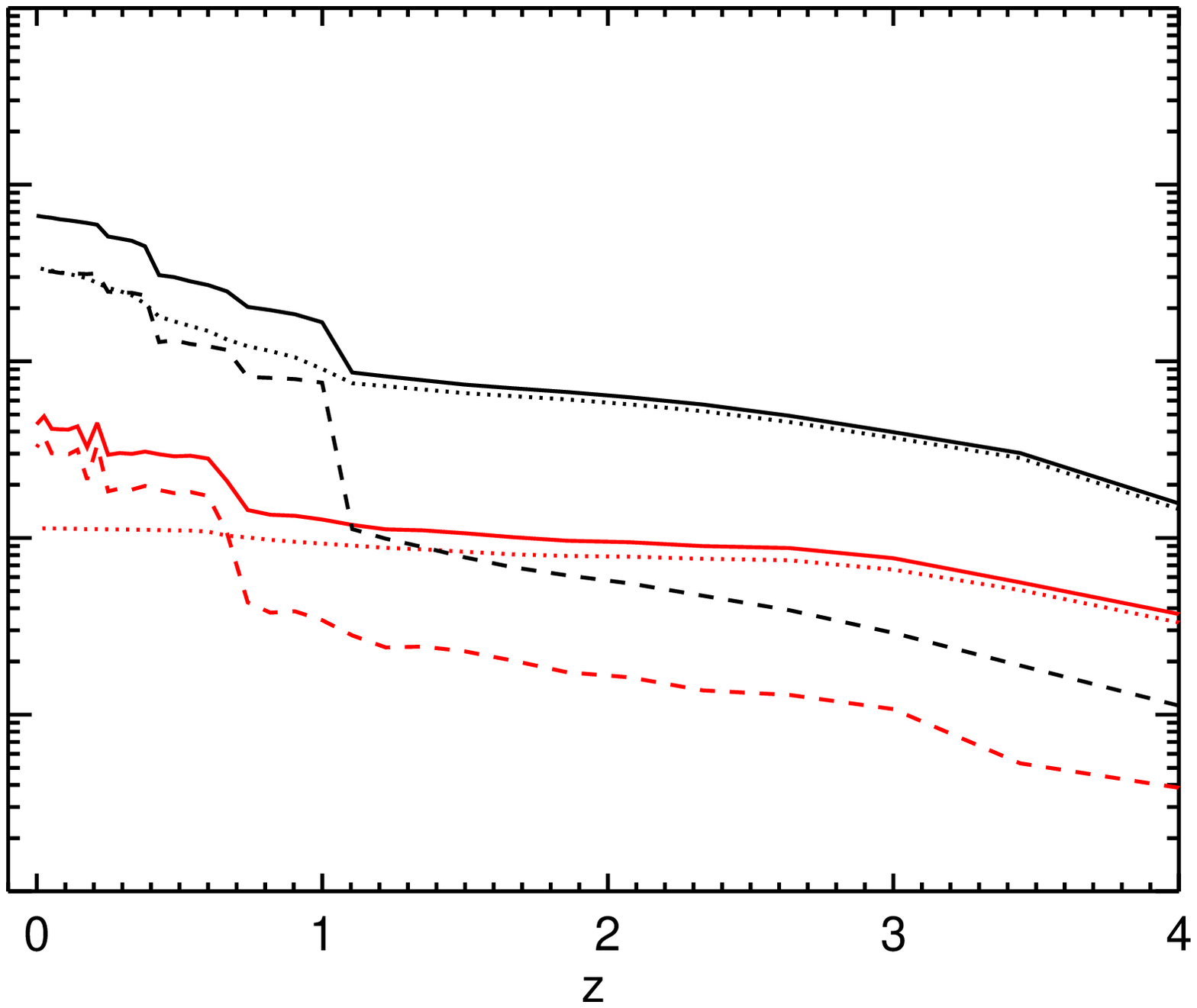}}}\hspace{-1.2cm}
  \centering{\resizebox*{!}{4.5cm}{\includegraphics{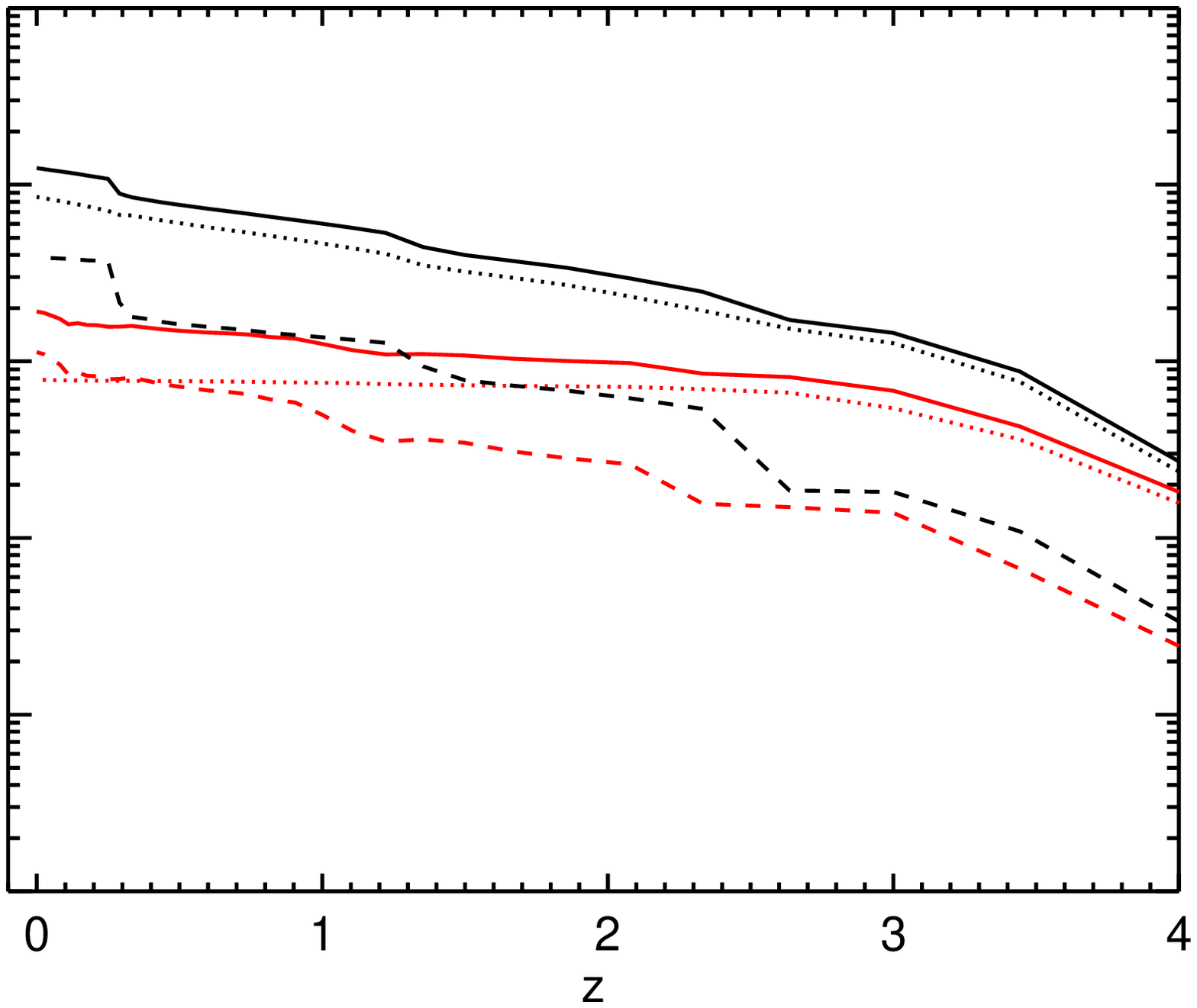}}}\vspace{-.3cm}
  \centering{\resizebox*{!}{3.cm}{\includegraphics{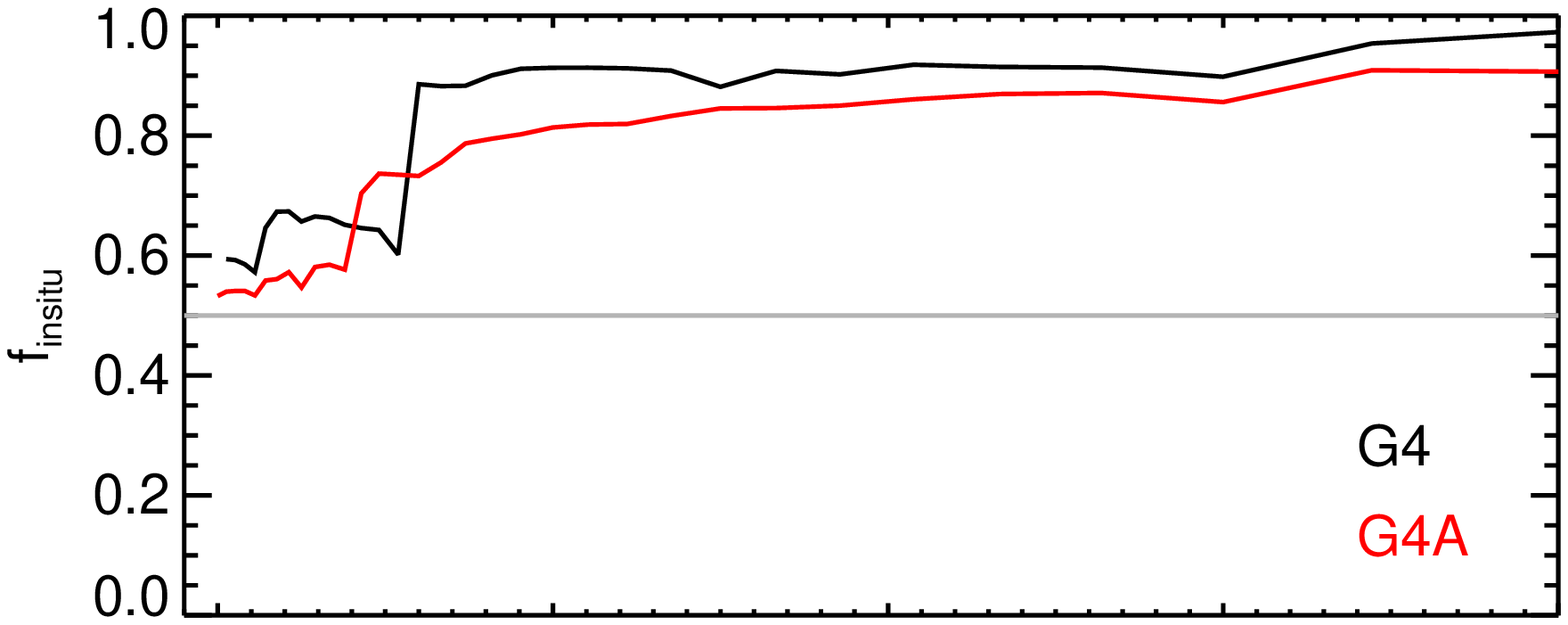}}}\hspace{-1.2cm}
  \centering{\resizebox*{!}{3.cm}{\includegraphics{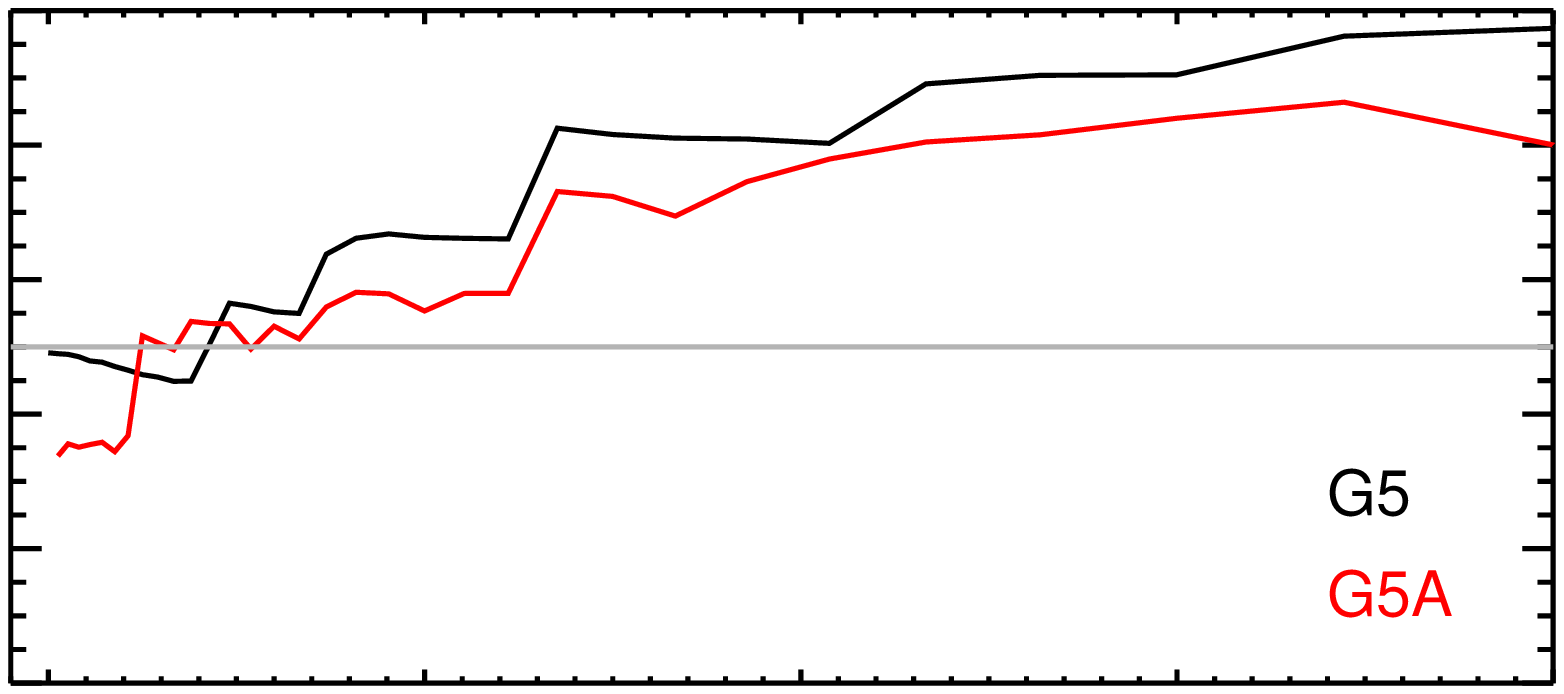}}}\hspace{-1.2cm}
  \centering{\resizebox*{!}{3.cm}{\includegraphics{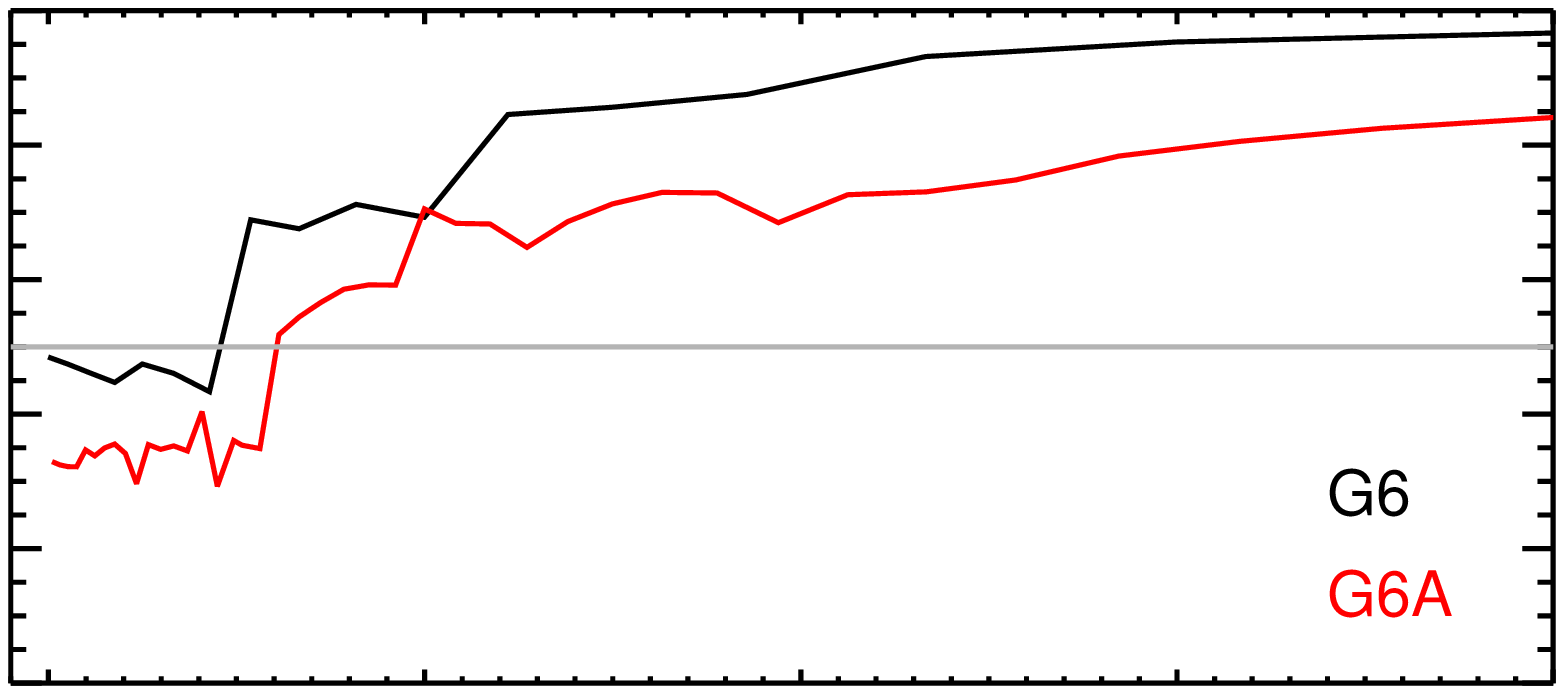}}}\vspace{-.7cm}
  \centering{\resizebox*{!}{4.5cm}{\includegraphics{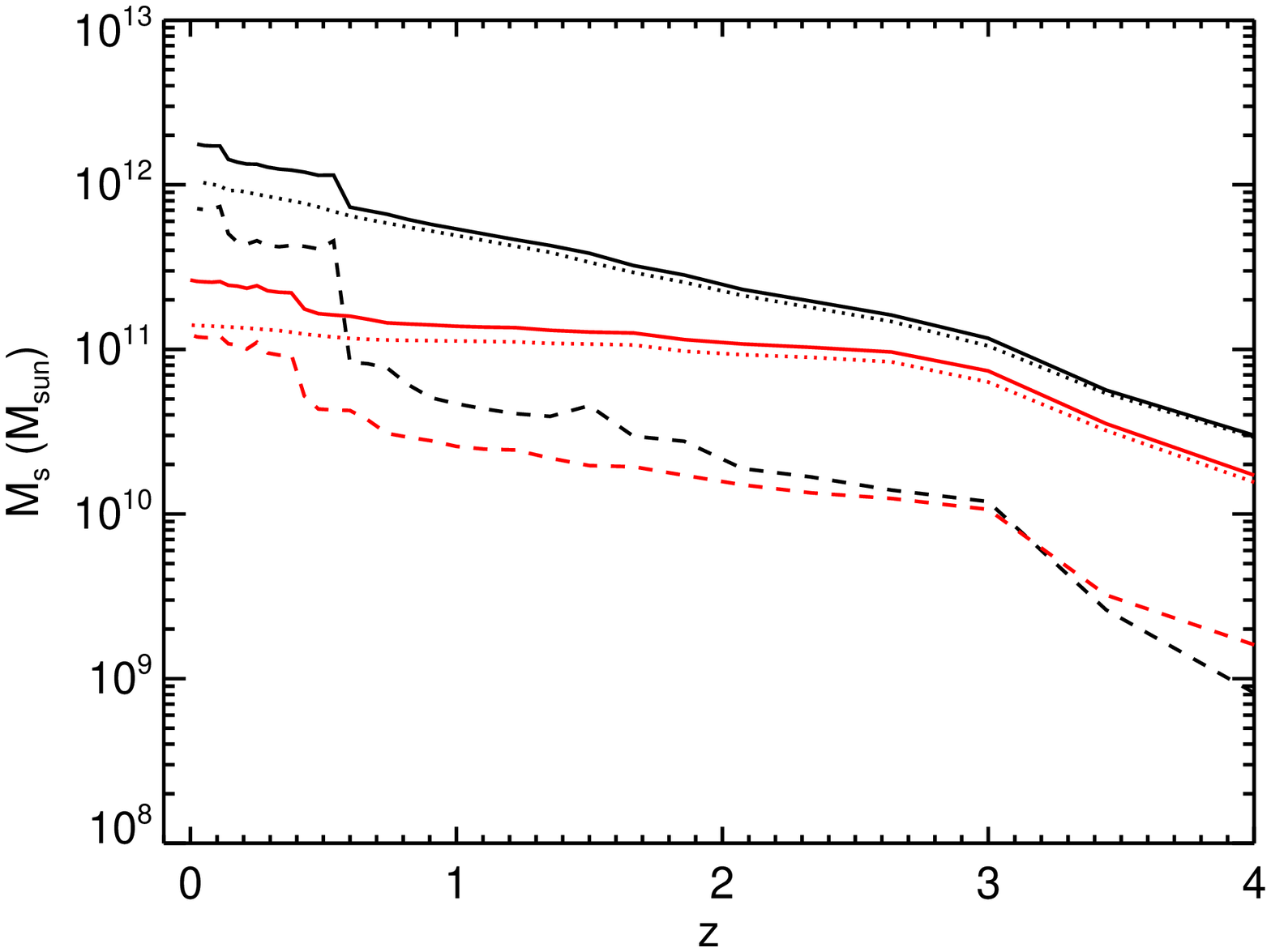}}}\hspace{-1.2cm}
  \centering{\resizebox*{!}{4.5cm}{\includegraphics{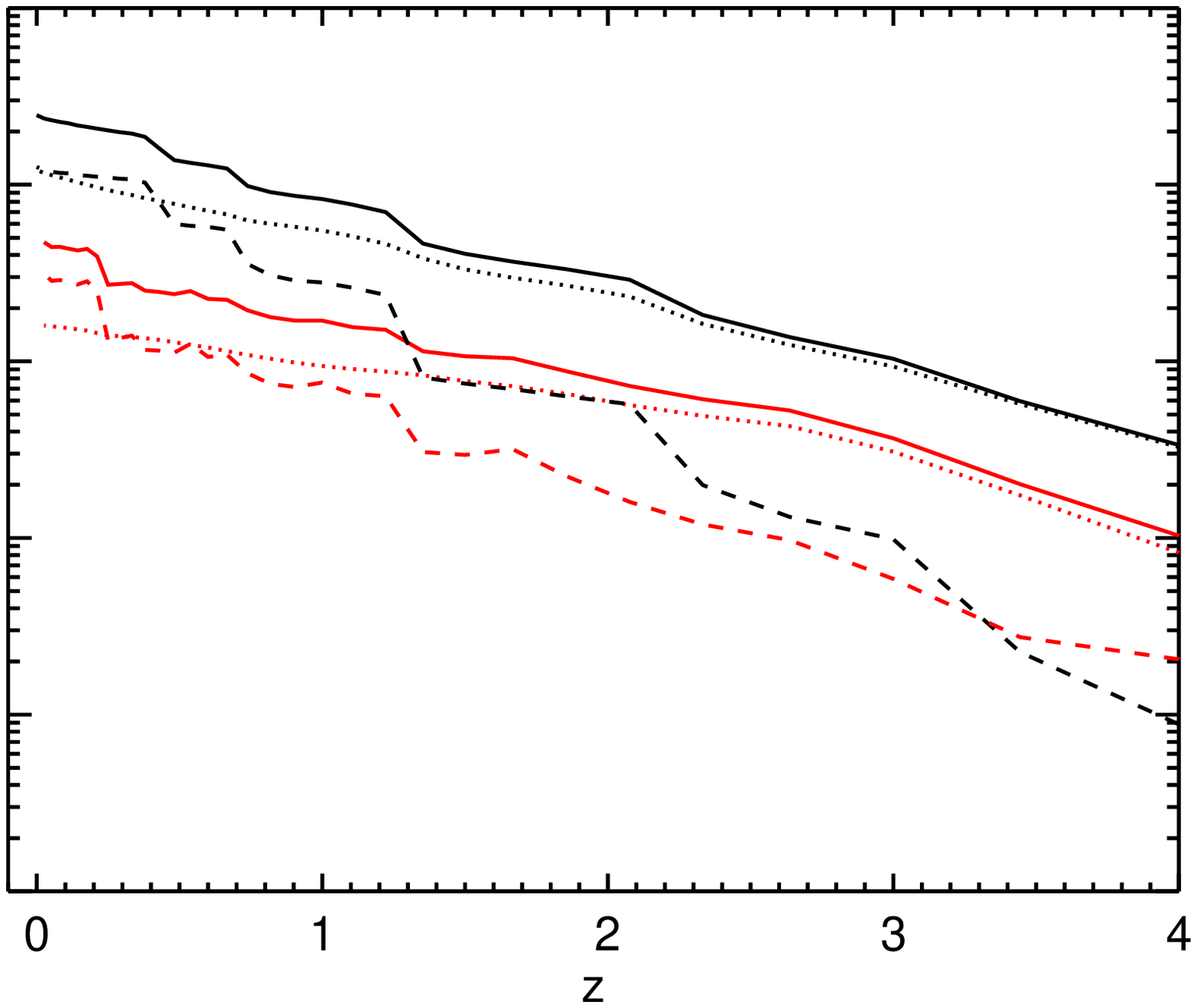}}}\hspace{-1.2cm}
  \centering{\resizebox*{!}{4.5cm}{\includegraphics{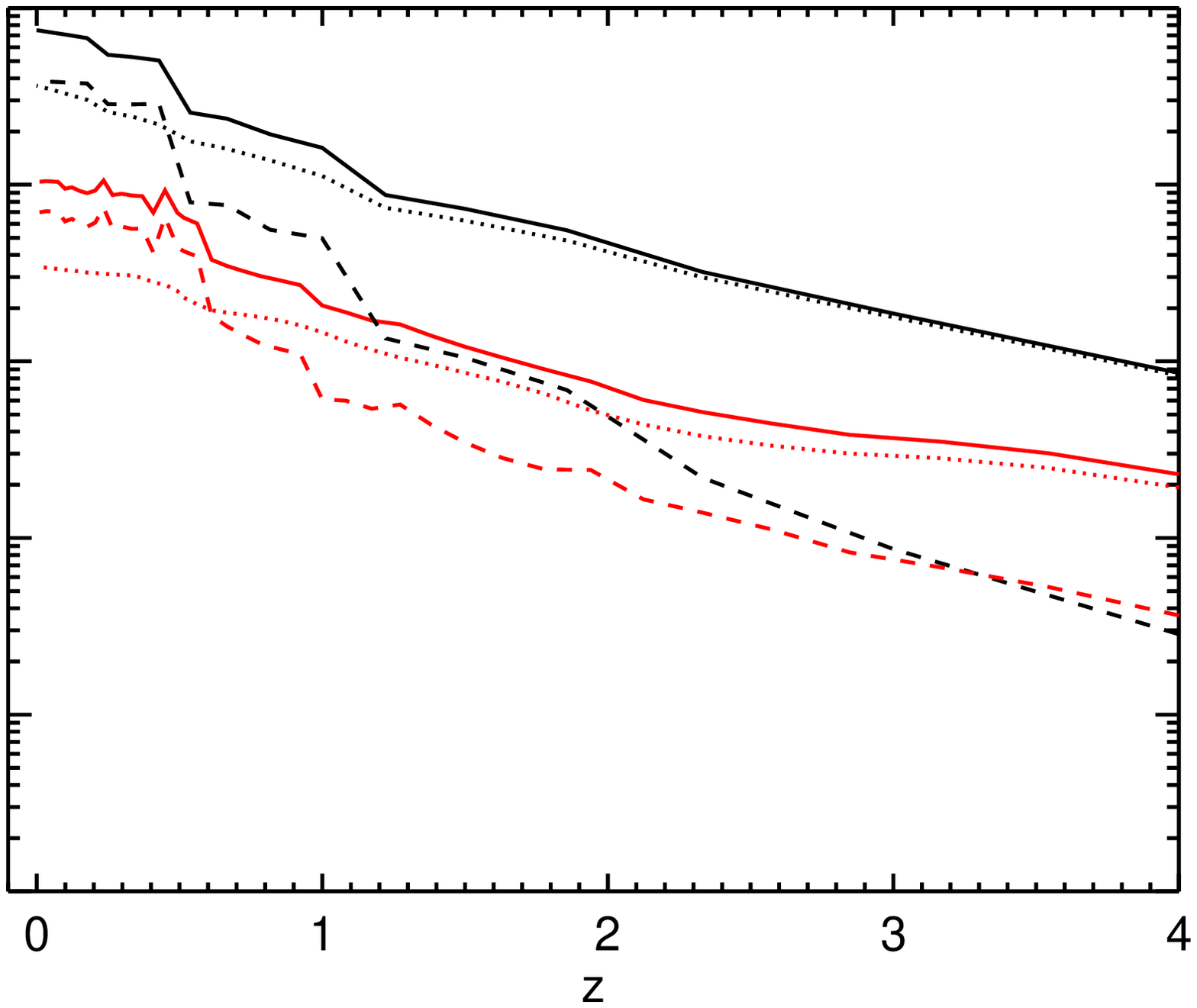}}}
  \caption{\emph{Top}: Fraction of in situ formed stars and, \emph{bottom}: total stellar mass (solid lines), mass of in situ formed stars (dotted lines) and of accreted stars (dashed lines) as a function of redshift for G1 (top left), G2 (top middle), G3 (top right), G4 (bottom left), G5 (bottom middle), and G6 (bottom right). The results for simulations without AGN feedback are in black (GX) and with AGN feedback in red (GXA). The grey horizontal lines indicate the 50 per cent level of in situ versus accreted stellar mass. }
    \label{fig:minsituacc}
\end{figure*}

To demonstrate the effect of AGN feedback on the stellar mass build-up, we define the \emph{in situ} formed stars as the stars formed within the main progenitor of the galaxy (within 10 per cent of the virial radius of the host halo), and the \emph{accreted} stars are defined to be the stars formed outside of the main progenitor of the galaxy and acquired through capture of satellites or by diffuse accretion.
Fig.~\ref{fig:minsituacc} shows the amount of stellar mass in the central massive galaxy formed in situ and the amount of stars that are accreted, for the different simulations, depending on the presence of AGN feedback.
Galaxies are dominated by the in situ growth of the stellar mass at high redshift due to the large fraction of baryons available in the form of cold star-forming gas~\citep{tacconietal10, daddietal10, duboisetal12}.
As a consequence, the specific star formation rates (star formation rate per galaxy stellar mass) are maintained at high values, even when AGN feedback is included~\citep{kimmetal12}.
\cite{conseliceetal12} have shown that observed massive galaxies $M_{\rm s}>10^{11}\, \rm M_\odot$ acquire the majority of their stars $61\pm21$~\% through the in situ star formation between redshift $1.5<z<3$ which is consistent with our findings.
As time goes on, the gas reservoir is consumed in the star formation process and ejected through the galactic winds produced by AGN feedback, and the specific star formation rates decline.
The result is that mergers have an increasing contribution to the mass build-up of these massive galaxies~\citep{delucia&blaizot07}, and the fraction of in situ-formed stars declines with time (decreasing redshift) as seen for all the six simulated halos independently of the presence of feedback.

\begin{figure}
  \centering{\resizebox*{!}{6.cm}{\includegraphics{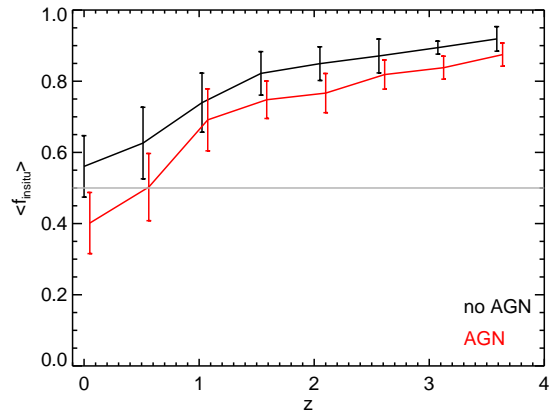}}}
  \caption{Average fraction of in situ formed stars as a function of redshift without AGN feedback (black) and with AGN feedback  (red). The grey horizontal lines indicate the 50 per cent level of in situ versus accreted stellar mass. The error bars indicate the standard deviation to the average value. The in situ fraction of stars decline with time, and the presence of AGN feedback reduces this fraction further down to turn galaxies into accretion dominated systems at $z=0$.}
    \label{fig:finsituave}
\end{figure}

The difference between the two simulation types (with or without AGN) is that, for the ``vanilla'' model (no AGN), the central galaxy mass is dominated by the in situ star formation (stars are formed locally) at $z=0$ with $f_{\rm insitu}\simeq 55$ per cent of the total stellar mass on average, while, for the AGN feedback runs, the stellar mass is dominated by the accretion of satellites at $z=0$ with $f_{\rm insitu}=40$ per cent on average, as seen in Fig.~\ref{fig:finsituave}.
It is important to stress that \emph{both} the mass of in situ formed stars and the mass of accreted stars at $z=0$ are decreased because of the presence of AGN feedback (Fig.~\ref{fig:minsituacc}), \emph{but} AGN feedback quenches the star formation more efficiently in the central galaxy than it does in the galaxy satellites.
Note that, as suggested by Fig.~\ref{fig:fstarvsz}, the total mass of stars in the central galaxy is already affected by the presence of AGN feedback early on ($z=3-4$), and that the fraction of in situ-formed stars is decreased early on when the activity of AGN is the strongest~\citep{duboisetal12}.

\begin{figure}
  \centering{\resizebox*{!}{6.cm}{\includegraphics{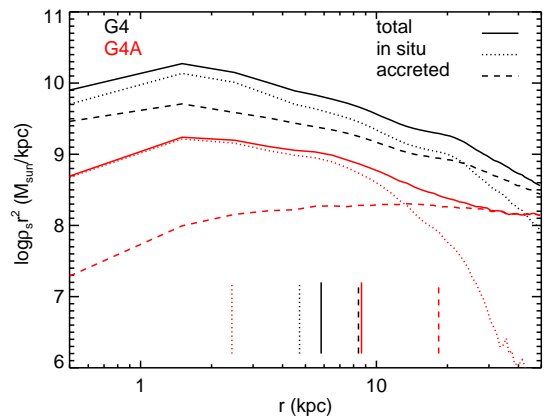}}}
  \caption{Total stellar density profiles multiplied by $r^2$ (solid lines) as a function of radius at $z=0$ for the central galaxy of G4 (black) and G4A (red), with the in situ selected stars (dotted lines) and accreted stars (dashed lines). The vertical bars indicate the position of the radius of half mass for the different components. The half-mass radius of the accreted component is larger than the half-mass radius of the in situ component. As the accreted material dominates the stellar density profiles at smaller radii when AGN feedback is turned on, the half-mass radius of the total stellar mass (in situ $+$ accreted) increases.}
    \label{fig:rhovsr_ia}
\end{figure}

As seen in the previous section (see Fig.~\ref{fig:velvsr_G1} and Fig.~\ref{fig:voversig}), the galaxies are transformed from rotationally supported (discs) into dispersion dominated (ellipticals) systems.
We will show that this is a direct consequence of the transformation of galaxies being build-up by in situ star formation (without AGN) into accretion dominated systems (with AGN), and that the transformation has important implications for the dynamical properties of these objects.
As an illustration, Fig.~\ref{fig:rhovsr_ia} shows the stellar density profiles of the central galaxy G4 and the central galaxy G4A at $z=0$, decomposed into in situ and accretion components.
The in situ component dominates the total stellar density profile in the core of the galaxy, whereas the accreted material dominates in the outskirts.
This is a natural outcome of the in situ stars forming in the depth of gravitational potentials, and the satellites being tidally stripped while falling into halos on elliptical orbits, thus depositing accreted stars far away from the centre~\citep{naabetal09, peiranietal10, hilzetal12}.

AGN feedback changes the radius at which the transition between an in situ-dominated density towards an accretion dominated density occurs.
As the proportion of accreted material is increased due to the efficient quenching of the in situ star formation by AGN feedback, the radius at which accreted material dominates is pushed towards the interior part of the halo (e.g. from 24 kpc to 14 kpc for the G4X halo at $z=0$, see Fig.~\ref{fig:rhovsr_ia}).
Note that the stellar density profiles are steeper than isothermal in both cases with a slightly flatter profile within 10 kpc for the AGN case than for the no AGN case.
The effective radius (or half-mass radius for 3-dimensional profiles) associated with the accreted material is larger than those of the in situ component.
It results that galaxies get puffed up due to the presence of AGN feedback because it enhances the fraction of accreted material by quenching the in situ star formation.
The size increase is orchestrated by the subsequent merger events and the scaling of size with mass depends also on the initial structural properties of the progenitor galaxies at higher redshift~\citep[see][]{laporteetal12, oseretal12}.

\begin{figure*}
  \centering{\resizebox*{!}{4.5cm}{\includegraphics{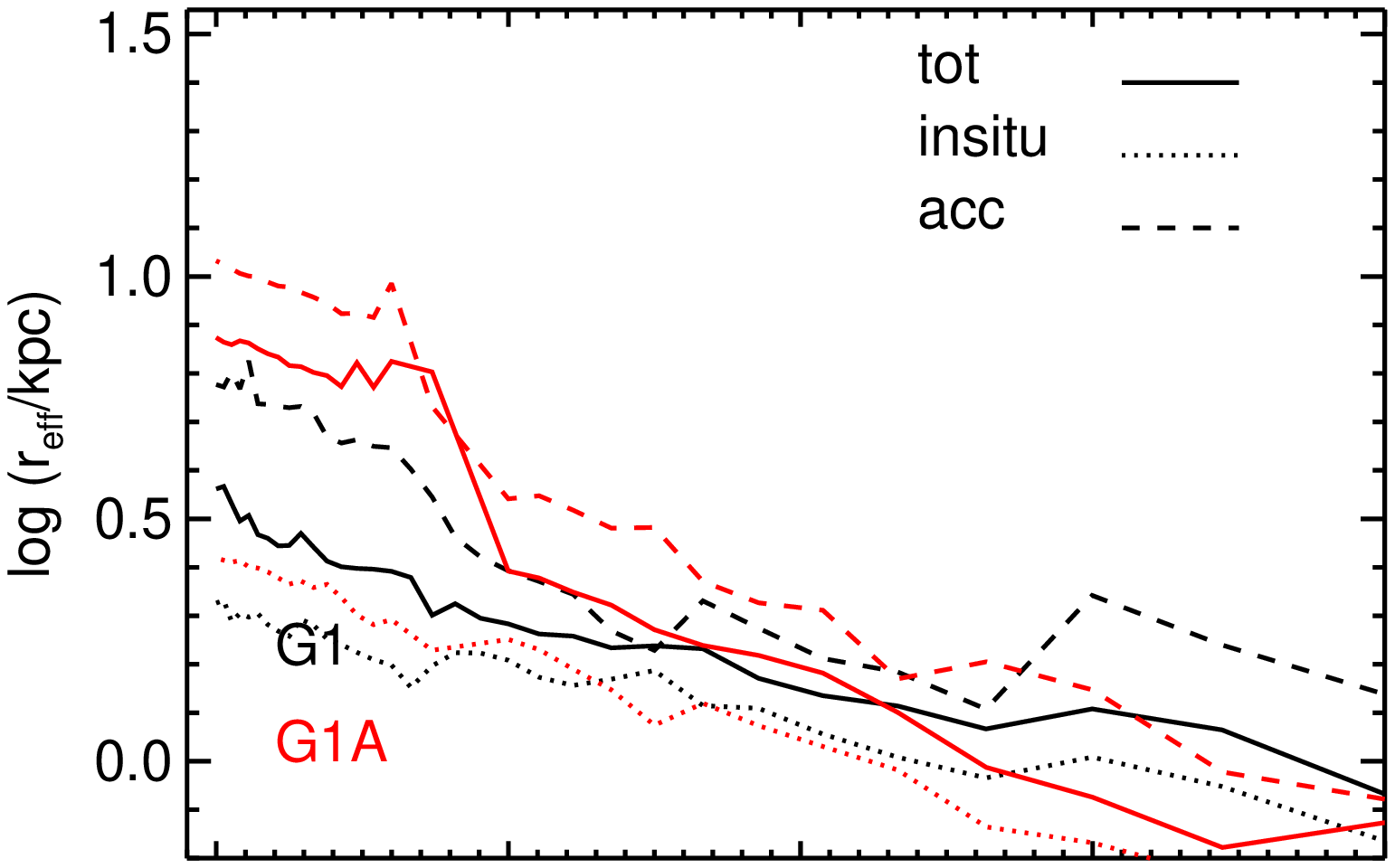}}}\hspace{-1.2cm}
  \centering{\resizebox*{!}{4.5cm}{\includegraphics{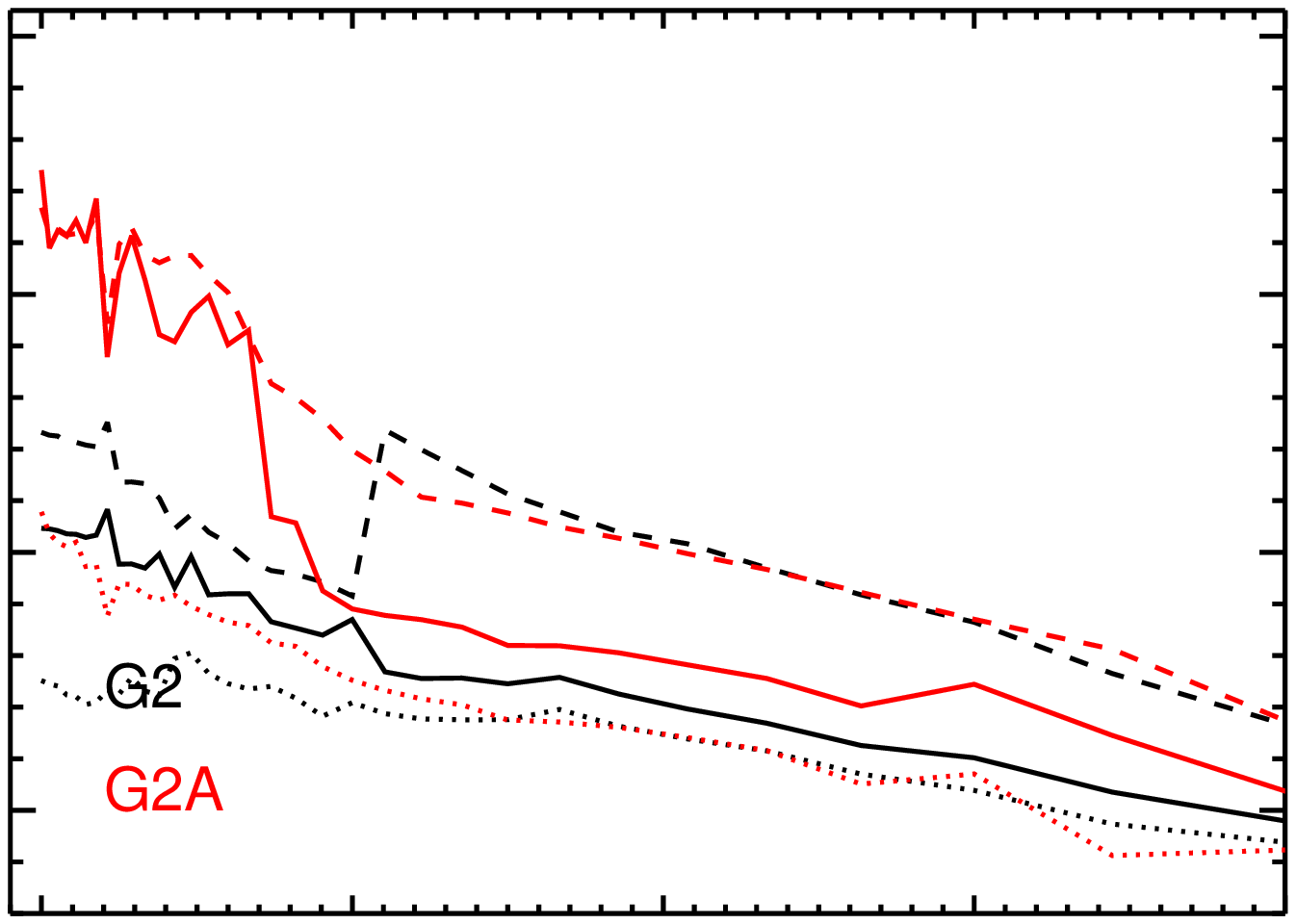}}}\hspace{-1.2cm}
  \centering{\resizebox*{!}{4.5cm}{\includegraphics{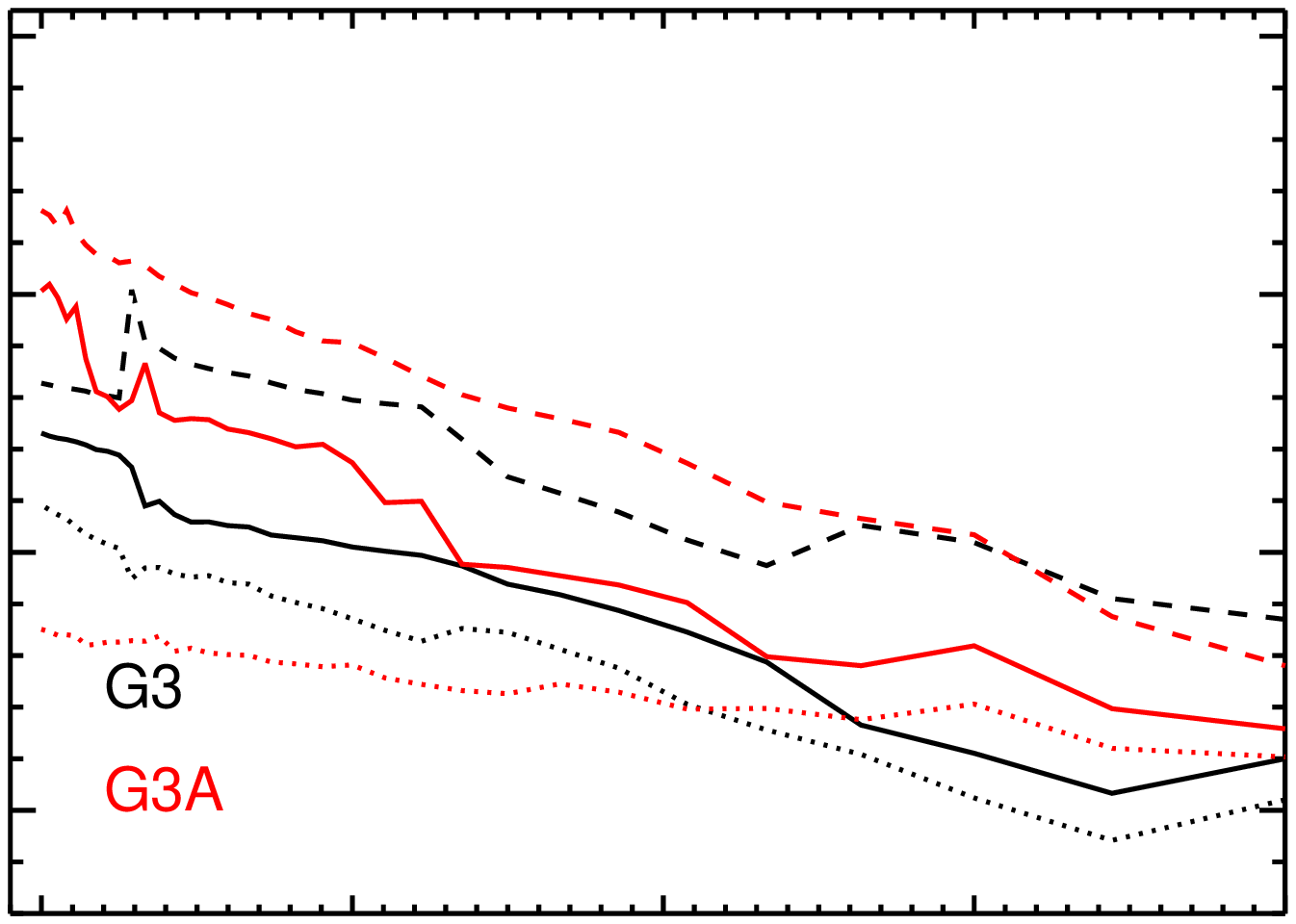}}}\vspace{-0.8cm}
  \centering{\resizebox*{!}{4.5cm}{\includegraphics{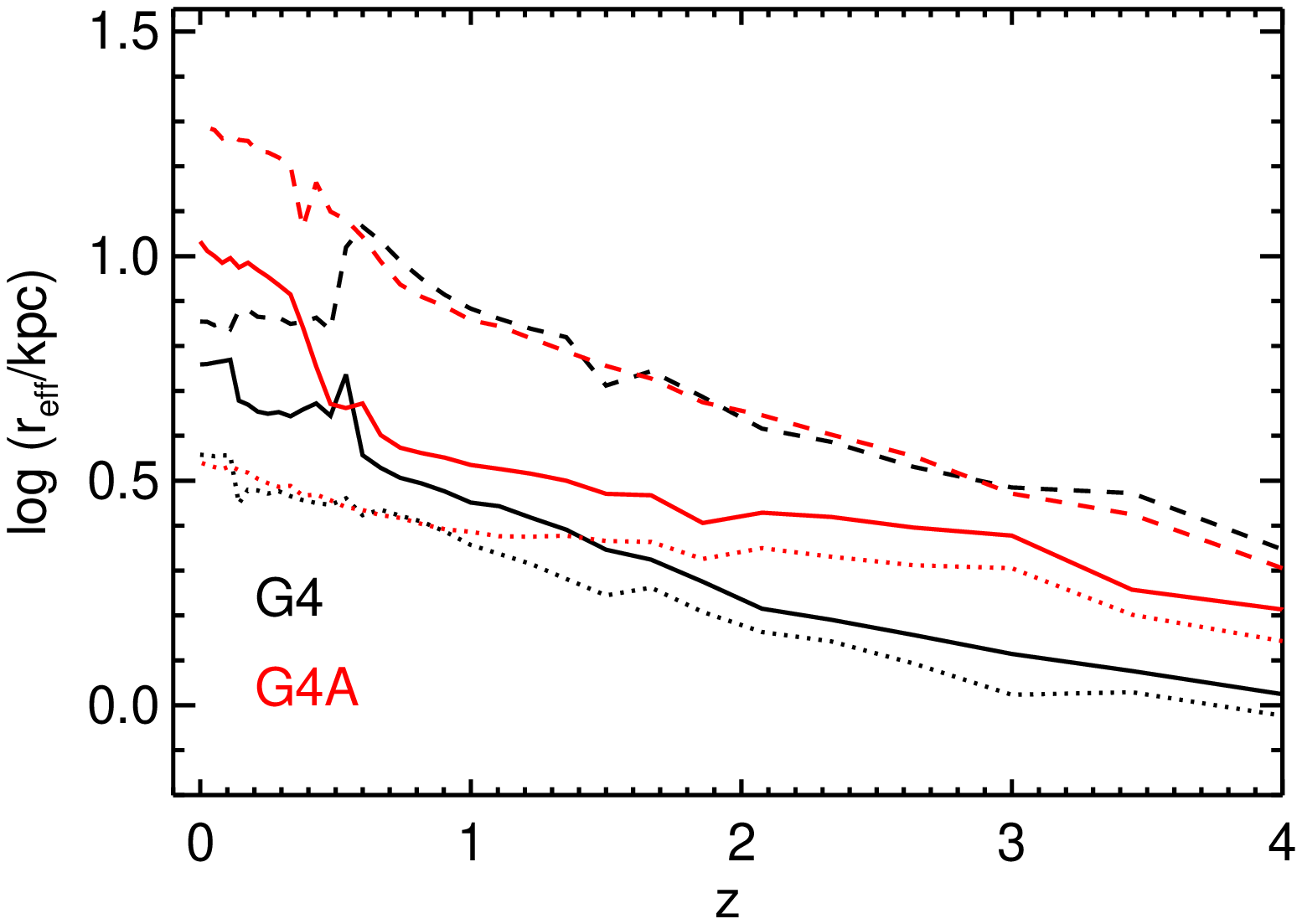}}}\hspace{-1.2cm}
  \centering{\resizebox*{!}{4.5cm}{\includegraphics{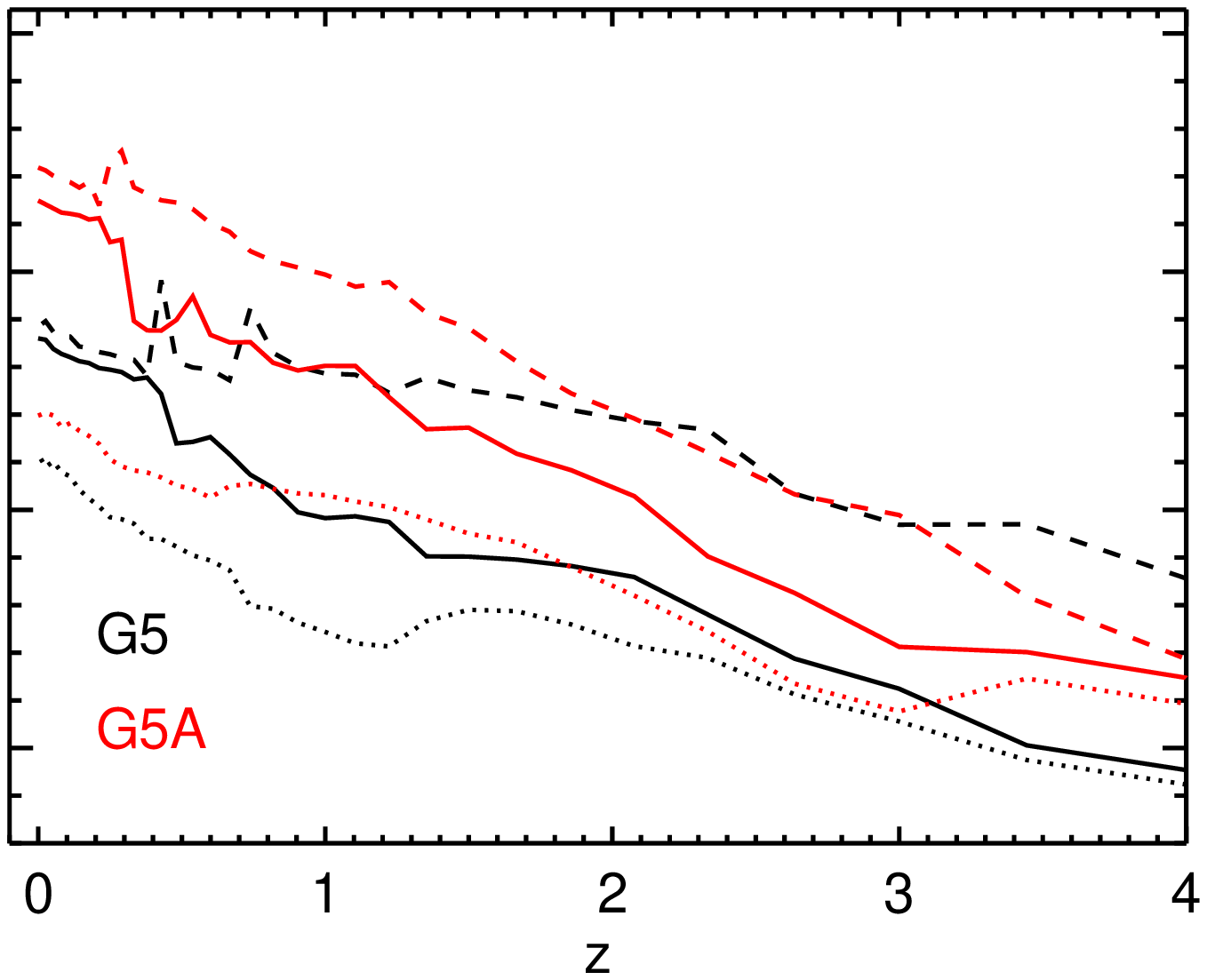}}}\hspace{-1.2cm}
  \centering{\resizebox*{!}{4.5cm}{\includegraphics{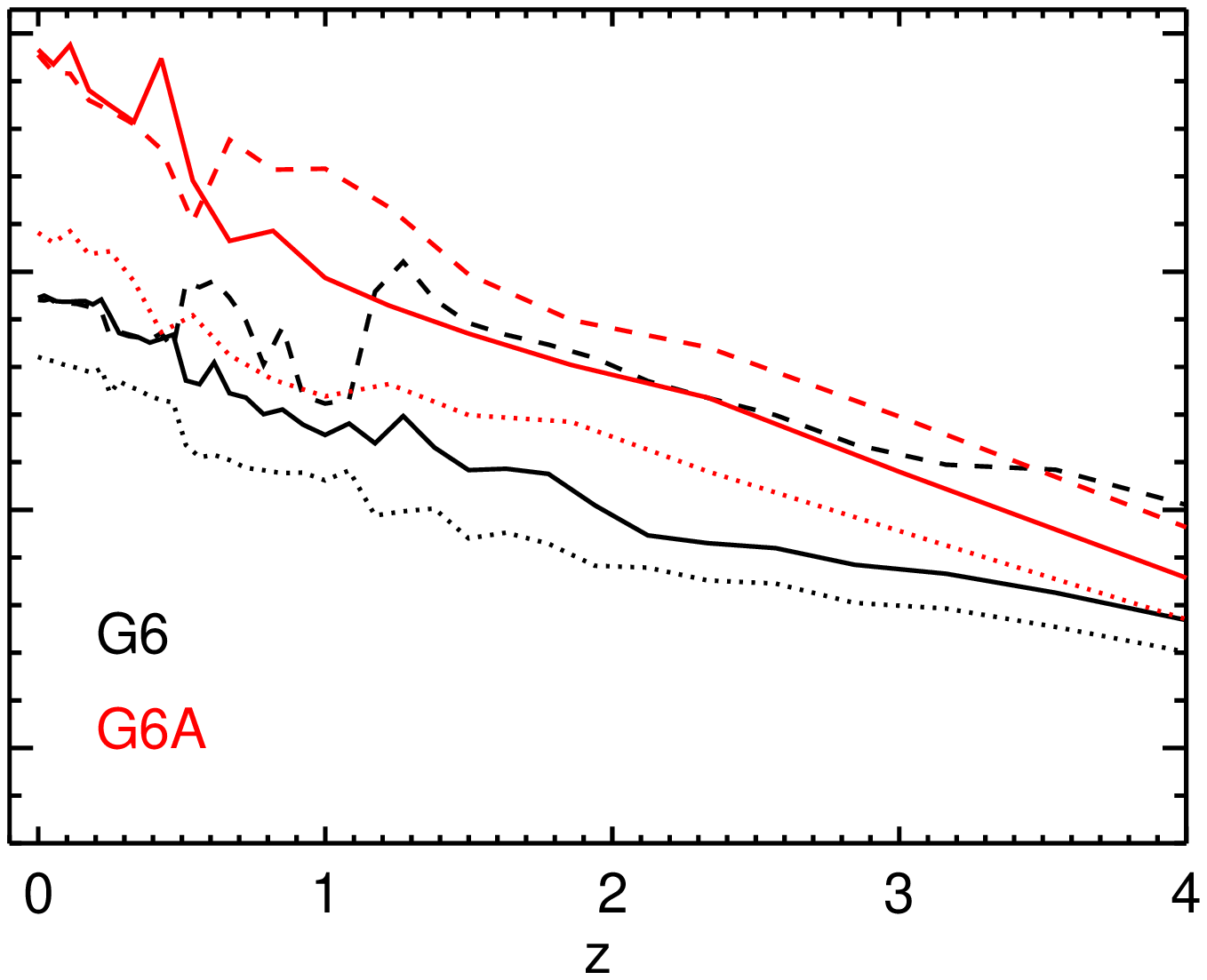}}}
  \caption{Effective radius of the central galaxy (solid), of the in situ component only (dotted) and of the accreted component only (dashed) as a function of redshift for the simulations GX without AGN feedback (black) or GXA with AGN feedback (red). All galaxies included AGN feedback show an increased value of the effective radius relative to the no AGN case.}
    \label{fig:reffinsitu}
\end{figure*}

Fig.~\ref{fig:reffinsitu} shows the evolution in redshift of the effective radius, and the in situ and accreted effective radii for all halos with or without AGN feedback.
All simulations show a larger value of their effective radii in the cases with AGN feedback, which is a direct consequence of the mechanism described above: as accretion has a more important role, galaxy sizes are increased.
This mechanism is also confirmed by the fact that high redshift effective radii are closer to their in situ effective radii, while low redshift effective radii are closer to the accreted effective radii (respectively when in situ, or accreted mass dominates, see Fig.~\ref{fig:minsituacc}).
The average logarithmic slope of the evolution of the effective radius with redshift ${\rm d}\log r_{\rm eff} /{\rm d} z$ measured between redshift 0 and 2 is equal to $-0.23$ in the case without AGN feedback, and to $-0.39$ in the case with AGN feedback.
Thus, the effective radius of galaxies simulated with AGN feedback evolves faster with time than galaxies without AGN.
This is a direct consequence of the galaxies having an increased fraction of accreted material in their mass assembly in the former case.

The effective radius of the accreted material is increased for all galaxies with the presence of AGN feedback between $0\le z \lesssim1$ (dashed lines in Fig.~\ref{fig:reffinsitu}), where the sharp transition from in situ to accreted dominated galaxies occurs (Fig.~\ref{fig:finsituave}).
The explanation is that tidal stripping becomes more efficient at removing stars from satellites.
The characteristic radius above which tidal stripping can remove material from a satellite is~\citep{king62} $r_{\rm ts}=D(0.5 M_{\rm sat}/M_{\rm s})^{0.5}$ where $D$ is the distance between the central galaxy and the satellite.
The distance at which satellites start losing stars is when $r_{\rm ts} < r_{\rm eff}$, or $D<r_{\rm eff} (2 M_{\rm s}/M_{\rm sat})^{0.5}$.
Thus, if galaxies are less compact (i.e. they have larger effective radii), satellites will be more subject to tidal stripping early on, and will lose their stars at larger distances.
This is compatible with the fact that galaxies are less compact with AGN feedback as observed from our simulations.

The behaviour of the effective radius of the in situ component together with the effect of AGN feedback shows more variations (dotted lines in Fig.~\ref{fig:reffinsitu}).
Some simulated galaxies show an increase of their in situ effective radius, while some other have a more compact in situ stellar content.
We identify two effects that can act in opposite directions.
As galaxies build from inside out in the classical picture of gas accretion, galactic discs of gas tend to be smaller due to strong AGN feedback as the accretion is quenched onto galaxies, an effect which can potentially decrease the size of the in situ component.
On the other hand, strong bursts of feedback have been identified as creating strong variations in the gravitational potential well, and can smooth the density profiles of both stars and DM in the most central parts~\citep{mashchenkoetal06, peiranietal08agn, pontzen&governato12, teyssieretal13}, in conjunction with the adiabatic expansion of DM and stars~\citep{blumenthaletal86, gnedin&zhao02, gnedinetal04} produced by the decrease of the gas density due to AGN feedback in the center of DM halos.
This effect can increase the size of galaxies and DM halo cores, but depends on the details of the accretion history onto the central galaxy~\citep{martizzietal12clumps}.
Some of the galaxies show a larger in situ effective radius in the AGN case compared to the no AGN case at high redshift where the wet and chaotic accretion takes place.
This can trigger strong bursts of AGN feedback that may be effective at flattening the density profiles by star formation quenching.
Also, we observe such flattening in the total mass density profiles (and in the DM component) and we will discuss this further in the next section by comparing our results from simulations to observations.

\begin{figure}
  \centering{\resizebox*{!}{6.cm}{\includegraphics{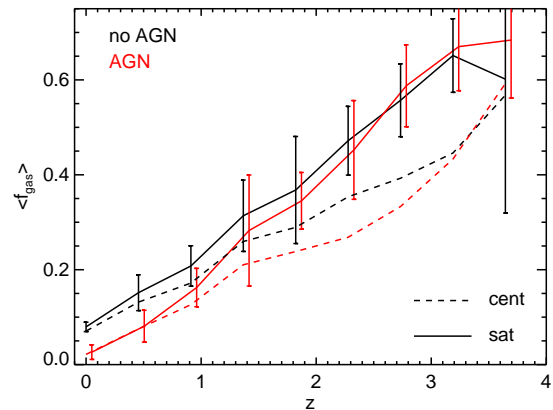}}}
  \caption{Average gas fraction as a function of redshift in the satellites (solid lines) and in the central galaxy (dashed lines) of the six simulated halos with (red) or without (black) AGN feedback, with the standard deviation (error bars). Galaxies are more gas-poor at low redshift than at high redshift and AGN feedback reduces further the fraction of cold gas.}
    \label{fig:fgas}
\end{figure}

The fraction of cold gas within galaxies $f_{\rm gas}=M_{\rm ISM}/(M_{\rm s}+M_{\rm ISM})$ (where $M_{\rm ISM}$ is the cold gas with gas density $n>n_0$ and within the galactic radius), both in central galaxy and in satellites ($M_{\rm ISM, s}=\Sigma_i M^i_{\rm ISM,s}$ summed over all satellites within the virial radius of the main halo), is declining with time with a gas fraction close to 0.1 for the no AGN case and 0.025 in the AGN case at $z=0$. 
The gas fractions are much larger for high redshift galaxies with values above 0.5, which are gas-dominated galaxies at $z=4$.
Thus, mergers between galaxies at high redshift are dissipative (wet), and are essentially dissipationless (dry) at low redshift.
The presence of AGN feedback leads to a mild variation on the gas fraction of galaxies at $z<2-3$, but even for runs including AGN feedback, galaxies above $z>3$ are still gas-dominated systems.
It confirms that AGN feedback exacerbates the effect of accretion of satellites on the effective radius at low redshift.
As AGN feedback reduces the gas fraction within galaxies at low redshift, dry mergers become more efficient at puffing up the size of galaxies.
However, at high redshift, the gas fraction between the AGN and no AGN runs are not very different, and it is the subsequent AGN activity that increases the effective radius of the, then, dominant in situ component (Fig.~\ref{fig:finsituave}), by blowing gas away from galaxies (see Fig.~\ref{fig:reffinsitu}).

\subsection{Implications for galaxy scaling relations and mass distribution}
\label{section:scalinglaws}

\begin{figure}
  \centering{\resizebox*{!}{6cm}{\includegraphics{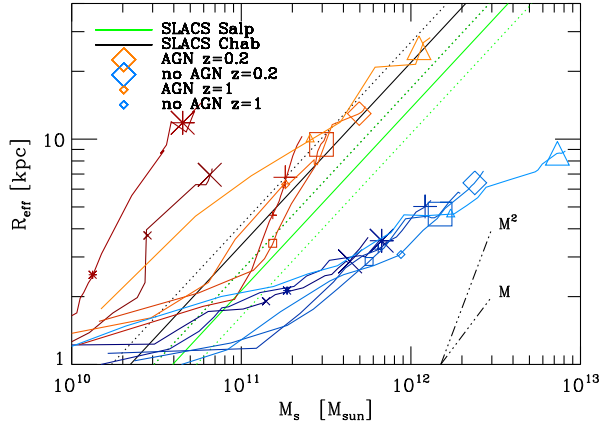}}}
  \caption{Effective radius $r_{\rm eff}$ versus stellar mass $M_{\rm s}$ for our six halos G1 (crosses), G2 (stars), G3 (pluses), G4 (squares), G5 (diamonds), and G6 (triangles) with (red) or without (blue) AGN feedback computed at different redshifts: $z=0.2$ are large symbols and $z=1$ are small symbols. The observed relations from the SLACS (at $z\sim0.2$) are overplotted as solid lines with their 2$\sigma$ standard deviation (dot) assuming either a Chabrier (black), or a Salpeter IMF (green). Galaxies simulated with AGN feedback are less compact than those simulated without AGN feedback, and are more compact at high redshift than they are at low redshift.}
    \label{fig:reffvsmstar}
\end{figure}

The scaling relations between mass and kinematics can put strong constraints on the modelling of the population of massive galaxies. 
In Fig.~\ref{fig:reffvsmstar}, we show the relations between the effective radius $r_{\rm eff}$ and the stellar mass of the central galaxy at different redshifts.
Independently of the adopted IMF, all galaxies simulated without AGN feedback lie below the $r_{\rm eff}$-$M_{\rm s}$ observational relation at $z=0.2$ from the SLACS data~\citep{augeretal10} that consists of ETGs within the mass range $10^{11}\lesssim M_{\rm s} \lesssim 6\times 10^{11}\, \rm M_\odot$.
These galaxies, suffering from over-cooling and being too massive, are also too compact.
This increased compactness is expected if the amount of in situ star formation is too large, as the in situ formed stars have small effective radius compared to the effective radius of the accreted stellar distribution (Fig.~\ref{fig:reffinsitu}).

The presence of AGN feedback increases the effective radius, while  at the same time it reduces  stellar masses and greatly improves the agreement between the simulations and observations for the four most massive galaxies (G3A, G4A, G5A, G6A).
The simulation points with AGN feedback stand close to the observational result for the four most massive galaxies at $z=0.2$, within the 2$\sigma$ error bars (for a Chabrier IMF).
If a Salpeter IMF is assumed, the AGN size-mass points of the four most massive galaxies at $z=0.2$ stand within the $5 \sigma$ relation.
The two least massive galaxies are largely above the observational relation, which makes these galaxies too puffy.
Such galaxies exist but are rare objects in the real Universe~\citep{williamsetal10}.
It is difficult to explain this particular behaviour, but one should note as they are the lightest galaxies, they are also the galaxies that are the most affected by the finite resolution.
At redshift $z=2$, the effective radius of these galaxies is very close to the resolution limit of the grid ($r_{\rm eff}\simeq2$ kpc against a $\Delta x=0.5$ kpc resolution), and the galaxy gets  extra support against gravity from the finite cell size.
At even higher redshifts, this effect is true for all simulated galaxies as they are smaller, and could have some non-negligible effects on the final size-mass relation at $z=0$.

Galaxies are evolving in size with time while they accumulate stars (Fig.~\ref{fig:reffinsitu}), and there is a small variation of the size-mass relation with redshift for the AGN feedback simulations, with galaxies being more compact at high redshift (i.e. the distribution of $r_{\rm eff}$ versus $M_{\rm s}$ at high redshift is below this relation at low redshift, Fig.~\ref{fig:reffvsmstar}).
Observations suggest such an evolution with time in the compactness of galaxies~\citep{trujiloetal07, cimattietal08, saraccoetal09, vandokkumetal10, huertasetal13}, and the effective radius is of the order of $\sim 1.5-2$ smaller at redshift 1 than at redshift 0 at constant mass for ETGs~\citep{huertasetal13}.
On the opposite, simulations points without AGN feedback show the opposite trend: galaxies tend to be less compact at high redshift.
From virial theorem arguments, it can be shown that galaxy growth driven by minor mergers should follow $r_{\rm eff}\propto M_{\rm s}^2$ and $r_{\rm eff}\propto M_{\rm s}$ through major mergers~\citep[see][]{hilzetal12}. For the AGN runs, mergers drive the size growth of the central galaxies, both minor and major. This leads to an intermediate relation $M_{\rm s} -M_{\rm s}^2$. However for the no AGN simulations, we clearly see that the size growth follows a shallower change with mass. This can be explained by the fact that these galaxies are dominated by the in situ star formation that concentrates stars in the innermost parts of the galaxies where the cold gas is deposited.
Note that there is likely an effect of limited resolution on the size evolution, that more importantly impacts the low-mass galaxies in their early stages (see the flattening of the curves at $z>1$ in Fig.~\ref{fig:reffvsmstar}). 
Because of additional numerical support we can expect their size to be artificially increased when they are close to the resolution limit.

\begin{figure}
  \centering{\resizebox*{!}{6cm}{\includegraphics{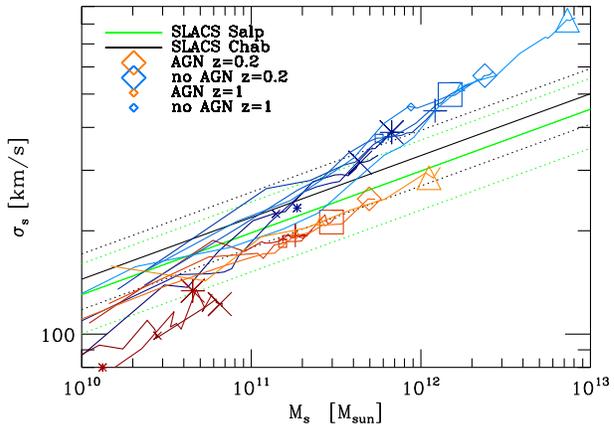}}}
  \caption{Same as Fig.~\ref{fig:reffvsmstar} for the velocity dispersion of stars $\sigma_{\rm s}$ within $r_{\rm eff}/2$ versus stellar mass $M_{\rm s}$. Galaxies simulated with AGN feedback have lower velocity dispersion values than for galaxies simulated without AGN feedback.}
    \label{fig:sigmavsmstar}
\end{figure}

\begin{figure}
  \centering{\resizebox*{!}{6cm}{\includegraphics{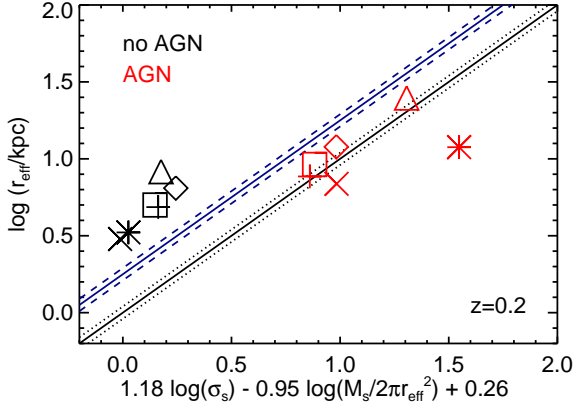}}}
  \caption{Stellar mass $M_{\rm s}$-plane at $z=0.2$ for galaxies with AGN feedback (red) and galaxies without AGN feedback (black). The observed relations from the SLACS (at $z\sim0.2$) are overplotted as solid lines with their 2$\sigma$ standard deviation (dot) assuming either a Chabrier (black), or a Salpeter IMF (blue). Galaxies simulated without AGN feedback lie above the $M_{\rm s}$-plane, while galaxies with AGN feedback are closer to $M_{\rm s}$-plane (favourising a Chabrier IMF).}
    \label{fig:mstarpplane}
\end{figure}

\begin{figure}
  \centering{\resizebox*{!}{6cm}{\includegraphics{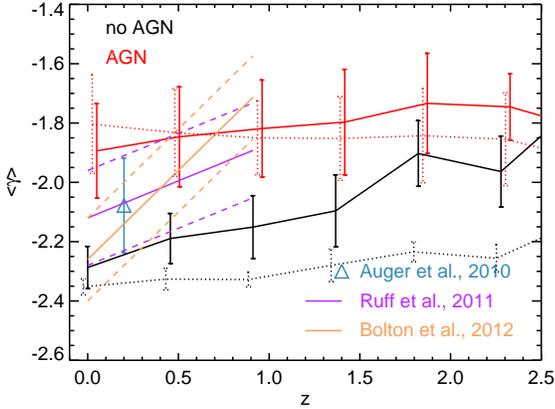}}}
  \caption{Average slope of the total mass density profile as a function of redshift for the six halos with (red) or without (black) AGN feedback. The slope are measured between $r_{\rm eff}/2$ and $r_{\rm eff}$ on the average mass density within radius $r$ (solid lines) similar to observations, or between $2 \Delta x$ and $0.1 r_{\rm vir}$ (dotted lines). Error bars correspond to the standard deviation. Observations from the SLACS sample~\citep{augeretal10} at $z=0.2$ are plotted in cyan. The observational trend of the slope with redshift found by~\citet{ruffetal11} and~\citet{boltonetal12} are plotted in violet and orange respectively with their intrinsic scatter. AGN feedback flattens the total density profiles. AGN feedback produces shallower total density profiles within galaxies.}
    \label{fig:slopeevol}
\end{figure}

\begin{figure}
  \centering{\resizebox*{!}{6cm}{\includegraphics{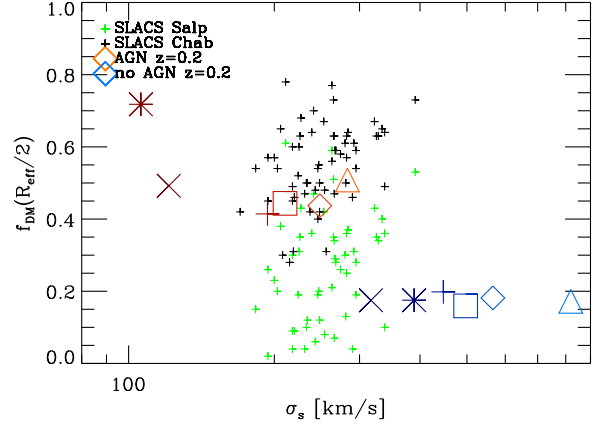}}}
  \caption{Same as Fig.~\ref{fig:reffvsmstar} for the DM fraction $f_{\rm DM}$ within $r_{\rm eff}/2$ versus the velocity dispersion of stars $\sigma_{\rm s}$. Galaxies simulated with AGN feedback have larger DM fraction within $r_{\rm eff}/2$ than galaxies simulated without AGN feedback.}
    \label{fig:fdm}
\end{figure}

We also measured the projected velocity dispersion of stars $\sigma_{\rm s}$ within $r_{\rm eff}/2$ relative to their galaxy mass $M_{\rm s}$ (Fig.~\ref{fig:sigmavsmstar}).
The galaxies without AGN feedback show large values of $\sigma_{\rm s}$.
This is the direct consequence of the size-mass relation: more compact galaxies need more dynamical support (velocity dispersion).
At $z=0.2$ the four most massive galaxies without AGN lie above 2$\sigma$ of the observational relation, while the two least massive lie within the relation (for a Chabrier IMF).
As can be predicted from the size-mass relation, the presence of AGN feedback produces more extended galaxies and, thus, galaxies with lower values of stellar velocity dispersion.
The four most massive galaxies with AGN have values of $\sigma_{\rm s}$-$M_{\rm s}$ in better agreement with observations than in the absence of AGN.
As for the size-mass relation, the $\sigma_{\rm s}$-$M_{\rm s}$ points for the two least massive galaxies are at more than 2$\sigma$  away from the observational relation at $z=0.2$ if AGN feedback is incorporated.
These low mass ETGs are above the size-mass observational fit and since they  need less kinematic support because they are too extended, they lie below the velocity dispersion-mass observational law.

An even more stringent observational constraint to reproduce is the fundamental plane of ETGs.
Here, we measure the mass equivalent, namely the $M_{\rm s}$-plane~\citep{boltonetal07, boltonetal08, augeretal10}, that links the galaxy size to the galaxy mass and velocity dispersion (Fig.~\ref{fig:mstarpplane}).
All galaxies simulated without AGN feedback stand above the $M_{\rm s}$-plane.
For the AGN feedback simulations, we find that the four most massive galaxies stand closer to the $M_{\rm s}$-plane than in the case without AGN feedback and they favour a Chabrier IMF.
For the two least massive galaxies including AGN, they stand below the $M_{\rm s}$-plane, with the G2A galaxy being at more than 10$\sigma$ away.

The slope of the total mass density profile measured by strong lensing provides a value close to isothermal $\gamma=-2.078$ ($\rho_{\rm tot}\propto r^{\gamma}$) at $z=0.2$ for ETGs~\citep{koopmansetal09, augeretal10, barnabeetal11}.
We measure the slope in our simulations for the average total density within a radius $r$ such that $\rho_{\rm tot}=M_{\rm tot}(<r)/(4/3\pi r^3)$.
We use two points to measure the average slope, at $r_{\rm eff}/2$ and $r_{\rm eff}$.
The first striking result is that the measured slopes are quite similar to runs with or without AGN, and are close to isothermal (see Fig.~\ref{fig:slopeevol}).
At $z=0$, galaxies without AGN feedback show an average slope close to $\gamma=-2.29\pm 0.07$, while for the galaxies with AGN feedback profiles are flatter with $\gamma=-1.89\pm 0.16$.
Note that the measured scatter in the AGN feedback case is in good agreement with that found in the observations.
As seen in Fig.~\ref{fig:rhovsr_ia}, stellar density profiles are steeper than isothermal for the G4 and G4A halos (but this is true for all zoomed halos). 
Thus, DM populates preferentially the outer part of the galaxy, so that the total density profiles flatten up to $\gamma$ above $-2$ for the AGN case.
The difference in the value of the slope already happens at $z=2$\footnote{Above this redshift the value of $r_{\rm eff}/2$ is too close to the resolution limit to be conclusive for the mass density slope.}.
Thus, AGN feedback at $z=2$ has been able to  significantly change the total distribution of matter.
Note that the observational points stand between the values of the runs with and without AGN feedback, and that the error bars are compatible with both values.
We also measure the slope between $2\Delta x$ and $0.1 r_{\rm vir}$, two points that are not affected by the presence of feedback, to clearly assess that the change in the slope is real and not just an effect of changing $r_{\rm eff}$ through feedback.
The slope within $0.1 r_{\rm vir}$ show the same features than the slope measured at $r_{\rm eff}$: AGN feedback flattens the total mass density profile.

Combining strong lensing data at higher redshift,~\cite{ruffetal11} and~\cite{boltonetal12} provide an estimation of the variation of the slope $\gamma$ with redshift.
They find that the slope is steeper at low redshift than it is at high redshift, with a variation of $0.25$ in~\cite{ruffetal11} and $0.60$ in~\cite{boltonetal12} with redshift.
Our simulations show such a mild evolution in the slope of the total density profile with redshift, and our measured averaged slope at high redshift $z\simeq0.5-1$ is in better agreement with observations for the AGN feedback runs.
Cosmological simulations from~\cite{remusetal13} show that massive ellipticals evolve towards profiles with $\gamma=-2$ slope at $z=0$ due to a combination of dry minor and major mergers, however they find a steeper profile at high redshift which is at odds with our findings and with the observations.

Likewise, when comparing in Fig.~\ref{fig:fdm}, the projected DM fraction within half the effective radius at $z\sim0.2$ in the SLACS data \citep{gavazzietal07,augeretal10} and in our simulations, we find a better agreement for the simulation with AGN feedback turned on ($f_{\rm dm} \sim 0.5$) than without ($f_{\rm dm} \sim 0.2$), mostly because the stellar mass (or equivalently the velocity dispersion) without AGN feedback is too large. In the former case, the dark matter fraction is slightly more consistent with observations assuming a Chabrier IMF.

\section{Comparison with previous work}
\label{section:comparison}

A key point of our work is that, in the absence of AGN feedback, the galaxies always exhibit a late-type morphology with a large disc of stars as can be observed in the  synthetic images in Fig.~\ref{fig:syntheticoptical}.
We find similar results to~\cite{lackneretal12} who use the {\sc enzo} (AMR code) to simulate some overdense and underdense regions of the universe: their stellar conversion efficiency is high $f_{\rm conv}=0.6$, and their most massive galaxies $M_{\rm s}> 10^{12}\, \rm M_\odot$ are dominated by  in situ star formation $f_{\rm insitu}=60$ per cent at $z=0$.
They use similar physics (gas cooling, low efficiency star formation, supernovae feedback and metal cooling) as for our simulations without AGN feedback.
In the work from~\cite{oseretal10} \citep[see also][]{naabetal07, naabetal09, oseretal12, johanssonetal12}, they employ the {\sc gadget} code~\citep{springel05} based on SPH, to study the formation of elliptical galaxies in group environments.
The same basic physical ingredients as for our simulations without AGN, and in~\cite{lackneretal12} are used in their work, except they use a larger value of the star formation efficiency $\epsilon_*\simeq 0.1$.
However the results of~\cite{oseretal10} are radically different from ours (and thus from~\citealp{lackneretal12}): they naturally obtain elliptical galaxies with low stellar conversion efficiency $0.1<f_{\rm conv}<0.45$ respectively for high mass and low mass halos in the mass range $10^{12} <M_{\rm h}<2.5\times 10^{13}\, \rm M_\odot$, and the growth of galaxies is largely dominated by the accretion of satellites with $f_{\rm in situ}=20$ per cent at $z=0$ for the halo mass range corresponding to ours.
Note that, in SPH simulations of $M_{\rm h}=3\times 10^{12} \,  \rm M_\odot$ halo by~\cite{khalatyanetal08} using SPH, they find little difference in their galaxy mass and morphology between the AGN case and the no-AGN case at $z=0$. The galaxy is more red with AGN feedback, and has a larger value of the Sersic index, but the galaxy is an elliptical in both cases and has the same low stellar conversion efficiency $f_{\rm conv}\simeq0.2$ at $z=0$.

There are several potential interpretations that can explain this difference in the results.
The first one is that the high star formation efficiency adopted can account for the difference seen in the accreted versus in situ star formation.
Higher efficiencies tend to form stars at higher redshift by rapidly converting gas into stars, thus, the fraction of accreted mass should increase at low redshift.
We make that simple test by running the G1 and G2 simulations with a $\epsilon_*=0.1$ efficiency (still without AGN feedback), and we found that the accreted fraction of the stellar mass has increased.
In line with the analysis performed by the decomposition of in situ versus accreted origin of stars, we indeed find that the galaxy becomes more spherical (supported by the velocity dispersion of stars as opposed to rotation).
However, the stellar mass obtained at $z=0$ is the same ($M_{\rm s}=5.4\times 10^{11}\rm \, M_\odot$, $f_{\rm conv}=0.76$ for G1; and $M_{\rm s}=7.0\times 10^{11}\rm \, M_\odot$, $f_{\rm conv}=0.66$ for G2) as for the low star formation efficiency case ($M_{\rm s}=5.6\times 10^{11}\rm \, M_\odot$, $f_{\rm conv}=0.79$ for G1; and $M_{\rm s}=7.3\times 10^{11}\rm \, M_\odot$, $f_{\rm conv}=0.70$ for G2), and cannot account for the difference observed with~\cite{oseretal10} ($f_{\rm conv}\simeq0.2$ for halo masses $M_{\rm h}\simeq 6\times 10^{12}\, \rm M_\odot$).
We observe that both DM mass resolution, and spatial resolution are of comparable values in~\cite{oseretal10} and here, even though it is difficult to compare a smoothing length with a minimum cell size.

Another much more worrying aspect is the different behavior of AMR and SPH codes as already underlined by~\cite{scannapiecoetal12} for Milky-Way halos, and then by~\cite{vogelsbergeretal12} who compared results from the {\sc arepo} code (moving mesh~\citealp{springel10}) with {\sc gadget} in a statistical sense.
With relatively similar physics, AMR codes (or moving mesh codes) tend to naturally produce overly massive disc-like galaxies, while SPH codes predict less massive galaxies that have already early-type shapes.
This is in line with our findings: the ``usual'' modelling of galaxy formation (i.e. not accounting for AGN feedback) with the {\sc ramses} code leads to overly massive disc-like galaxies even in the centres of massive groups of galaxies.
This is potentially driven by a more accurate treatment of the gas instabilities in grid-based codes as demonstrated by~\cite{agertzetal07}.
It can have critical consequences on the amount of gas mixing between satellites (cold phase) and the hot ambient medium in massive halos which can lead to more efficient gas cooling~\citep{sijackietal12}.
Moreover, the treatment of  blast wave explosions (SN or AGN) with thin discontinuities can have problematic behaviors if the adaptive time-stepping becomes too aggressive~\citep{durier&dallavecchia12}.

A careful inspection of the numerical treatment of the hydrodynamics and its consequences for the formation of massive galaxies remains to be conducted.
This is especially important as it can lead to an underestimate of the role played by AGN feedback in shaping the masses, colours, and morphologies of massive ETGs.
We reserve this numerical comparison for future work. 

Using a different technique, semi-analytical models of galaxy formation have also produced predictions for the amount of in situ versus accreted stellar material in massive galaxies. \cite{hirschmannetal12} show that AGN feedback plays an important role to reduce the fraction of in situ formed stars down to $f_{\rm insitu}\simeq0.5$ at $z=0$ for there most massive galaxies equivalent to the mass range probed in this work, while this fraction goes up to $f_{\rm insitu}\simeq0.65-0.75$ in the absence of AGN feedback. \cite{lee&yi13} find even lower values of $f_{\rm insitu}\simeq0.2$ using their semi-analytical model with AGN feedback, while they obtain $f_{\rm insitu}\simeq0.7$ if no AGN feedback is considered (private communication with the authors).
In both cases, the absence of AGN feedback leads to in situ-dominated galaxies with levels similar to what we find in our hydrodynamical simulations~\citep[as in][]{lackneretal12}. 
However, they find different values of the in situ fraction of stars for their semi-analytical models including AGN feedback, possibly because of the different implementations of AGN feedback.

\section{Conclusion and Discussion}
\label{section:conclusion}

By means of high resolution hydrodynamical cosmological simulations, we have studied the impact of AGN feedback on the formation and evolution of massive galaxies in the centre of groups of galaxies.
The main finding of this work is that we have demonstrated that AGN feedback can turn blue massive late-type disc galaxies into red early-type galaxies.
This process happens because AGN feedback reduces the in situ star formation in the central most massive galaxy, and turns it into a stellar accretion-dominated system.
In more detail, we find that:
\begin{itemize}
\item{The conversion efficiency of baryons into stars in the central galaxy is reduced by a factor 7 from $f_{\rm conv}=0.7$ to $f_{\rm conv}=0.1$ at $z=0$ due to AGN feedback.}
\item{Galaxies are intrinsically redder with AGN feedback than without, with, respectively, $g-r=0.65$ and $g-r=0.45$.}
\item{The DM mass fraction is increased (from $\langle f_{\rm DM}\rangle=0.53$ to $\langle f_{\rm DM} \rangle=0.83$) with AGN at $z=0$, and stellar (from $\langle f_{\rm s}\rangle= 0.42$ to $\langle f_{\rm s} \rangle= 0.15$) and gas (from $\langle f_{\rm g} \rangle= 0.04$ to $\langle f_{\rm g} \rangle= 0.016$) fractions are reduced within the core of halos (in $0.1 r_{\rm vir}$) at $z=0$ by the presence of AGN feedback.}
\item{Central galaxies are rotation-dominated (discs with $v/\sigma>1$) without AGN feedback, and dispersion dominated (ellipticals with $v/\sigma<1$) with AGN feedback.}
\item{The fraction of stellar mass in the halo locked into satellites is larger if AGN feedback is present. It leads to the formation of central ETGs dominated by accretion of stars rather than by in situ star formation at $z=0$.}
\item{Galaxies are dominated by in situ star formation at high redshift because they are gas-rich systems that rapidly convert gas into stars. As their reservoir of gas decreases with time and the star formation rate is quenched, the proportion of accreted mass increases. }
\item{Accreted stars are deposited far away from the central galaxy, while in situ formed stars are close to the centre of the halo. As a consequence, the effective radius of the accreted distribution of stars is larger than that of the in situ component.}
\item{Because of the increased fraction of accreted material in the AGN simulations and the enhanced dissipationless nature of low-redshift mergers, the effective radius of central galaxies is larger, and galaxies are less compact. }
\item{The presence of AGN feedback provides a better fit to the observed dynamical scaling relations (size-mass, velocity-mass, $M_{\rm s}$-plane) than without AGN feedback. }
\item{With AGN feedback, galaxies are smaller at high redshift at constant mass, while they are larger without AGN feedback.}
\item{The total mass density profiles in galaxies are affected by AGN feedback, and they get shallower.}
\end{itemize}

We note finally that the conclusions of this work could be mitigated by some possible artefacts of our methodology.
We employ a finite cell size resolution of 0.5 kpc, which can possibly affect the compactness of the simulated galaxies in particular at high redshift, where their sizes approaches the resolution limit.
The stellar-halo mass relation at $z=0$ is compatible with a Salpeter IMF for the most massive galaxies and with a Kroupa (or Chabrier) IMF for the least massive galaxies.
In contrast, the size-mass relation favours a Kroupa IMF for the most massive galaxies, while the two least massive ones are excluded from the observational constraints (Kroupa or Salpeter). 
This leads to an inconsistency that can be the outcome of a lack of resolution, or wrong/missing modelling of the galactic physics, that remains to be explored further.
As the multiphase structure of the gas is not resolved, the formation of massive clumps of gas produced by the disc instabilities at high redshift is not triggered~\citep{agertzetal09, dekeletal09clumps} and any morphological quenching associated with this phenomenon is missed~\citep{martigetal09}.
However, the increase in resolution is not enough to prevent the galaxies from being too massive and strong feedback processes are still required~\citep{duboisetal12agnhighz}.
Also, because of finite computational time, we limited our sample to six objects over one dynamical range of halo mass.
More halos spanning a broader mass range have to be simulated to obtain  better statistical significance on our preliminary theoretical predictions both in terms of mean average trends and intrinsic scatter.
We stress that, without AGN feedback, {\em none} of the observational constraints can be reproduced regardless of the choice of the IMF.

As already highlighted by many authors~\citep[e.g.][ and references therin]{khochfar&silk06, boylankolcinetal06, malleretal06, naabetal06, naabetal07, delucia&blaizot07, bournaudetal07, guo&white08, hopkinsetal09, nipotietal09, feldmannetal10, shankaretal13}, the role of the mergers of central galaxies with their satellites, and particularly of dry mergers, is crucial to understanding the size and velocity dispersion of galaxies and their evolution with time.
We intentionally did not discuss in details the singular role of minor versus major mergers, or wet versus dry mergers, but instead we emphasised the particular effects of AGN feedback in transforming in-situ dominated galaxies into accretion-dominated systems.
However, we note that dry mergers become increasingly more dominant at low redshift, driving the low redshift evolution of galaxy sizes, and that AGN feedback enhances this effect further.
In future work we will perform a dedicated study of these effects by employing a more statistically significant sample of galaxies.

Finally, when comparing our results, without AGN feedback, to the existing literature employing hydrodynamical cosmological simulations and the formation of ETGs (mostly without using AGN feedback), we found our results to be consistent with galaxies simulated with an AMR technique (blue massive discs dominated by in situ star formation), and inconsistent with those obtained with SPH (red ellipticals dominated by accretion).
This conflict  calls for a thorough analysis of the respective merits of the different numerical techniques and their consequences for the formation and evolution of massive galaxies. 

\section*{Acknowledgments}
We are grateful to Julien Devriendt, Christophe Pichon, Francesco Shankar, Romain Teyssier and Tommaso Treu for stimulating discussions.
We thank the anonymous referee for his useful comments that improved the quality of the paper.
The simulations presented here were run on the DiRAC facility jointly funded by STFC, the Large Facilities Capital Fund of BIS and the University of Oxford. 
This research is part of the Horizon-UK project. 
This research was supported in part by ERC project  267117 (DARK) hosted by  Universit\'e Pierre et  Marie Curie - Paris 6.
RG acknowledges the CNES for its financial support and the PNCG for support in the organisation of a workshop on ETGs in Paris, July 2012. 
YD acknowledges Fabio Acero and St\'ephanie Gain for the hospitality provided during the redaction of this paper.

\bibliographystyle{mn2e}
\bibliography{author}

\label{lastpage}

\appendix

\end{document}